\documentclass[useAMS,usenatbib]{mn2e}
\usepackage{hyperref}
\usepackage{times}
\usepackage{natbib}
\usepackage{aas_macros}
\usepackage{amsmath}
\usepackage{amssymb}
\usepackage{rotating}
\usepackage[labelfont=bf]{caption}
\usepackage{supertabular}
\usepackage{enumitem}
\usepackage{color}

\newcommand{\ergs}{$\mathrm{erg\;s^{-1}}$}

\newcommand{\msun}{\mathcal{M}_\odot}
\newcommand{\Msun}{$\msun$}
\newcommand{\mstel}{\mathcal{M}_*}
\newcommand{\Mstel}{$\mstel$}
\newcommand{\lx}{L_\mathrm{X}}
\newcommand{\LX}{$\lx$}
\newcommand{\lmode}{L_\mathrm{X}^\mathrm{mode}}
\newcommand{\Lmode}{$\lmode$}
\newcommand{\giv}{\;|\;}
\newcommand{\sfrsed}{\mathrm{SFR_{SED}}}
\newcommand{\sfruvir}{\mathrm{SFR_{UV+IR}}}
\newcommand{\SFRSED}{$\sfrsed$}
\newcommand{\SFRUVIR}{$\sfruvir$}
\newcommand{\Plx}{$p(\log \lx \giv \mstel,z)$}
\newcommand{\newt}[1]{{#1}}
\newcommand{\more}[1]{{#1}}
\newcommand{\refresp}[1]{{#1}}

\begin{document}

\title[The main sequence of star formation in X-rays]{X-rays across the galaxy population I: tracing the main sequence of star formation}
\author[J. Aird et al.]{J. Aird$^{1}$\thanks{jaird@ast.cam.ac.uk}, A. L. Coil$^2$ and A. Georgakakis$^{3,4}$\\
$^1$Institute of Astronomy,
University of Cambridge,
Madingley Road,
Cambridge,
CB3 0HA\\
$^2$Center for Astrophysics and Space Sciences (CASS), Department of Physics, University of California, San Diego, CA 92093, USA\\
$^3$Max Planck Institute f\"{u}r Extraterrestrische Physik, Giessenbachstrasse, 85748 Garching, Germany\\
$^4$IAASARS, National Observatory of Athens, GR-15236 Penteli, Greece}
\pagerange{\pageref{firstpage}--\pageref{lastpage}} \pubyear{2016}

\date{Accepted 2016 Nov 10. Received 2016 Oct 19; in original form 2016 July 01.}

\maketitle
\label{firstpage}
\begin{abstract}
We use deep \textit{Chandra} imaging to measure the distribution of X-ray luminosities ($L_\mathrm{X}$) for samples of star-forming galaxies 
as a function of stellar mass and redshift, using a Bayesian method to push below the nominal X-ray detection limits.
Our luminosity distributions all show narrow peaks at $L_\mathrm{X}\lesssim 10^{42}$~erg~s$^{-1}$ that we associate with star formation, as opposed to AGN that are traced by a broad tail to higher $L_\mathrm{X}$. Tracking the luminosity of these peaks as a function of stellar mass reveals an ``X-ray main sequence" with a constant slope $\approx0.63\pm0.03$ over $8.5 \lesssim \log \mathcal{M}_*/\mathcal{M}_\odot \lesssim 11.5$ and $0.1 \lesssim z \lesssim 4$, with a normalization that increases with redshift as $(1+z)^{3.79\pm0.12}$. 
We also compare the peak X-ray luminosities with UV-to-IR tracers of star formation rates (SFRs) to calibrate the scaling between $L_\mathrm{X}$ and SFR. 
We find that \mbox{$L_\mathrm{X} \propto$ SFR$^{0.83} \times (1+z)^{1.3}$}, 
where the redshift evolution and non-linearity likely reflect changes in high-mass X-ray binary populations of star-forming galaxies.
Using galaxies with a broader range of SFR, we also constrain a stellar-mass-dependent contribution to $L_\mathrm{X}$, likely related to low-mass X-ray binaries. Using this calibration, we convert our X-ray main sequence to SFRs and measure a star-forming main sequence with a constant slope $\approx 0.76\pm0.06$ and a normalization that evolves with redshift as $(1+z)^{2.95\pm0.33}$. Based on the X-ray emission, there is no evidence for a break in the main sequence at high stellar masses, although we cannot rule out a turnover given the uncertainties in the scaling of $L_\mathrm{X}$ to SFR.

\end{abstract}

\section{Introduction}
\label{sec:intro}

It is now well established that most galaxies can be divided into two, fairly distinct populations: \emph{quiescent} galaxies, which appear red due to their passively evolving stellar populations and generally have elliptical morphologies; and \emph{star-forming} galaxies, which have blue colours due to the ongoing formation of new stars and tend to have ``late-type" morphologies with significant disk components \citep[e.g.][]{Bell04,Faber07,Blanton09}. 
Over cosmic time, the total combined mass of quiescent galaxies is found to increase \citep[e.g.][]{Brown07,Moustakas13}, indicating that some process is shutting down star formation and transforming massive star-forming galaxies into quiescent systems. 

Star-forming galaxies, on the other hand, are found to follow a relatively tight correlation between the current star-formation rate (SFR) and stellar mass (\Mstel), referred to as the ``main sequence of star formation" \citep[][]{Noeske07}.
The average SFR increases with \Mstel, following approximately a power law of the form SFR~$\propto\mstel^m$, where $m$ denotes the logarithmic ``slope" of the main sequence and is generally found in the range $m\sim0.6-1$ \citep[e.g.][]{Elbaz07,Karim11,Speagle14}.
The tightness of this correlation, with an intrinsic scatter of $\sim0.3-0.5$~dex \citep[e.g.][]{Rodighiero11,Ilbert15,Schreiber15}, indicates that the majority of star-forming galaxies build up their stellar populations in a smooth and continuous manner.
The main sequence evolves strongly with redshift, shifting to lower SFRs (at a fixed \Mstel) as redshift decreases \citep[e.g.][]{Noeske07,Daddi07b,Elbaz11}, indicating a uniform reduction in the rate at which star formation is fuelled as cosmic time progresses.
This drop in the normalization of the main sequence drives the rapid decline in the overall density of star formation in the Universe between $z\sim2$ and the present day. 

Given the fundamental constraints that the main sequence places on our understanding of galaxy evolution, there have been substantial efforts to measure the slope and normalization over a wide range of stellar masses and redshifts (e.g. \citealt{Karim11,Whitaker12,Schreiber15}, see \citealt{Speagle14} for an overview).
A roughly linear relation ($m\approx1$) is found for low stellar masses ($\mstel \lesssim 10^{10} \mathcal{M}_\odot$) at all redshifts \citep[e.g.][]{Whitaker14,Schreiber15,Tomczak16}.
At higher stellar masses ($\mstel \gtrsim 10^{10} \mathcal{M}_\odot$), however, the slope appears to flatten as redshift decreases \citep[e.g.][]{Whitaker14,Schreiber15,Tomczak16}.
The slope of the main sequence reflects the efficiency at which new stars are formed relative to the sum of past star formation (i.e. the total stellar mass of the galaxy).
A turnover at high stellar masses and late cosmic times could indicate a reduction in the star formation efficiency at the highest stellar masses or the onset of any overall quenching process.

One of the major challenges in studies of the main sequence is to measure the SFRs of galaxies with sufficient accuracy over a wide dynamic range in stellar mass and redshift (see \citealt{Kennicutt12} and \citealt{Madau14} for recent reviews of SFR estimators). 
The SFR can be inferred from the rest-frame UV emission, which directly traces the emission from massive stars 
\newt{formed within the last $\sim$10--100~Myr} \citep[e.g.][]{Kennicutt98}.
However, the UV emission is extremely sensitive to the effects of dust; thus large (and in many cases uncertain) corrections are required to recover the total SFR \citep[e.g.][]{Gordon00,Salim07,Hao11}.
Alternatively, obscured star formation \newt{(occuring within the last $\sim100$~Myr)} can be traced more directly via the re-radiated emission from dust at infrared (IR) wavelengths of $\sim$8--1000\micron. 
With the advent of \emph{Herschel} it has become possible to trace the peak of the far-IR emission (at rest-frame $\sim$ 60--100\micron) out to $z\sim4$ \citep[e.g.][]{Elbaz11}, although the limited sensitivity means only highly star-forming galaxies are detected and a stacking analysis is usually required to probe the typical SFRs of main sequence galaxies \citep[e.g.][]{Schreiber15}.
Fainter galaxies can be detected in 24\micron\ imaging from \textit{Spitzer} and used to infer the dust emission at longer wavelengths, although stacking analyses are still required to reach the very lowest SFRs \citep[e.g.][]{Whitaker14}.
\newt{While such stacking analyses are powerful, they can be affected by selection effects (when defining galaxy samples for stacking), 
are sensitive to background subtraction and blending effects, and provide an average that may not accurately represent the underlying distribution \citep[see e.g.][and references therein]{Viero13,Schreiber15}.}

Alternatively, dust-corrected SFRs can be estimated by modelling the broadband photometry of galaxies (spanning UV, optical and IR wavelengths) with stellar population synthesis models \citep[e.g.][]{Wuyts11,Conroy13b}, although this method is sensitive to the assumed models and the corrections for dust extinction.
Furthermore, the same data are often used to estimate the galaxy stellar mass, which can introduce correlated uncertainties between parameters that could bias measurements of the main sequence \citep[e.g.][]{Reddy12}.
The SFR can also be probed via nebular line emission such as H$\alpha$ \newt{(providing a prompt tracer of star formation occuring within the last $\sim3$--10~Myr)}, although dust corrections must be applied and large spectroscopically observed galaxy samples are required \citep[e.g.][]{Moustakas06,Shivaei15}.
Radio emission also provides a probe of the SFR, although again a stacking analysis is required to probe the bulk of the star-forming galaxy population out to high redshifts \citep[e.g.][]{Karim11}.

X-ray emission also provides a tracer of the SFR. 
The overall X-ray emission from a galaxy is due to the combined emission from high-mass X-ray binaries (HMXBs) and low-mass X-ray binaries (LMXBs) within the galaxy, hot gas throughout the galaxy, and accretion onto the central supermassive black hole that powers an Active Galactic Nucleus (AGN), if present \citep[e.g.][]{Fabbiano89}.
HMXBs are expected to form promptly after a burst of star-formation (within $\sim$5~Myr) and have relatively short lives \citep[$\sim100$--300~Myr e.g.][]{Fragos13}.
Thus, HMXBs should be a relatively direct tracer of recent star-formation and provide the dominant contribution to the X-ray luminosity in normal star-forming galaxies (i.e. with negligible current AGN activity). 
LMXBs, on the other hand, tend to form later ($\sim100$--300~Myr after a burst of star formation) and 
\newt{then gradually fade over $\sim$Gyr time-scales}
\more{
Thus, the X-ray luminosity from LMXBs will provide a delayed tracer of the SFR \citep[e.g.][]{Ghosh01} and may} scale with the overall stellar mass of the galaxy, rather than recent star formation \citep[e.g.][]{Lehmer10}. 
Diffuse, hot, ionized gas, thought to be heated by supernovae and winds from massive stars, is also observed in star-forming galaxies, with the total luminosity scaling with SFR and contributing up to a third of the overall emission at softer X-ray energies \citep[e.g.][]{Mineo12b}. 
The X-ray luminosity from an AGN, if present, will typically dominate over any other X-ray emission from a galaxy \citep[e.g.][]{Brandt15}.

A number of studies have explored the use of the X-ray luminosity as a tracer of the SFR.
In the local universe, it is possible to directly detect moderate-to-high SFR galaxies, which have been used to calibrate the X-ray luminosity as a SFR tracer \citep[e.g.][]{David92,Ranalli03,Rovilos09,Lehmer10}.
The deepest \textit{Chandra} surveys are able to detect the highest SFR galaxies (e.g. ultraluminous infrared galaxies) out to $z\sim1$ and thus extend such studies of the X-ray properties to higher redshifts \citep[e.g.][]{Symeonidis11,Symeonidis14,Mineo14}.
Stacking analyses have been used to push to fainter limits and probe more typical, high-redshift galaxy populations \citep[e.g.][]{Nandra02,Laird06,Basu-Zych07,Lehmer16}, providing crucial constraints on the relation between the physical properties of galaxies and their X-ray binary populations.
However, extremely deep X-ray data are needed to directly detect star-forming galaxies, as well as reliably identify and exclude AGNs from any stacking analyses.
Such studies are further complicated by the variation in the depths of X-ray survey data across different fields and within a single field. 
Thus, the adoption of X-ray tracers of SFRs has been somewhat limited, with most studies restricted to the small areas covered by the \textit{Chandra} Deep Fields \citep[e.g.][]{Lehmer08,Symeonidis14,Lehmer16}. 

In this paper we develop a new approach to probe the X-ray emission from large samples of star-forming galaxies, combining data from a number of \textit{Chandra} surveys of varying depths.
We determine the \emph{distribution} of X-ray luminosities for samples of star-forming galaxies as a function of redshift and stellar mass, spanning $z\sim0-4$ and $\mstel \sim 10^{8.5-11.5} \msun$, using a Bayesian technique that allows us to push substantially below the nominal detection limits in a given field.
Our measured distributions allow us to identify and separate two different origins for the X-ray emission:  
1) galactic emission (predominantly from high-mass X-ray binaries), which produces a peak in the distribution at low luminosities that traces the average star formation rate for galaxies of a given stellar mass and redshift and is the focus of this paper; 
and
2) the emission from active galactic nuclei (AGN), tracing the distribution of supermassive black hole accretion activity within galaxies of a given stellar mass and redshift, which is explored in a companion paper \citep[in preparation, hereafter Paper II]{Aird16b}.

In Section~\ref{sec:data} we describe our data and define our samples. 
We construct large, stellar-mass limited samples of star-forming galaxies from near-infrared selected catalogues from the CANDELS/3D-HST and UltraVISTA surveys.
We extract X-ray information for all galaxies in our samples from the available \textit{Chandra} imaging.
Section~\ref{sec:lumdist} presents our measurements of the intrinsic distributions of X-ray luminosities within star-forming galaxies as a function of stellar mass and redshift. 
We identify peaks in these distributions at low luminosities that we relate to star formation processes. 
In Section~\ref{sec:msofsf} we measure the position of these peaks as a function of stellar mass and redshift, revealing the ``X-ray main sequence''.
In Section~\ref{sec:xraysfr} we compare with UV-to-IR SFR estimates to constrain the relation between X-ray luminosity and SFR, allowing us to relate the X-ray main sequence to the main sequence of star formation.  
Section~\ref{sec:discuss} discusses our findings and compares to previous work.
We summarize our findings in Section~\ref{sec:summary}.
We adopt a flat cosmology with $\Omega_\Lambda = 0.7$ and $H_0 = 70$~km~s$^{-1}$~Mpc$^{-1}$ and assume a \cite{Chabrier03} stellar initial mass function (IMF) throughout this paper.

\section{Data and sample selection}
\label{sec:data}

\subsection{Near-infrared selected photometric catalogues}
\label{sec:photocats}

We construct a large sample of star-forming galaxies by combining data from four of the CANDELS survey fields \citep{Grogin11,Koekemoer11} and the UltraVISTA survey of the wider COSMOS field \citep{McCracken12}. 
We adopt catalogues of objects selected at near-infrared (NIR) wavelengths, which allows us to define a sample of galaxies in an homogenous manner over a wide range of redshifts and down to a well-defined stellar mass completeness limit (see Section \ref{sec:sampledef} below for details).

The CANDELS survey is an ultra-deep NIR imaging campaign with the Wide Field Camera 3 (WFC3) on the \textit{Hubble Space Telescope} (\textit{HST}), covering $\sim$800~arcmin$^2$ across five premier multiwavelength survey fields, four of which are used in this study (GOODS-N, GOODS-S, AEGIS, and COSMOS).
The 3D-HST survey \citep{Brammer12} complements the deep CANDELS imaging with low-resolution near-infrared spectroscopy, primarily using the the WFC3 G141 grism on \textit{HST}. 
In this paper, we use the v4.1.5 photometric and redshift catalogues provided by the 3D-HST collaboration \citep{Skelton14,Momcheva16}.
The photometric catalogues are based on detection in $F125W + F140W +F160W$ combined images. 
PSF-matched photometry is then obtained from a wide range of ground- and space-based imaging campaigns in the various fields, covering the UV ($\sim0.4$\micron) to mid-infrared (MIR: $\sim8$\micron)
 with up to 44 broad or medium band filters \citep[full details are given by][hereafter S14]{Skelton14}.
Deblended photometry at 24\micron\ from \textit{Spitzer}/MIPS is also provided by \citet{Whitaker14}.
The combined catalogue consists of $\sim$150,000 objects 
over the four CANDELS/3DHST fields.

The UltraVISTA survey is a deep, wide-area near-infrared imaging survey, covering $\sim$1.6 deg$^2$ in the COSMOS field \citep{Scoville07} and thus complements the deeper CANDELS imaging, providing a better sampling of galaxies with lower redshifts ($z\lesssim1$) and higher stellar masses ($\mathcal{M}_*\gtrsim10^{10}$\Msun). Near-infrared imaging in the $Y$, $J$, $H$ and $K_S$ bands was obtained as part of the deepest component of the ESO public survey programme with the Visible and Infrared Survey Telescope for Astronomy (VISTA).
In this paper, we adopt the $K_S$-selected catalogues provided by \citet[hereafter M13]{Muzzin13}.  
This catalogue provides PSF-matched photometry in 30 photometric bands, covering $0.15-24$\micron, and includes the data from $GALEX$ \citep{Martin05}, CFHT/Subaru \citep{Capak07}, UltraVISTA \citep{McCracken12}, and S-COSMOS \citep{Sanders07}.
We mask out any areas contaminated by bright stars, as defined by M13.
The CANDELS WFC3 imaging of the COSMOS field is fully contained within the larger UltraVISTA survey area; in this paper we adopt the deeper CANDELS/3D-HST photometric catalogues when available and thus also mask out this area from the UltraVISTA catalogues.
Our UltraVISTA catalogue thus contains $\sim$230,000 $K_S$-selected sources.

\subsection{\textit{Chandra} X-ray data}
\label{sec:xraydata}

\textit{Chandra} X-ray imaging of varying depths is available in our four CANDELS fields, with exposure times of $\sim$160~ks in COSMOS \citep{Elvis09}, $\sim$800~ks in AEGIS \citep{Nandra15}, $\sim$2~Ms in GOODS-N \citep{Alexander03} and $\sim$4~Ms in GOODS-S \citep{Xue11}.\footnote{Recent \textit{Chandra} imaging of the GOODS-S field (that will provide a total depth $\sim7$~Ms, P.I. Brandt) and the $\sim$600~ks \textit{Chandra} imaging of the fifth CANDELS field (UDS, P.I. Hasinger) are not considered in this work.}
\textit{Chandra} imaging of $\sim$50 -- 160~ks depth of the entire UltraVISTA field has also recently become available as part of the COSMOS-Legacy programme \citep{Civano16,Marchesi16}, which we exploit in this paper.

All of the \textit{Chandra} observations were analysed using our own data reduction and source detection procedure \citep[see][for details]{Laird09,Nandra15}, providing homogenous and well-characterised data products that are required for our statistical analysis \citep[see also][]{Georgakakis08,Georgakakis14,Georgakakis15,Rangel13,Aird08,Aird10,Aird15}.
We apply the point source detection procedure described by \citet{Laird09} with a Poisson false probability threshold of $<4\times10^{-6}$ in the soft (0.5--2~keV), hard (2--7~keV), full (0.5--7~keV) and ultrahard (4-7~keV) energy bands, combining the source lists to create a merged point source catalogue. 
This provides samples of 914 significant X-ray detections in the area of our four CANDELS fields and 2851 significant X-ray detections in our UltraVISTA area. 

We cross-match our sample of significant X-ray detections with the NIR-selected photometric catalogues using the likelihood ratio technique \citep[see ][]{Ciliegi03,Brusa07}, applying a threshold that maximizes the sum of the completeness and reliability (as described by \citealt{Luo10}, see also \citealt{Aird15}).
Using this technique, we are able to robustly associate an NIR-selected object with 831 (91 per cent) of the detected X-ray sources in the CANDELS fields and 2663 (93 per cent) of the X-ray sources in the UltraVISTA area.

We extract X-ray information for the remaining NIR-selected objects within our fields that are not already matched with a significant X-ray detection.
We first remove any NIR objects from our catalogues that lie close to a significant X-ray source (but are not associated with the X-ray source according to our likelihood ratio matching), removing objects within 1.5 times the radius corresponding to the 90~per cent enclosed energy fraction (EEF) of the exposure-weighted \textit{Chandra} PSF.
This cut removes between 5 and 10 per cent of objects from our NIR catalogues in  an unbiased manner (assuming the NIR catalogues are not strongly clustered at $\lesssim$5 arcsecond scales).
We then extract the total X-ray counts in the full, soft and hard energy bands from within a circular aperture at the position of every remaining NIR object. 
The radius of the aperture is based on the exposure-weighted PSF and corresponds to an EEF of 70~per cent. 
We also estimate the background counts within the same aperture from our background maps\footnote{Background maps are generated by replacing counts in the region of detected sources with values from a local background region and smoothing to create an overall background map. See \citet{Georgakakis08} for full details.} and the effective exposure at the position of each object. 
Our measurements of the total and background counts are identical in our source detection procedure but are applied here to the larger catalogue of NIR-selected objects. 
The extracted total counts, background counts and exposures contain information for sources that fall below our direct detection threshold but may correspond to a lower significance detection.
The overall sensitivity of our X-ray observations is also described by these data; i.e. information on the fraction of sources that are \emph{not} directly detected in the X-ray images.
All of this information is used by our Bayesian analysis method, described in Appendix~\ref{app:bayesmix} and implemented in Section~\ref{sec:lumdist} below.

\subsection{Redshifts, rest-frame colours, stellar masses and star-formation rates}
\label{sec:redshifts}

The 3DHST and UltraVISTA catalogues provide spectroscopic redshifts, compiled from a number of ground-based spectroscopic follow-up campaigns (see S14 and M13 and references therein for details).
We also include additional spectroscopic redshifts from the compilations described by \citet{Aird15} and \citet{Marchesi16} and references therein, which include follow-up campaigns focused on the X-ray source populations.
In addition, we include recently obtained spectroscopic redshifts from the first data releases of the MOSDEF survey \citep{Kriek15} and VIMOS Ultra Deep Survey \citep[VUDS:][Tasca et al. in preparation]{LeFevre15}.
If a higher resolution spectroscopic redshift is unavailable, we adopt redshifts  based on the combination of the low-resolution WFC3 grism spectroscopy and the photometry in the CANDELS fields \citep{Momcheva16}
or the low-resolution spectroscopic redshifts of large numbers of galaxies obtained with PRIMUS in the UltraVISTA field \citep{Coil11,Cool13}.
We match the spectroscopic redshift catalogues to our NIR-selected catalogues using a 2\arcsec\ search radius, correct for any global offset, then re-match using a stricter 0.5\arcsec\ search radius.
In total, we are able to assign high- or low-resolution spectroscopic redshifts to 17,692 sources in our CANDELS catalogue and 13,860 sources in the UltraVISTA catalogue, including 680 (82 per cent) and 1648 (59 per cent) of the X-ray detected sources in CANDELS and UltraVISTA, respectively.

For the sources without spectroscopic redshifts, we must resort to photometric redshift estimates (photo-$z$). 
For the bulk of our sources, we adopt the photo-$z$ provided by S14 and M13.
These photo-$z$ are calculated using the \textit{EaZY} code \citep{Brammer08} with a standard set of galaxy templates and achieve an accuracy of $\sigma\lesssim 0.02$\footnote{Photo-$z$ accuracy, $\sigma$, is given in terms of the median absolute deviation, normalized relative to $(1+z_\mathrm{spec})$.} and outlier rates of $\sim$2 -- 5 per cent when compared with high-quality spectroscopic galaxy redshift samples. 
For X-ray detected sources, we instead adopt photo-$z$ that are optimised for sources which may have significant AGN contributions to the optical and IR light.
We adopt high-quality AGN photo-$z$ from \citet{Hsu14}, \citet{Nandra15}, or \citet{Marchesi16} for the GOODS-S, AEGIS and COSMOS/UltraVISTA fields, respectively, all of which were calculated using techniques developed by \citet{Salvato09,Salvato11} and achieve accuracies of $\sigma\lesssim0.03$ with outlier rates of $\sim$5 per cent. 
If a high-quality AGN photo-$z$ is not available from these catalogues (e.g. in the GOODS-N field or due to slight differences in the X-ray source lists), we re-calculate the photo-$z$ based on the 3DHST or UltraVISTA photometry using the method described by \citet{Aird15}. 
These photo-$z$ have a lower accuracy ($\sigma\sim0.05$) and higher outlier rate ($\sim$10-15 per cent) than the work by Salvato and collaborators but are only adopted for a small fraction ($\sim$ 7 per cent) of our X-ray sources.
Given our high spectroscopic completeness for X-ray sources and the high quality of the bulk of our galaxy and AGN photo-$z$, we neglect uncertainties in the photo-$z$ for this work.

We calculate rest-frame colours in the standard $U$, $V$ and $J$ band passes for all sources in our catalogues. 
We use \emph{EaZY} with the standard set of galaxy templates to interpolate from the observed photometry \citep[see][]{Brammer11}, fixing the redshift at the spectroscopic value (if available) or our best photo-$z$ estimate. 
We do not make any corrections for an AGN contribution to the spectral energy distributions (SEDs); thus our estimates correspond to the \emph{observed} rest-frame colours and will be contaminated by bright AGNs in some cases (see further discussion in Section~\ref{sec:sampledef} below).

We use the FAST code \citep{Kriek09} to fit the SEDs of our sources with stellar population synthesis models and estimate both stellar masses and SFRs. 
We apply a number of important modifications in our implementation of the FAST code to ensure we recover accurate and physically meaningful stellar population parameters across our wide redshift range.
We adopt delayed-$\tau$ models of the star-formation histories but apply a redshift-dependent \emph{minimum} age, ensuring our code adopts realistic galaxy templates at both low and high redshifts. 
We also use a $\chi^2$-weighted average over the grid of stellar population parameters assumed in FAST when estimating stellar masses and SFRs, which reduces the effects of the relatively coarse grid and provides a fairer weighting of the range of viable models, removing some clear outliers (e.g. star-forming galaxies with blue optical colours that can be assigned steeply declining star-formation histories and erroneously low SFRs).
Further details are given in Appendix \ref{app:sedfits} below. 
We also account for an AGN contribution to the SED, as described in the appendix to Paper II, although at the low X-ray luminosities probed in this paper the AGN contribution to the optical-to-MIR SED is generally negligible and thus does not have an impact on the results presented here.  
Our final SED-based SFR estimates use an ``SFR ladder" \citep[e.g.][]{Wuyts11}: 
when a 24\micron\ detection is available we use UV+IR SFRs that sum the unobscured and obscured star formation seen in the UV and far-IR respectively\footnote{For X-ray detected sources, we apply a correction to allow for an AGN contribution in the UV and IR; see the appendix of Paper II for full details.}; for sources that are not detected at 24\micron\ we instead adopt the dust-corrected estimates of SFRs from the fits to the UV-to-MIR SEDs.
Changing the details of the SED modelling (such as adopting different star formation histories or neglecting our $\chi^2$ weighting scheme) has a negligible impact on our estimates of the total galaxy stellar mass.
Thus, we are able to estimate accurate stellar masses for the bulk of our sample (including the majority of our X-ray detected sources). 
Our final SED-based SFR estimates are more sensitive to the assumptions in the SED modelling and are thus less certain for an individual sources. Nonetheless, our consistent calculations of SFRs ensure that when combining large samples of galaxies we are able to accurately recover overall trends between SFR and X-ray luminosity. 
Furthermore, the main results of this paper (see Section~\ref{sec:msofsf}) do not rely on these multiwavelength measurements of the SFR.

\subsection{Star-forming galaxy sample definition}
\label{sec:sampledef}

\begin{figure*}
\begin{center}
\includegraphics[width=\textwidth,trim=0 0 0 0]{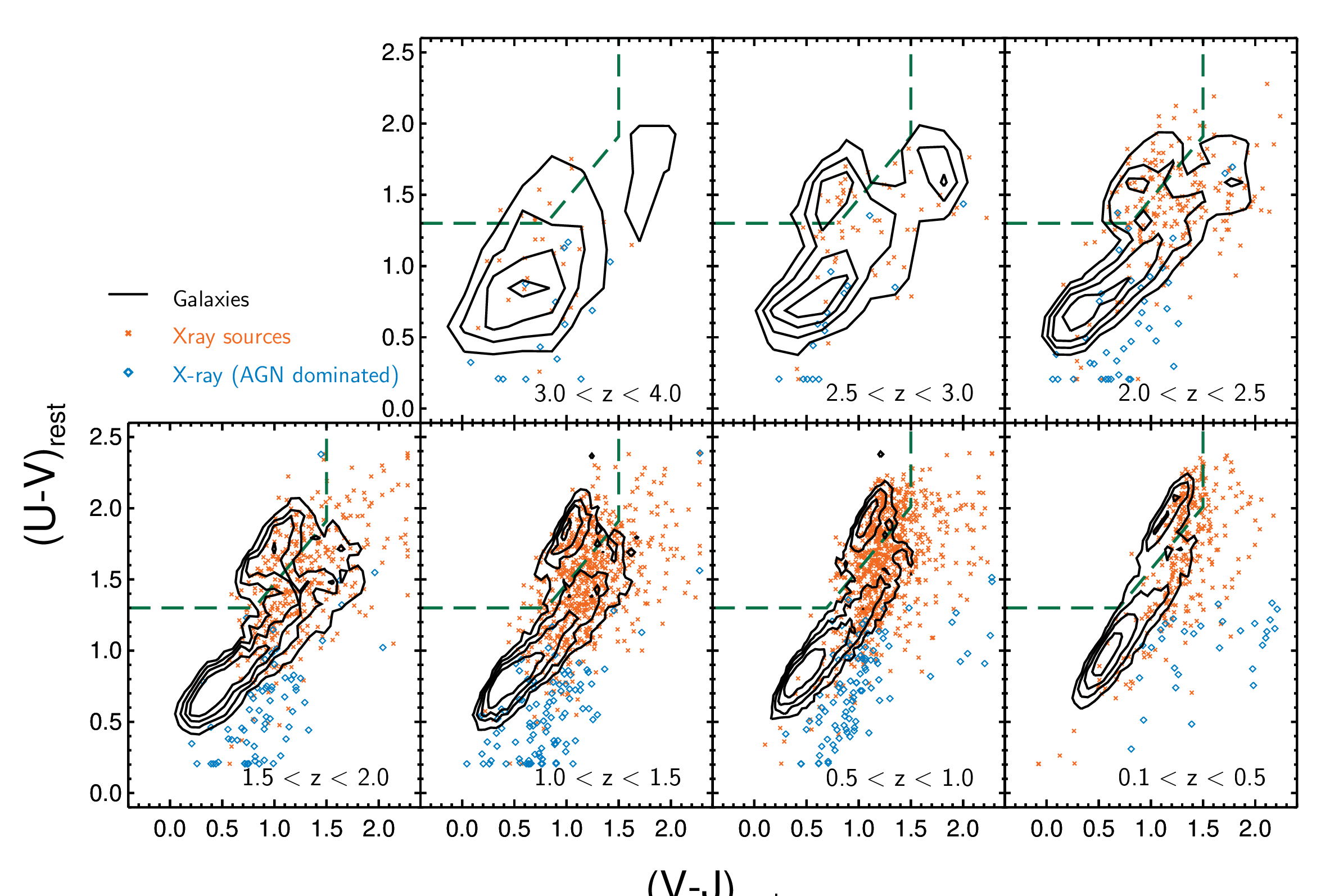}
\end{center}
\caption{
Rest-frame $U-V$ versus $V-J$ colours of galaxies (black contours) in our combined stellar-mass-limited CANDELS and UltraVISTA samples. 
A clear bimodality---seen out to at least $z\sim2.5$---allows us to separate quiescent and star-forming galaxies (including dust-reddened systems) using the colour-colour cuts defined by Equation~\ref{eq:uvjcolcut} (green dashed lines). 
X-ray detected objects (orange crosses) span the galaxy population, although they are biased toward the dusty star-forming region due to the selection bias toward high stellar mass host galaxies in X-ray selected samples \citep{Aird12}. 
Blue diamonds indicate X-ray sources where our two-component SED fitting indicates that the optical light is dominated by the AGN resulting in blue, contaminated colours; these sources are excluded from the sample of star-forming galaxies used in this paper.
}
\label{fig:uvj}
\end{figure*}

\begin{figure*}
\includegraphics[width=0.7\textwidth,trim=40 10 0 0]{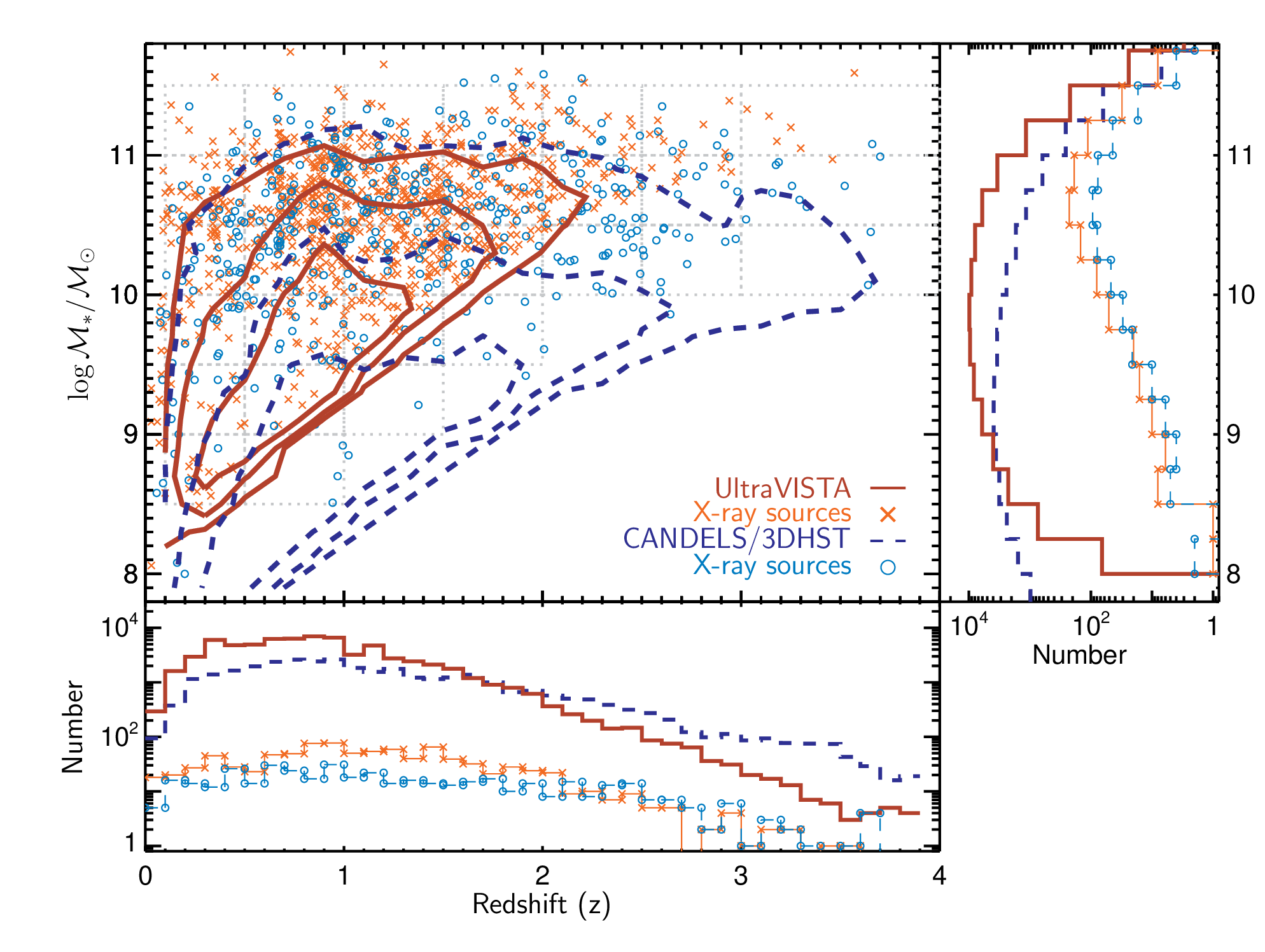}
\caption{
Distributions of stellar mass and redshift for the CANDELS/3DHST (dashed blue contours) and UltraVISTA (solid red contours) star-forming galaxy samples.
Contours enclose 50, 80 and 95 per cent of the star-forming galaxies in each survey.
Points (light blue circles and orange crosses for CANDELS and UltraVISTA, respectively) indicate sources with soft band (0.5--2~keV) X-ray detections. 
The grey dotted lines indicate our bins of stellar mass and redshift, which are required to lie wholly above the stellar mass completeness limits of the CANDELS fields. 
}
\label{fig:masscont}
\end{figure*}
Our study requires well-defined, unbiased samples of the star-forming galaxy population down to a given stellar mass limit. 
To construct such samples, we first generate magnitude-limited catalogues by applying limits of $H_\mathrm{F160W}<25.1$ and $K_\mathrm{S}<23.4$ to our CANDELS and UltraVISTA catalogues, respectively. 
These limits correspond to the 90 per cent completeness limits for point sources given by S14 and M13.

Next, we remove any sources that are spectroscopically identified as stars.
We also remove objects that are photometrically flagged as stars based on the positions in the size-magnitude plane (in the S14 catalogues of the CANDELS fields) or a cut in the $u^* - J$ vs. $J-K_\mathrm{S}$ colour-colour space (in the M13 catalogues of the UltraVISTA field).

We then apply stellar mass completeness limits. 
For the UltraVISTA field we adopt the empirically determined 95 per cent stellar mass completeness limits (as a function of redshift) that were calculated by \citet{Muzzin13b} by comparing  the distribution of stellar masses in the UltraVISTA catalogue to deeper NIR-selected catalogues. 
We shift the mass limits down by 0.04 dex as we adopt a \citet{Chabrier03} initial mass function in our SED fitting. 
We restrict our UltraVISTA sample to galaxies lying above this stellar mass completeness limit at a given redshift.
For the CANDELS fields we shift the UltraVISTA completeness limits down by an additional order of magnitude (1 dex), allowing for the significantly deeper NIR imaging.
This provides a completeness limit as a function of redshift that is agreement with the empirically determined 95 per cent completeness limits calculated by \citet{Tal14} for the CANDELS fields at $z\lesssim1.5$ and is more conservative at higher redshifts.

We restrict our samples to include only star-forming galaxies by applying a cut based on the rest-frame $U-V$ versus $V-J$ colour-colour diagram (hereafter the ``UVJ diagram'').
This diagnostic divides the galaxy population according to the well-established bimodality between actively star-forming galaxies and quiescent galaxies, with the added advantage that dust-reddened star-forming systems can be separated from the quiescent galaxy population and classified correctly \citep{Labbe05,Wuyts07,Williams09,Whitaker11}.
Figure~\ref{fig:uvj} shows the UVJ diagram for the combination of our stellar-mass-limited samples from CANDELS and UltraVISTA (grey contours). 
We observe a clear bimodality in our galaxy sample up to at least $z\sim2.5$ that allows us to separate quiescent galaxies from the star-forming population.
We adopt colour-colour cuts from \citet{Muzzin13b}, where quiescent galaxies satisfy
\begin{eqnarray}
(U-V) > 1.3,\;\; (V-J) < 1.5 && \left(\mathrm{all\; redshifts}\right)\nonumber\\
(U-V) > 0.88 \times (V-J) + 0.69 && \left(0.0<z<1.0\right)\\  
(U-V) > 0.88 \times (V-J) + 0.59 && \left(1.0<z<4.0\right)\nonumber
\label{eq:uvjcolcut}
\end{eqnarray}
which were adapted from the colour cuts originally proposed by \citet{Williams09}.

The orange crosses in Figure~\ref{fig:uvj} indicate the colours of sources with significant X-ray detections. 
\more{As previously seen by \citet{Georgakakis14}}, the X-ray detections are distributed across the quiescent and star-forming galaxy populations, although there is a strong trend toward the dusty star-forming region of the UVJ diagram (redder $U-V$ and $V-J$ colours but below the quiescent galaxy selection box).
This trend is due to the selection bias toward high stellar mass host galaxies in X-ray selected AGN samples \citep{Aird12} and the fact that higher stellar mass galaxies are known to contain higher levels of dust (e.g. \citealt{Santini14}, see also \citealt{Georgakakis14}, \citealt{Cowley16}, Azadi et al. in preparation).
The blue diamonds indicate X-ray detections where our two-component SED fitting finds that more than 50 per cent of the light at rest-frame 5000~\AA\ is coming from the AGN component; thus, the rest-frame UVJ colours are contaminated by the AGN light, leading to the very blue colours in the UVJ diagram and making it impossible to classify the host galaxy as star-forming or quiescent. 
These sources could constitute at most 0.5 per cent of our total star-forming galaxy sample (assuming they all had star-forming hosts) and their X-ray emission is dominated by the AGN (rather than tracking the SFR, which is the aim of this paper). 
Thus, we exclude these AGN-dominated sources from the sample of star-forming galaxies used in this paper. 

Our final sample consists of 32,927 star-forming galaxies in the CANDELS fields and 68,718 star-forming galaxies in the UltraVISTA field, of which 422 and 881 are directly detected in the soft (0.5--2~keV) X-ray band in CANDELS and UltraVISTA respectively.
Figure~\ref{fig:masscont} shows the distributions of stellar masses and redshifts of the galaxy samples and the X-ray detected sources from the two surveys.
Grey dotted lines indicate the bins of stellar mass and redshift that are used in Section~\ref{sec:lumdist} below and are required to lie wholly above the stellar mass completeness limit of CANDELS.
The higher stellar mass completeness limit applied to the UltraVISTA catalogue, however, cuts through some of our stellar mass bins, and thus galaxies with lower stellar masses and higher redshifts may be under-represented in some bins.
Removing sources from one part of a bin due to the mass completeness limits should not have an impact on our results under the reasonable assumption (implicit in our analysis) that the true distribution of luminosities does not change signicantly over a stellar mass--redshift bin. 
The numbers of galaxies in each stellar mass--redshift bin (and the number with significant soft X-ray detections) are provided in each of the panels of Figure~\ref{fig:plxm_soft}.

\section{The intrinsic distributions of X-ray luminosities within samples of star-forming galaxies}
\label{sec:lumdist}

\begin{figure*}
\includegraphics[width=\textwidth]{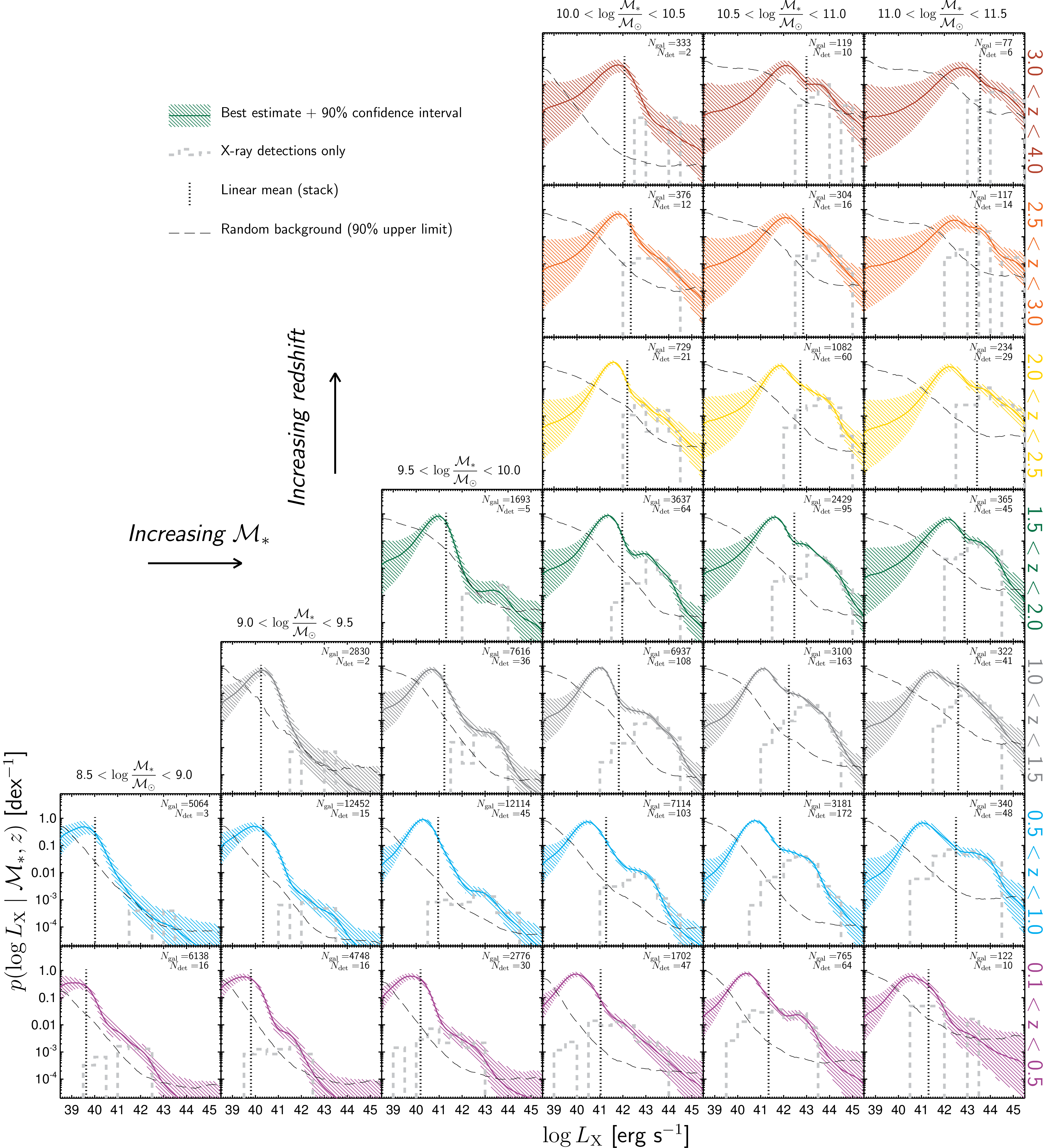}
\caption{
Measurements of the intrinsic probability distributions of X-ray luminosities for samples of star-forming galaxies, binned by stellar mass and redshift. 
The thick coloured lines show the best estimate of $p(\log \lx \giv \mstel,z)$ using our flexible Bayesian mixture modelling approach (shaded regions indicate the 90 per cent confidence intervals). 
The grey dashed histograms show the observed distributions of X-ray luminosities for galaxies that satisfy the nominal X-ray detection criterion in the 0.5--2~keV band. 
Our best estimates of $p(\log \lx \giv \mstel,z)$ use data from all galaxies in a sample (including non-detections) and account for the varying depths of the X-ray imaging to recover the intrinsic distributions, which differ significantly from these observed distributions.
The total number of galaxies and the total number of X-ray detections in each bin is given in the legend of each panel.  
The vertical dotted lines indicate the mean X-ray luminosity from a simple stacking analysis (including both detected X-ray sources and non-detections).
The thin black dashed curves are calculated by randomly shifting the positions of all galaxies in a given stellar mass--redshift bin and repeating our Bayesian analysis for this random sampling of the \textit{Chandra} background (the curve indicates the 90 per cent upper limit on the recovered distribution).
In all panels, we observe a significant peak  at low luminosities ($\lx \lesssim 10^{42}$~\ergs) in our best estimates of $p(\log \lx \giv \mstel,z)$, which traces the distribution of SFRs in each galaxy sub-sample, and a tail to higher luminosities that traces the distribution of AGN accretion activity.
}
\label{fig:plxm_soft}
\end{figure*}

In this section we measure the distributions of X-ray luminosities within our star-forming galaxy sample, divided into bins of stellar mass and redshift. 
We use a non-parametric Bayesian mixture modelling approach---described fully in Appendix~\ref{app:bayesmix} below---to recover $p(\log \lx \giv \mstel,z)$, the underlying probability distribution function of X-ray luminosities for galaxies of a given stellar mass and redshift, with limited assumptions on the shape of the distribution \citep[other than a prior that prefers smoothly varying distributions, see also][]{Buchner15}.
We derive rest-frame 2--10~keV luminosities (hereafter, \LX) based on the observed count rates in the soft (0.5--2~keV) \textit{Chandra} energy band, adopting a fixed conversion factor that assumes an intrinsic X-ray spectral shape with photon index $\Gamma=1.9$, subject to Galactic absorption only\footnote{
We do not apply any corrections for \emph{intrinsic} absorption, local to the X-ray source when estimating luminosities and thus our distribution functions track the \emph{observed} luminosity.
The X-ray emission that traces the overall SFR of a galaxy, coming from the integrated emission from X-ray binaries throughout the host galaxy, is not expected to be subject to significant line-of-sight absorption (even in relatively dusty galaxies) and thus can be tracked by the observed X-ray luminosity.
}
We use data from the 0.5--2~keV \textit{Chandra} band 
as previous studies have shown that this energy range is most sensitive to the star-forming galaxy population \citep[e.g.][]{Lehmer12,Aird15}.
Our method accounts for the Poisson nature of the X-ray data, thus allowing for uncertainties in the X-ray flux measurements and naturally correcting for the effects of the Eddington bias \citep[see also][]{Laird09,Aird10,Aird15}.

Figure~\ref{fig:plxm_soft} presents our estimates of the intrinsic distributions of \LX\ within sub-samples of the star-forming galaxy population, divided into bins of stellar mass (increasing to the right) and redshift (increasing towards higher panels). 
The solid coloured lines show our best estimates of the intrinsic distributions; the shaded regions indicate the 90 per cent confidence intervals. 
Our estimates of the intrinsic distributions differ significantly from the observed distributions for X-ray detected sources, shown by the grey histograms in each panel. 
Our method accounts for the varying sensitivity of the X-ray imaging over the CANDELS and UltraVISTA fields (due to both the overall differences in exposure times between fields and the variation in sensitivity \emph{within} a field) and thus applies a completeness correction to the X-ray detected populations.\footnote{
\refresp{A additional concern is the difference in the stellar mass completeness limits of the CANDELS and UltraVISTA fields.
We have verified that our results are not significantly altered if we exclude galaxies from the UltraVISTA field in stellar mass--redshift bins that are intersected by the completeness limits of UltraVISTA (see Figure~\ref{fig:masscont}) and thus only use galaxies from the deeper CANDELS/3DHST fields for these bins.}}

Additionally, our method uses the X-ray data for every galaxy in our sample. 
We are thus able to use information from sources that fall just below the nominal X-ray detection threshold and the combined X-ray information from the large numbers of galaxies in our samples, which may not individually produce detectable X-ray emission but when combined provide important constraints.  
Fully utilising all of this information allows us, in many cases, to probe substantially below the luminosity of the faintest detected source and constrain the shape of $p(\log \lx \giv \mstel,z)$.
	
In each panel of Figure~\ref{fig:plxm_soft}, our recovered distributions show a relatively narrow peak at lower luminosities ($\lx\lesssim10^{42}$ \ergs) with a broad tail to higher luminosities and possibly a second higher-luminosity peak.
We propose that this observed structure is related to the two different processes that produce the X-ray emission in star-forming galaxies: 
1) the integrated emission from X-ray binaries, hot gas and supernovae, which roughly trace the SFR and results in the low-luminosity peak; 
and 2) AGN accretion activity, which produces the high luminosity tail. 
Our interpretation is consistent with the expected X-ray luminosities due to star formation in normal galaxies and the previous identification of star-forming galaxies at faint X-ray fluxes in deep survey fields \citep[e.g.][]{Georgakakis06b,Lehmer12,Aird15}.

We note that in some cases the position of the low-luminosity peak falls within the range of luminosities probed by the directly detected X-ray populations  (shown by the grey dashed histograms).
To further assess whether these peaks are real features, rather than an artefact of our analysis, we randomly shift the positions of all galaxies in a given stellar mass--redshift bin by 30--60\arcsec\ and repeat our full analysis: extracting the total counts (from the \textit{Chandra} images), background estimates (from the background maps) and exposures at the new position of each galaxy; then applying our Bayesian analysis to produce an estimate of $p(\log \lx \giv \mstel,z)$ for this random sampling of the \textit{Chandra} background. 
The thin black dashed curves in Figure~\ref{fig:plxm_soft} show the upper 90 per cent confidence interval 
from the random background samples.\footnote{
An \emph{a priori} assumption that there is an underlying distribution of \LX\ is implicit in our Bayesian modelling.
See also panel e) of Figure~\ref{fig:plx_sims}.}
In some panels, the 90 per cent upper limits from the background samples exceed the upper limits from the real data at the very lowest luminosities (below the peaks observed in the real data). 
These differences reflect our requirement that $p(\log \lx \giv \mstel,z)$ must integrate to unity; the existence of sources at higher luminosities in the real data thus requires a smaller fraction of galaxies with the lowest luminosities. 
While our nominal uncertainties are larger at these very low luminosities, the exact shape and normalization of our estimates of $p(\log \lx \giv \mstel,z)$ should be treated with caution in this regime.
Nonetheless, the low-luminosity peaks in the observed distributions all lie significantly above the upper limits from the background samples in all panels. 
Furthermore, we do \emph{not} observe a low-luminosity peak in our random background samples, indicating that this feature is not due to uncertainties in our modelling of the \emph{Chandra} background. 
\refresp{We also note that the existence of a distinct, narrow peak in the distribution of \LX\ goes against the underlying prior in our Bayesian method which prefers a smoothly varying distribution.}
We are therefore confident that the observed low-luminosity peaks are real features 
\refresp{in our data} and that we are able to track their positions accurately over a wide range of redshifts, luminosities and stellar masses.
\refresp{The validity of our Bayesian method and our ability to recover the position of \Lmode\ in the different stellar mass and redshift bins is explored further using extensive simulations described in Appendix~\ref{app:sims} below.}

The vertical black dotted lines in Figure~\ref{fig:plxm_soft} show the results of a simple X-ray stacking analysis for the galaxies in each stellar mass--redshift bin (including X-ray detections). 
Stacking recovers the linear mean of the distributions but loses the information on the shape of the underlying distributions that is recovered by our Bayesian method. 
To trace the low-luminosity peak via a stacking analysis would require the bulk of higher luminosity AGN to be excluded to ensure the signal is dominated by the galactic emission. Our results show that appropriate threshold depends on stellar mass and redshift. Furthermore, it is difficult with a stacking analysis to remove the signal from AGNs that fall below the nominal X-ray detection thresholds, which are traced correctly and separated in our analysis.

The positions of the low-luminosity peaks provide estimates of the \emph{mode} of the distribution of X-ray luminosities from star formation. 
By tracing the position of the peak as a function of stellar mass and redshift, we are able to track the mode of the main sequence of star formation via the X-ray emission.
\more{
We note that a distinct low-luminosity peak is seen up to high redshifts ($z\sim3-4$), even in lower mass galaxies ($\mstel\lesssim10^{10.5}\msun$), indicating that with our advanced method we are able to probe star formation via the X-ray emission even for the faintest star-forming galaxies in the CANDELS fields.}

\section{The X-ray main sequence in star-forming galaxies}
\label{sec:msofsf}

\begin{figure}
\includegraphics[width=\columnwidth,trim=35 10 10 0]{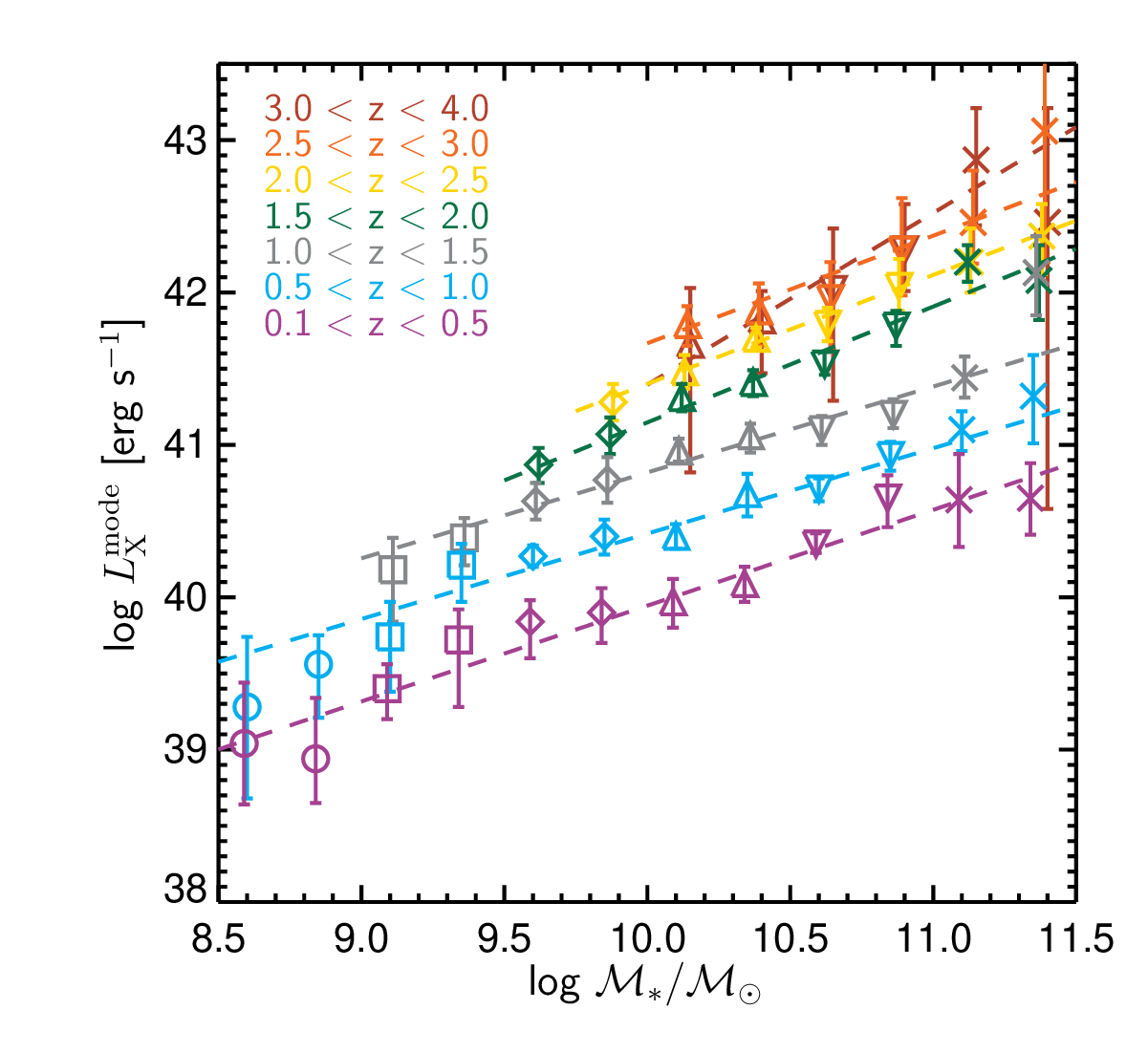}
\caption{
The peak of the distribution of X-ray luminosities from star-forming galaxies, \Lmode, as a function of stellar mass and redshift, revealing the ``X-ray main sequence.''
Errors on \Lmode\ are $1\sigma$ equivalent uncertainties, based on the combination of our jack-knife resampling and the statistical uncertainty from the posterior distributions of \Lmode\ in our Bayesian analysis.
In each redshift bin we find a linear relationship between $\log \lmode$ and $\log \mstel$, which we fit using Equation \ref{eq:lxms_indz} (dashed lines are the fits in each redshift bin). 
We find that the X-ray main sequence has an approximately constant slope of $\approx0.6$ and a normalization that increases with redshift as $\sim(1+z)^{3.79\pm0.12}$ (see Figure~\ref{fig:normslope_vs_z} and Section~\ref{sec:msofsf} for further details).
}
\label{fig:lx_ms}
\end{figure}

\begin{figure}
\includegraphics[width=\columnwidth,trim=20 10 15 0]{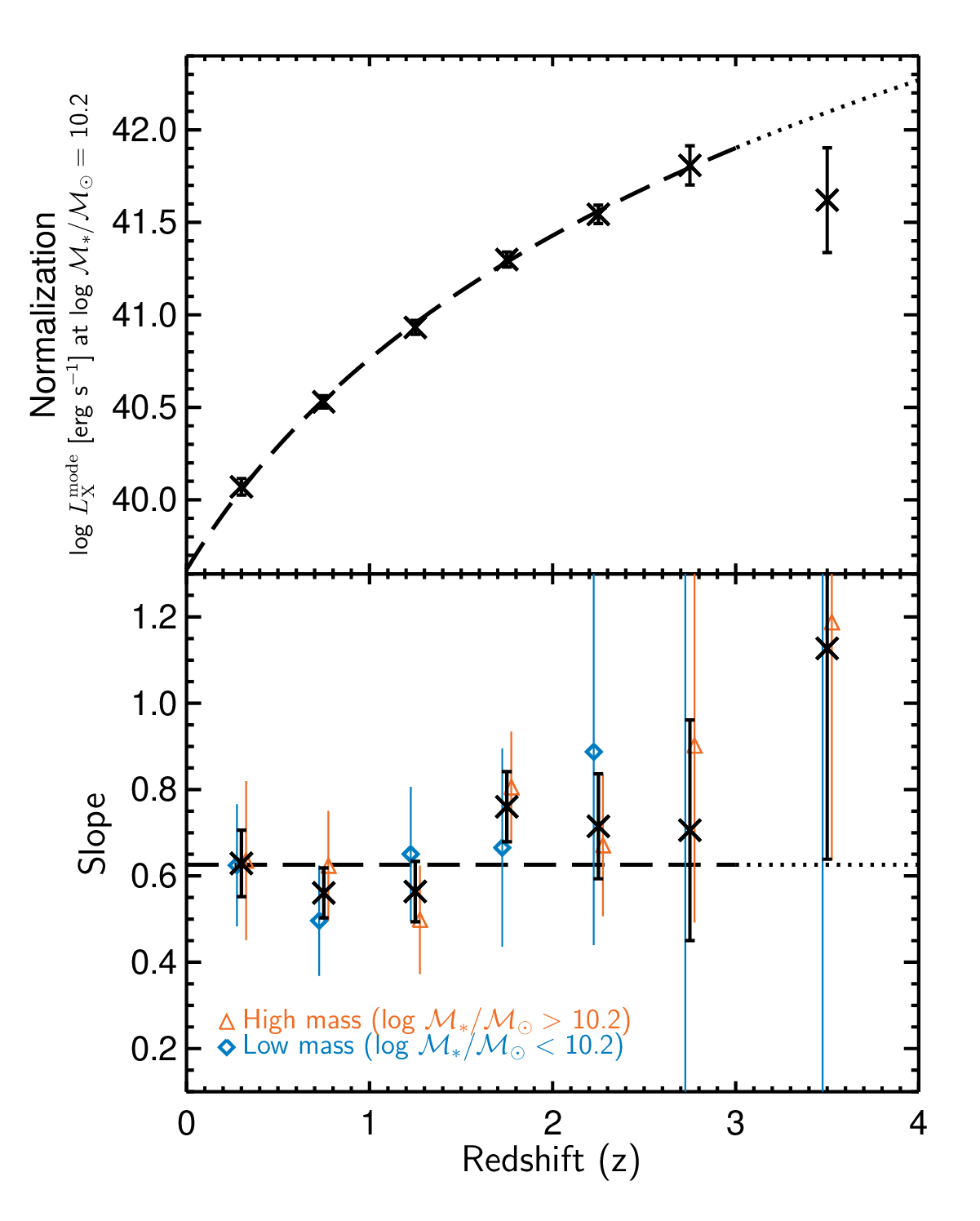}
\caption{
Normalization and slope of the X-ray main sequence as a function of redshift. 
Black crosses indicate the normalization and slope from fits to Equation~\ref{eq:lxms_indz} in each individual redshift bin.
The black dashed line is the result of the overall fit, allowing for evolution in the normalization and assuming a constant slope at all redshifts (see Equation~\ref{eq:lxms_all}). The dotted black line shows the fit at $z>3$ where our parametrization does not provide a good fit to the evolution. 
In the bottom panel we also show individual constraints for the high- and low-mass slopes when fitting with a broken power-law function in an individual redshift bin. 
There is no evidence for a difference between the high- and low-mass slopes at any redshift, although the low-mass slope is very poorly constrained at $z\gtrsim2$.
}
\label{fig:normslope_vs_z}
\end{figure}

In this section we use the peaks in the X-ray luminosity distributions identified in Section~\ref{sec:lumdist} above (and shown in Figure~\ref{fig:plxm_soft}) to constrain the ``X-ray main sequence" as a function of stellar mass and redshift. 
Figure~\ref{fig:lx_ms} shows our measurements of the position of the low-luminosity peak, tracking the mode of the distribution of X-ray luminosity due to star formation processes (hereafter \Lmode), as a function of stellar mass for our various redshift bins. 
We have further sub-divided our star-forming galaxy sample into 0.25~dex wide bins of stellar mass, measuring \Lmode\ for each bin, to accurately track the stellar mass dependence.
\more{We note that our flexible Bayesian modelling allows us to isolate the low luminosity peak from the broader distribution of luminosities related to AGNs and thus AGN contamination should not have a significant impact on our measurements of \Lmode, in contrast to simpler stacking analyses (see Section~\ref{sec:discuss_agn} for further discussion).}

To estimate the uncertainty in \Lmode\ we adopt a jack-knife re-sampling approach \citep[e.g.][]{Lupton93}, whereby we sequentially remove an individual field and repeat our full analysis on the remaining fields. 
For the jack-knife re-sampling, we split the large UltraVISTA field into
eight sub-fields of approximately equal size such that each sub-field contains a similar number of star-forming galaxies as an individual CANDELS field (to within $\sim$20 per cent). 
The jack-knife re-sampling approach allows us to assess any impact from ``cosmic variance" (the sample variance due to the existence of large-scale cosmological structures in our different fields) and to check for any systematic variations due to the differing data quality over our fields. 
We are able to identify a low-luminosity peak in $p(\log \lx \giv \mstel,z)$ in every bin for each of our jack-knife samples. 
Most reassuringly, the existence and position of the recovered peak is not significantly altered by the exclusion of the GOODS-S field, which has the deepest X-ray data, indicating that our results are not dominated by this field. 
We find that the dispersion in \Lmode\ between jack-knife samples is generally smaller than the statistical error (based on the posterior distributions from our Bayesian analysis) for an individual jack-knife sample.
We therefore combine the posterior distributions from each jack-knife sample (giving each equal weight) and determine the 68 per cent central confidence interval (1$\sigma$ equivalent) for \Lmode\ from the combined posterior distribution. 
Our errors on \Lmode\ in Figure~\ref{fig:lx_ms} are therefore conservative, allowing for both the statistical uncertainty and field-to-field variations. 
Our best estimates of \Lmode\ and the uncertainties are given in Table~\ref{tab:xrayms}.

Figure~\ref{fig:lx_ms} reveals a clear trend whereby \Lmode\ is higher for higher stellar mass galaxies---revealing a main sequence in X-rays---and the overall normalization of this relation increases with redshift.\footnote{\newt{Hints of such a relation were previously identified by \citet{Lehmer08} but over a much more limited range in \Mstel\ and $z$.}}
We fit the main sequence in an individual redshift bin with a log-linear relationship, 
\begin{equation}
\log \lx^\mathrm{mode}(\mstel) \; [\mathrm{erg\;s^{-1}}]= a  + b\left(\log \frac{\mstel}{\msun} - 10.2\right).
\label{eq:lxms_indz}
\end{equation}
We adopt a standard $\chi^2$ fitting approach, using our estimates of the uncertainties in \Lmode\ described above. 
A fully Bayesian analysis of the slope and normalization of the main sequence is beyond 
the scope of this paper and is deferred to future studies. 
The dashed lines in Figure~\ref{fig:lx_ms} indicate our log-linear fits in each redshift bin.

Figure~\ref{fig:normslope_vs_z} shows the normalization (top panel) and slope (bottom panel) of our fits to the X-ray main sequence as a function of redshift.
Our results are consistent with a constant slope of $b\approx 0.6$ at all redshifts. 
We also attempt to fit a broken power-law relation with a different slope above and below $\log \mstel/\msun=10.2$ \citep[following][]{Whitaker14}, shown by the red and blue points in the lower panel of Figure~\ref{fig:normslope_vs_z} for the high- and low-mass slopes, respectively. 
We find no significant evidence for a difference between the high- and low-mass slopes at any redshift for our X-ray main sequence.

We observe a strong evolution in the overall normalization of the X-ray main sequence as a function of redshift out to at least $z\sim 3$. 
We attempt to fit this overall evolution as
\begin{align}
\log \lx^\mathrm{mode}&(\mstel,z) \;[\mathrm{erg\;s^{-1}}] =  
\label{eq:lxms_all}
\\
  &a + b\left(\log \frac{\mstel}{\msun} - 10.2 \right) + c\log \frac{1+z}{1+z_0}\nonumber
\end{align}
where we fix $z_0=1.0$ and assume the same slope of the X-ray main sequence, $b$, at all redshifts. 
Our overall fit to the X-ray main sequence gives $a=40.76\pm0.02$ with a constant slope of $b=0.63\pm0.03$ and redshift evolution described by $c=3.79\pm0.12$.
The black dashed lines in Figure~\ref{fig:normslope_vs_z} show our best overall fit for the evolving normalization and constant slope. 
We note that our parametrization may not be a good fit to the evolution at $z\gtrsim3$ (indicated by the dotted line in Figure~\ref{fig:normslope_vs_z}), where there is weak evidence ($\sim 2\sigma$) that the normalization does not increase further toward higher redshifts \citep[see also][]{Speagle14}.

We discuss these findings and the implications for the evolution of the galaxy population in Section~\ref{sec:discuss_msevol} below.

%%%%%%%%%%%%%%%%%%%%%%%%%%%%%%%%
% Table of LX vs Mass
\begin{table}
\caption{Measurements of the X-ray main sequence (see Figures~\ref{fig:lx_ms} and \ref{fig:ms_vs_literature}).}
\label{tab:xrayms}
\centerline{
\resizebox{1.04\columnwidth}{!}{
\begin{tabular}{l l l l l}
\hline
Redshift & $\log \mstel$    & $\log \lmode$ & $\log$~SFR$_\mathrm{X}$ $^{(a)}$ & $\log$~SFR$_\mathrm{X}^\mathrm{lim}$ $^{(b)}$ \\
($z$)     &     [\Msun]      &   [erg s$^{-1}$]  & [\Msun\ yr$^{-1}$] &  [\Msun\ yr$^{-1}$]  \\
\hline
 0.10 --  0.50 &  8.50 --  8.75 & $ 39.04^{+0.40}_{-0.40}$ & $            -0.71^{+0.48}_{-0.48}$ & $\phantom{-}$---\\
 0.10 --  0.50 &  8.75 --  9.00 & $ 38.94^{+0.40}_{-0.29}$ & $            -0.83^{+0.48}_{-0.35}$ & $\phantom{-}$---\\
 0.10 --  0.50 &  9.00 --  9.25 & $ 39.40^{+0.16}_{-0.20}$ & $            -0.28^{+0.19}_{-0.24}$ & $            -0.55$\\
 0.10 --  0.50 &  9.25 --  9.50 & $ 39.72^{+0.20}_{-0.44}$ & $\phantom{-}  0.11^{+0.24}_{-0.53}$ & $            -0.50$\\
 0.10 --  0.50 &  9.50 --  9.75 & $ 39.84^{+0.14}_{-0.24}$ & $\phantom{-}  0.25^{+0.17}_{-0.29}$ & $            -0.08$\\
 0.10 --  0.50 &  9.75 -- 10.00 & $ 39.90^{+0.16}_{-0.20}$ & $\phantom{-}  0.32^{+0.19}_{-0.24}$ & $            -0.02$\\
 0.10 --  0.50 & 10.00 -- 10.25 & $ 39.97^{+0.15}_{-0.17}$ & $\phantom{-}  0.41^{+0.18}_{-0.20}$ & $\phantom{-}  0.01$\\
 0.10 --  0.50 & 10.25 -- 10.50 & $ 40.09^{+0.11}_{-0.12}$ & $\phantom{-}  0.55^{+0.13}_{-0.14}$ & $\phantom{-}  0.14$\\
 0.10 --  0.50 & 10.50 -- 10.75 & $ 40.35^{+0.07}_{-0.06}$ & $\phantom{-}  0.86^{+0.08}_{-0.07}$ & $\phantom{-}  0.61$\\
 0.10 --  0.50 & 10.75 -- 11.00 & $ 40.65^{+0.15}_{-0.19}$ & $\phantom{-}  1.23^{+0.18}_{-0.23}$ & $\phantom{-}  0.73$\\
 0.10 --  0.50 & 11.00 -- 11.25 & $ 40.64^{+0.30}_{-0.31}$ & $\phantom{-}  1.21^{+0.36}_{-0.37}$ & $            -1.00$\\
 0.10 --  0.50 & 11.25 -- 11.50 & $ 40.65^{+0.23}_{-0.24}$ & $\phantom{-}  1.23^{+0.28}_{-0.29}$ & $            -1.00$\vspace{3pt}\\
 0.50 --  1.00 &  8.50 --  8.75 & $ 39.28^{+0.46}_{-0.60}$ & $            -0.63^{+0.55}_{-0.72}$ & $\phantom{-}$---\\
 0.50 --  1.00 &  8.75 --  9.00 & $ 39.56^{+0.19}_{-0.35}$ & $            -0.30^{+0.23}_{-0.42}$ & $            -0.77$\\
 0.50 --  1.00 &  9.00 --  9.25 & $ 39.74^{+0.23}_{-0.36}$ & $            -0.08^{+0.28}_{-0.43}$ & $            -0.60$\\
 0.50 --  1.00 &  9.25 --  9.50 & $ 40.21^{+0.14}_{-0.24}$ & $\phantom{-}  0.49^{+0.17}_{-0.29}$ & $\phantom{-}$---\\
 0.50 --  1.00 &  9.50 --  9.75 & $ 40.27^{+0.07}_{-0.07}$ & $\phantom{-}  0.56^{+0.08}_{-0.08}$ & $\phantom{-}$---\\
 0.50 --  1.00 &  9.75 -- 10.00 & $ 40.40^{+0.11}_{-0.12}$ & $\phantom{-}  0.72^{+0.13}_{-0.14}$ & $\phantom{-}$---\\
 0.50 --  1.00 & 10.00 -- 10.25 & $ 40.40^{+0.08}_{-0.08}$ & $\phantom{-}  0.72^{+0.10}_{-0.10}$ & $\phantom{-}  0.53$\\
 0.50 --  1.00 & 10.25 -- 10.50 & $ 40.68^{+0.13}_{-0.15}$ & $\phantom{-}  1.05^{+0.16}_{-0.18}$ & $\phantom{-}  0.76$\\
 0.50 --  1.00 & 10.50 -- 10.75 & $ 40.71^{+0.07}_{-0.08}$ & $\phantom{-}  1.09^{+0.08}_{-0.10}$ & $\phantom{-}  0.72$\\
 0.50 --  1.00 & 10.75 -- 11.00 & $ 40.93^{+0.09}_{-0.10}$ & $\phantom{-}  1.35^{+0.11}_{-0.12}$ & $\phantom{-}  0.88$\\
 0.50 --  1.00 & 11.00 -- 11.25 & $ 41.10^{+0.12}_{-0.14}$ & $\phantom{-}  1.56^{+0.14}_{-0.17}$ & $\phantom{-}  0.67$\\
 0.50 --  1.00 & 11.25 -- 11.50 & $ 41.32^{+0.27}_{-0.31}$ & $\phantom{-}  1.82^{+0.33}_{-0.37}$ & $            -1.00$ \vspace{3pt}\\
 1.00 --  1.50 &  9.00 --  9.25 & $ 40.18^{+0.21}_{-0.34}$ & $\phantom{-}  0.27^{+0.25}_{-0.41}$ & $\phantom{-}$---\\
 1.00 --  1.50 &  9.25 --  9.50 & $ 40.39^{+0.13}_{-0.18}$ & $\phantom{-}  0.53^{+0.16}_{-0.22}$ & $\phantom{-}$---\\
 1.00 --  1.50 &  9.50 --  9.75 & $ 40.63^{+0.12}_{-0.12}$ & $\phantom{-}  0.82^{+0.14}_{-0.14}$ & $\phantom{-}$---\\
 1.00 --  1.50 &  9.75 -- 10.00 & $ 40.77^{+0.15}_{-0.15}$ & $\phantom{-}  0.99^{+0.18}_{-0.18}$ & $\phantom{-}$---\\
 1.00 --  1.50 & 10.00 -- 10.25 & $ 40.96^{+0.08}_{-0.07}$ & $\phantom{-}  1.21^{+0.10}_{-0.08}$ & $\phantom{-}$---\\
 1.00 --  1.50 & 10.25 -- 10.50 & $ 41.06^{+0.08}_{-0.11}$ & $\phantom{-}  1.34^{+0.10}_{-0.13}$ & $\phantom{-}  1.16$\\
 1.00 --  1.50 & 10.50 -- 10.75 & $ 41.10^{+0.09}_{-0.10}$ & $\phantom{-}  1.38^{+0.11}_{-0.12}$ & $\phantom{-}  1.00$\\
 1.00 --  1.50 & 10.75 -- 11.00 & $ 41.20^{+0.10}_{-0.09}$ & $\phantom{-}  1.50^{+0.12}_{-0.11}$ & $\phantom{-}  0.75$\\
 1.00 --  1.50 & 11.00 -- 11.25 & $ 41.44^{+0.14}_{-0.13}$ & $\phantom{-}  1.79^{+0.17}_{-0.16}$ & $\phantom{-}  0.66$\\
 1.00 --  1.50 & 11.25 -- 11.50 & $ 42.13^{+0.24}_{-0.28}$ & $\phantom{-}  2.62^{+0.29}_{-0.34}$ & $\phantom{-}  2.18$ \vspace{3pt}\\
 1.50 --  2.00 &  9.50 --  9.75 & $ 40.87^{+0.11}_{-0.12}$ & $\phantom{-}  0.97^{+0.13}_{-0.14}$ & $\phantom{-}$---\\
 1.50 --  2.00 &  9.75 -- 10.00 & $ 41.07^{+0.11}_{-0.13}$ & $\phantom{-}  1.21^{+0.13}_{-0.16}$ & $\phantom{-}$---\\
 1.50 --  2.00 & 10.00 -- 10.25 & $ 41.31^{+0.09}_{-0.09}$ & $\phantom{-}  1.50^{+0.11}_{-0.11}$ & $\phantom{-}$---\\
 1.50 --  2.00 & 10.25 -- 10.50 & $ 41.41^{+0.08}_{-0.09}$ & $\phantom{-}  1.62^{+0.10}_{-0.11}$ & $\phantom{-}$---\\
 1.50 --  2.00 & 10.50 -- 10.75 & $ 41.53^{+0.08}_{-0.07}$ & $\phantom{-}  1.76^{+0.10}_{-0.08}$ & $\phantom{-}  1.63$\\
 1.50 --  2.00 & 10.75 -- 11.00 & $ 41.78^{+0.10}_{-0.13}$ & $\phantom{-}  2.06^{+0.12}_{-0.16}$ & $\phantom{-}  1.81$\\
 1.50 --  2.00 & 11.00 -- 11.25 & $ 42.20^{+0.11}_{-0.13}$ & $\phantom{-}  2.57^{+0.13}_{-0.16}$ & $\phantom{-}$---\\
 1.50 --  2.00 & 11.25 -- 11.50 & $ 42.08^{+0.23}_{-0.26}$ & $\phantom{-}  2.42^{+0.28}_{-0.31}$ & $            -1.00$ \vspace{3pt}\\
 2.00 --  2.50 &  9.75 -- 10.00 & $ 41.28^{+0.12}_{-0.12}$ & $\phantom{-}  1.34^{+0.14}_{-0.14}$ & $\phantom{-}$---\\
 2.00 --  2.50 & 10.00 -- 10.25 & $ 41.48^{+0.11}_{-0.11}$ & $\phantom{-}  1.58^{+0.13}_{-0.13}$ & $\phantom{-}$---\\
 2.00 --  2.50 & 10.25 -- 10.50 & $ 41.70^{+0.07}_{-0.06}$ & $\phantom{-}  1.85^{+0.08}_{-0.07}$ & $\phantom{-}$---\\
 2.00 --  2.50 & 10.50 -- 10.75 & $ 41.79^{+0.11}_{-0.11}$ & $\phantom{-}  1.96^{+0.13}_{-0.13}$ & $\phantom{-}  1.78$\\
 2.00 --  2.50 & 10.75 -- 11.00 & $ 42.04^{+0.18}_{-0.16}$ & $\phantom{-}  2.26^{+0.22}_{-0.19}$ & $\phantom{-}  1.97$\\
 2.00 --  2.50 & 11.00 -- 11.25 & $ 42.20^{+0.22}_{-0.20}$ & $\phantom{-}  2.45^{+0.27}_{-0.24}$ & $\phantom{-}  1.88$\\
 2.00 --  2.50 & 11.25 -- 11.50 & $ 42.37^{+0.21}_{-0.25}$ & $\phantom{-}  2.66^{+0.25}_{-0.30}$ & $\phantom{-}  1.30$ \vspace{3pt}\\
 2.50 --  3.00 & 10.00 -- 10.25 & $ 41.79^{+0.12}_{-0.14}$ & $\phantom{-}  1.86^{+0.14}_{-0.17}$ & $\phantom{-}$---\\
 2.50 --  3.00 & 10.25 -- 10.50 & $ 41.88^{+0.18}_{-0.23}$ & $\phantom{-}  1.96^{+0.22}_{-0.28}$ & $\phantom{-}  1.66$\\
 2.50 --  3.00 & 10.50 -- 10.75 & $ 41.96^{+0.24}_{-0.34}$ & $\phantom{-}  2.06^{+0.29}_{-0.41}$ & $\phantom{-}  0.88$\\
 2.50 --  3.00 & 10.75 -- 11.00 & $ 42.28^{+0.34}_{-0.30}$ & $\phantom{-}  2.45^{+0.41}_{-0.36}$ & $\phantom{-}  1.79$\\
 2.50 --  3.00 & 11.00 -- 11.25 & $ 42.46^{+0.34}_{-0.27}$ & $\phantom{-}  2.66^{+0.41}_{-0.33}$ & $\phantom{-}  1.94$\\
 2.50 --  3.00 & 11.25 -- 11.50 & $ 43.06^{+0.46}_{-0.86}$ & $\phantom{-}  3.39^{+0.55}_{-1.04}$ & $            -1.00$ \vspace{3pt}\\
 3.00 --  4.00 & 10.00 -- 10.25 & $ 41.66^{+0.37}_{-0.84}$ & $\phantom{-}  1.57^{+0.45}_{-1.01}$ & $            -1.00$\\
 3.00 --  4.00 & 10.25 -- 10.50 & $ 41.82^{+0.19}_{-0.35}$ & $\phantom{-}  1.76^{+0.23}_{-0.42}$ & $            -1.00$\\
 3.00 --  4.00 & 10.50 -- 10.75 & $ 42.02^{+0.40}_{-0.73}$ & $\phantom{-}  2.01^{+0.48}_{-0.88}$ & $            -1.00$\\
 3.00 --  4.00 & 10.75 -- 11.00 & $ 42.28^{+0.30}_{-0.27}$ & $\phantom{-}  2.32^{+0.36}_{-0.33}$ & $            -1.00$\\
 3.00 --  4.00 & 11.00 -- 11.25 & $ 42.87^{+0.34}_{-0.43}$ & $\phantom{-}  3.03^{+0.41}_{-0.52}$ & $\phantom{-}  2.00$\\
 3.00 --  4.00 & 11.25 -- 11.50 & $ 42.46^{+0.75}_{-1.88}$ & $\phantom{-}  2.54^{+0.90}_{-2.27}$ & $            -1.00$ \\
\hline
\end{tabular}
}}
{\raggedright \footnotesize
$^{(a)}$ Estimate of the SFR corresponding to the X-ray main sequence, converted from the observed \Lmode\ using Model 4 in Table~\ref{tab:lxsfrmodels}.\\
$^{(b)}$ Lower limit on the SFR converted from the lower limit on the observed \Lmode\, allowing for a stellar-mass-dependent contribution from LMXBs to the X-ray luminosity (i.e. using Model 6 in Table~\ref{tab:lxsfrmodels}).
}
\end{table} 
%%%%%%%%%%%%%%%%%%%%%%%%%%%%%%%%%%%%%%%%%%%%%%%%%%

\section{The relationship between X-ray luminosity and star formation rate}
\label{sec:xraysfr}

Our results reveal an ``X-ray main sequence" in star-forming galaxies over a wide range of redshifts and stellar masses. 
To interpret this relation, we need to understand how the observed X-ray luminosities are related to the SFRs of galaxies.
As discussed in Section~\ref{sec:intro}, the overall X-ray emission from normal star-forming galaxies (i.e. with negligible current AGN activity) is predominantly due to the combined emission from HMXBs, LMXBs and diffuse hot gas.
All of these components will, to some extent, be related to the star formation history of the galaxy. 
The hot gas and HMXB components are expected to trace relatively recent star formation (and thus track the overall SFR), \more{whereas the LMXB population provides a delayed tracer of star formation that persists over much longer time-scales (thus the total LMXB luminosity may instead track the total stellar mass of a galaxy rather than the current SFR).}

\begin{figure*}
\includegraphics[width=1.1\columnwidth,trim=60 10 30 30]{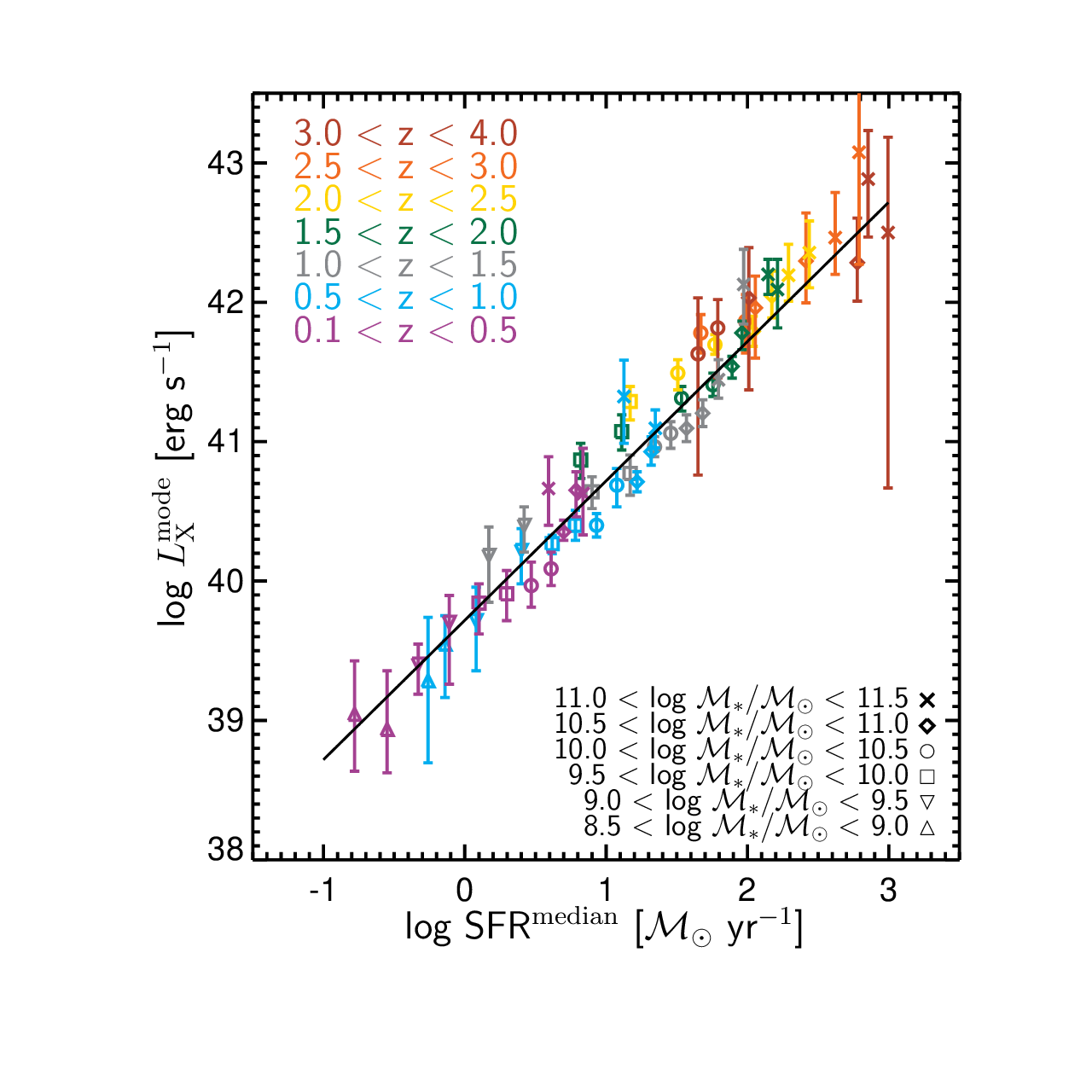}
\includegraphics[width=0.9\columnwidth,trim=80 50 50 60]{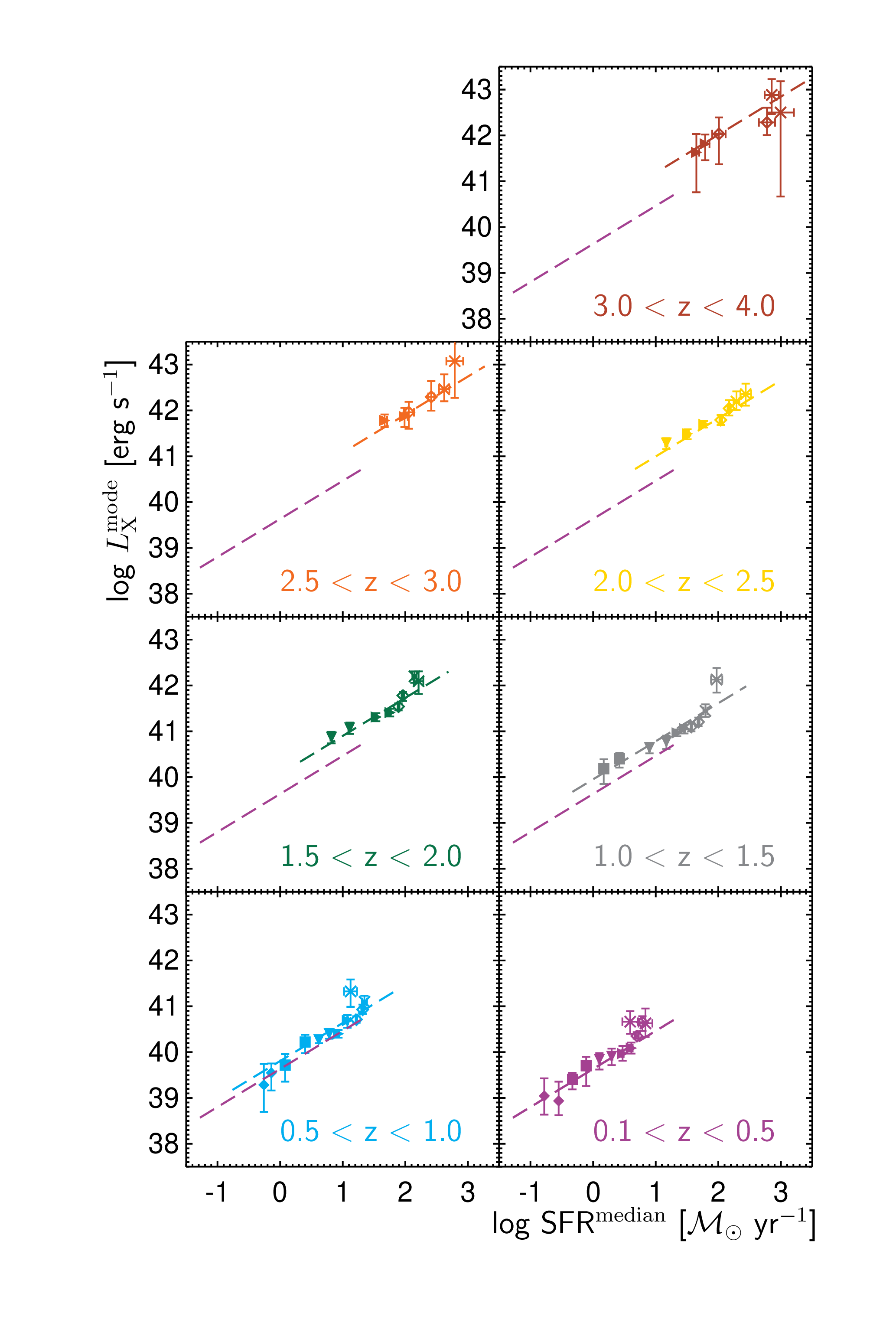}
\caption{
\refresp{
The peak of the distribution of X-ray luminosities from star-forming galaxies, \Lmode, 
compared to the median SFR of each sample (based on either the optical SED or UV+IR estimates, see Appendix~\ref{app:sedfits}). 
The large left panel shows the data from all redshift bins (indicated by the different colours) and stellar mass bins (indicated by the symbol types), compared to a simple linear scaling (black line, Model 1 in Table~\ref{tab:lxsfrmodels}).
The smaller right panels show the same data in each redshift bin, compared to our best fitting relation with a redshift dependence and a non-linear slope (Model 4 in Table~\ref{tab:lxsfrmodels}). The model evaluated at $z=0.3$ is reproduced in every panel (purple dashed line) to illustrate the evolution.
Errors on \Lmode\ are $1\sigma$ equivalent, based on the jack-knife resampling and Bayesian posteriors as in Figure~\ref{fig:lx_ms}.
}
}
\label{fig:lx_vs_sfr}
\end{figure*}

To assess how well our measurements of \Lmode\ are tracing the average SFRs in our star-forming galaxy samples, in Figure~\ref{fig:lx_vs_sfr} we compare our measurements of \Lmode\
to the median SFR (estimated using our multiwavelength ``SFR ladder", see Appendix~\ref{app:sedfits}) for galaxies in each of the stellar mass--redshift bins used in Figure~\ref{fig:lx_ms}.
We observe a roughly linear correlation between \Lmode\ and the median SFR, indicating that the X-ray luminosity is tracing, on average, the SFRs of the star-forming galaxy population.  
Performing a simple least-squares fit (with errors based on our jack-knife resampling, as described in Section~\ref{sec:msofsf} above), we find a linear relation of the form 
\begin{equation}
\log \lx [\mathrm{erg\;s^{-1}}] = 39.72\pm0.02 + \log \mathrm{SFR} [\mathcal{M}_\odot\;\mathrm{yr^{-1}}]
\label{eq:lxtosfr}
\end{equation}
or equivalently $\lx\approx 5.2\times10^{39} \times $SFR (with units given in Equation~\ref{eq:lxtosfr}), which is shown by the thin black line in \refresp{the large left panel of} Figure~\ref{fig:lx_vs_sfr}.
\refresp{We note that this overall correlation will be partly driven by changes in luminosity distance over the wide redshift range covered by our sample.}
Furthermore, the fit has a relatively large reduced $\chi^2$ ($\chi^2/\nu=1.96$) and thus does not provide a satisfactory description of the data. 
We therefore attempt to find a more general functional form to describe the relationship between SFR and \LX.
The different functional forms, along with the best-fitting parameters and the resulting $\chi^2$ values, are given in Table~\ref{tab:lxsfrmodels}.

First, we allowed for a non-linear slope in Equation~\ref{eq:lxtosfr} (denoted as Model 2 in Table~\ref{tab:lxsfrmodels}), which does not significantly improve the fit.
Next, we allowed for a redshift dependence of the form $\lx\propto (1+z)^C$ with both a linear and non-linear slope (Models 3 and 4, respectively). 
Allowing for a redshift dependence (Model 3) provides a significant improvement in $\chi^2$ (at the $>3\sigma$ level, based on an $F$ test) over the basic relation (Model 1). 
Allowing for a non-linear SFR-dependence (Model 4) subsequently improves the fit (at $>3\sigma$) compared to Model 3 and provides a statistically acceptable fit ($\chi^2/\nu\approx1$).  
The evolving relationship described by Model 4 is shown by the coloured lines in the \refresp{right panels} of Figure~\ref{fig:lx_vs_sfr}.
\refresp{The existence of a non-linear, redshift-dependent correlation between \Lmode\ and $\mathrm{SFR^{median}}$ is robust and not driven by the luminosity distance as it is observed \emph{at a fixed redshift} and the median redshift of galaxies in each data point does not change within a given redshift bin (i.e. within a single small panel of Figure~\ref{fig:lx_vs_sfr}).}

\begin{table*}
\caption{Summary of fits for the relationship between the average X-ray luminosity (\Lmode) and SED-based estimates of the SFR and stellar mass (\Mstel). Units of \LX\ in \ergs, SFR in \Msun~yr$^{-1}$, and \Mstel\ in \Msun\ are assumed for all fits.
}
\label{tab:lxsfrmodels}
\begin{tabular}{l l r l l l l l l l}
\hline
Model   &  Description  & Parameter & Value & \multicolumn{3}{l}{Star-forming galaxies} & \multicolumn{3}{l}{All galaxies, binned by SFR}\\
			&						&					&			& $\chi^2$ & $\nu$ & $\chi^2/\nu$ &  $\chi^2$ & $\nu$ & $\chi^2/\nu$ \\
\hline

1			& $\log \lx = A + \log$ SFR &  $A$    & $39.72\pm0.02$ & 117.36 & 60 & 1.96 & --- & --- & --- \vspace{10pt}\\

2			& $\log \lx = A + B\log$ SFR &  $A$   & $39.62\pm0.04$ & 111.03 & 59 & 1.88 & --- & --- & --- \\
			&											  &	$B$  & $\;\; 1.07\pm0.03$  &  &  & & & &\vspace{10pt}\\

3			& $\log \lx = A + \log$ SFR$ + C\log(1+z)$
														& $A$ & $39.46\pm0.04$   & 79.83 & 59 & 1.35 & --- & --- & --- \\
			&											& $C$ & $\;\; 0.76\pm0.12$     &  &  & & & &\vspace{10pt}\\

4  	 	& $\log \lx = A + B\log$ SFR$ + C\log(1+z)$
												& $A$ & $39.48\pm0.05$   & 66.14 & 58 & 1.14 & --- & --- & --- \\
			&							        & $B$ & $\;\; 0.83\pm0.05$     &  &  & & & &\\		
			&									& $C$ & $\;\; 1.34\pm0.20$     &  &  & & & &\vspace{10pt}\\				
\hline
5			& $\lx = \alpha(1+z)^\gamma \mstel + \beta(1+z)^\delta $SFR$^\theta$ 
												& $\log \alpha$ & $28.81\pm0.08$ &  (58.97)$^a$  & (56) & (1.05) & 92.86 & 88 & 1.06\\
			&									& $\gamma$     & $\;\; 3.90\pm0.36$ & &  & & & &\\		
			&									& $\log \beta$   & $39.50\pm0.06$  & &  & & & &\\	
			&									& $\delta$		 & $\;\;0.67\pm0.31$ & &  & & & &\\
			&									& $\theta$        & $\;\; 0.86\pm0.05$  &  &  & & & &\\
\hline
\end{tabular}\\
\raggedright
$^a$ For Model 5 the fit is performed using data from all galaxies at $0.1<z<2$, binned by SFR, $z$ and \Mstel; here we report the resulting $\chi^2$ when applying this function to the original star-forming galaxy samples (binned by $z$ and \Mstel).
\end{table*}

\begin{figure*}
\includegraphics[width=\textwidth,trim=35 10 10 0]{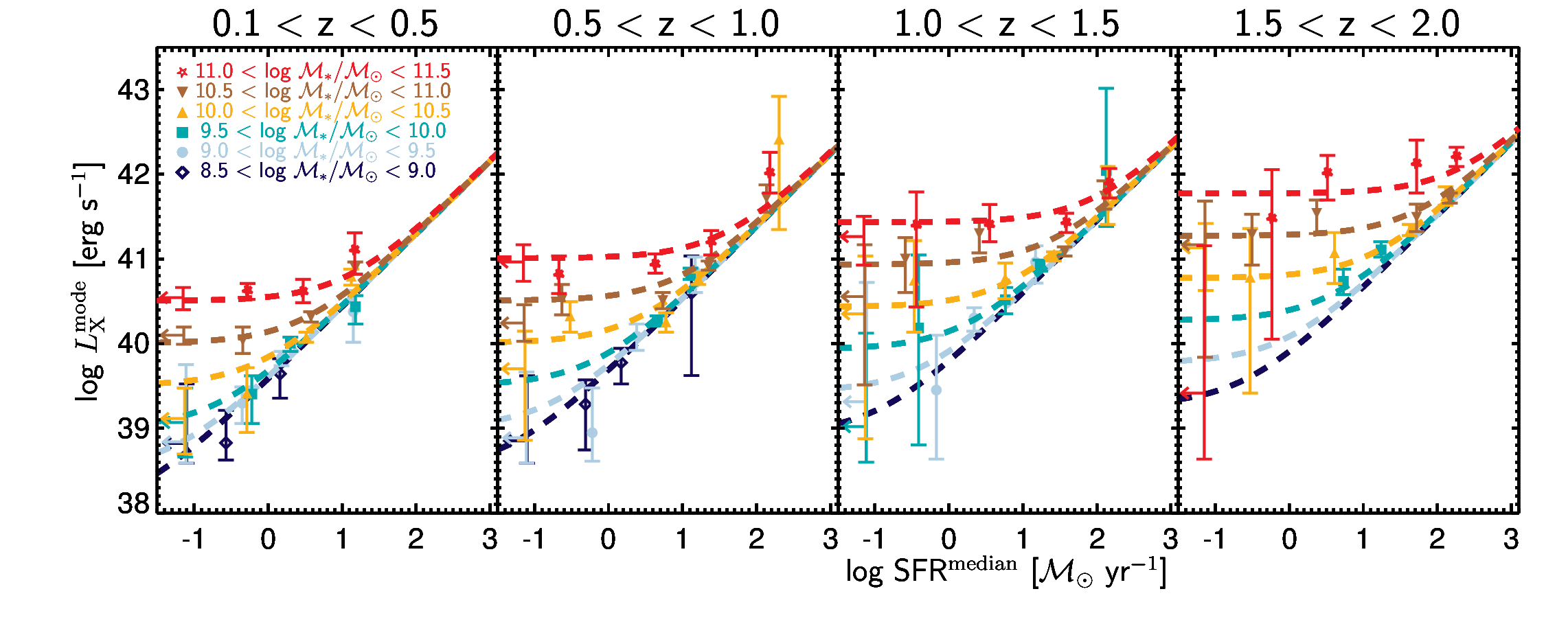}
\caption{
The peak of the distribution of X-ray luminosities, \Lmode, for all galaxies---divided into bins of redshift, stellar mass, and SFR---versus the median SFR of galaxies in a given bin (data points). 
Colours and symbol types correspond to the stellar mass bin, as indicated.
A single SFR bin is used for galaxies with SFR $<10^{-1}$~\Msun~yr$^{-1}$ and is indicated by the arrow at $\log$ SFR [$\msun$ yr$^{-1}$]$\approx-1$. 
The dashed lines indicate the best overall fit allowing for a stellar-mass-dependent contribution to the X-ray luminosity, as described by Model 5 in Table~\ref{tab:lxsfrmodels}, leading to a plateau in the \LX-SFR relation at low SFR for high stellar mass galaxies. 
}
\label{fig:lx_vs_sfr_sfrbins}
\end{figure*}

These results indicate that a redshift-dependent, linear scaling between $\log \lx$ and $\log$~SFR holds, \emph{on average}, for star-forming galaxies over the range of stellar masses and redshifts probed by our study.
We note that our SFR-\LX\ relation is for the properties of star-forming galaxies averaged over large numbers of sources at a given stellar mass and redshift. A significantly larger intrinsic scatter and potential systematic variations may exist for \emph{individual} galaxies but would not be revealed by our work.
\refresp{
In addition, while our SFR estimates appear robust, the derived \LX --SFR scaling relation could be affected by any remaining systematic biases (e.g. that depend on stellar mass, redshift, or a related parameter such as metallicity) in either the UV$+$IR or SED-based estimates that form our SFR ladder.}

In Appendix~\ref{app:qugals} we also find evidence for X-ray emission from quiescent galaxies with comparable luminosities to our ``X-ray main sequence" (at $\log \mstel/\msun\gtrsim10.5$), despite the much lower SFRs of these quiescent galaxies.
\more{
These results indicate the presence of LMXB populations and thus the overall X-ray emission may have a component that roughly scales with the total stellar mass of a galaxy, as well as a component due to HMXBs that directly scales with the current SFR \citep[see also][]{Lehmer10,Fragos13,Lehmer16}}.
To investigate this relation further, we divide our \emph{entire} galaxy sample (combining those classified as either star-forming or quiescent in the UVJ diagram) into bins of stellar mass, redshift, and SFR.
We measure $p(\log \lx \giv \mstel, z, \mathrm{SFR})$ for each of these sub-samples and determine the position of the low-luminosity peak, \Lmode. 
The data points in Figure~\ref{fig:lx_vs_sfr_sfrbins} show our results for $z<2$ (where our galaxy samples are large enough to provide meaningful constraints), with colours indicating different stellar mass bins. 
Our measurements show a clear plateau at low SFRs (for fixed \Mstel), related to the LMXB contribution to the overall X-ray emission.
The level of this plateau evolves with redshift (indicating a higher LMXB luminosity per unit galaxy stellar mass at higher redshifts). 
We fit these data with the following function (Model 5 in Table~\ref{tab:lxsfrmodels}):
\begin{equation}
\lx\; [\mathrm{erg\; s^{-1}}] 
= \alpha(1+z)^\gamma \mstel + \beta(1+z)^\delta \mathrm{SFR}^\theta
\label{eq:lx_vs_sfr_model5}
\end{equation}
where the total X-ray luminosity is assumed to be the sum of the contribution from LMXBs and HMXBs that scale with the total stellar mass and the SFR, respectively. 
We allow for a redshift dependence in each of the scale factors and a non-linear scaling with the SFR with exponent $\theta$, motivated by our Model 4.
We fit for $\alpha$, $\beta$, $\gamma$, $\delta$ and $\theta$ using the data from all galaxies at $0.1<z<2$ (binned by SFR, \Mstel\ and $z$, as shown in Figure~\ref{fig:lx_vs_sfr_sfrbins}) and report the best fit values in Table~\ref{tab:lxsfrmodels}. 
We also apply Model 5 to our original data for star-forming galaxies only (shown in Figure~\ref{fig:lx_vs_sfr}), fixing the parameters to the best-fitting values for all galaxies. 
The resulting $\chi^2/\nu=1.05$ indicates that this model also provides a good description of the data for star-forming galaxies.
In addition, the parameters describing the SFR-dependence in Model 5 ($\log \beta$, $\delta$, and $\theta$) are statistically consistent with the equivalent parameters in Model 4 (to within $2\sigma$). 
Fixing $\theta=1.0$ gives a worse fit to the SFR-binned data from all galaxies \newt{(although the other parameters do not change significantly compared with their uncertainties)} and a significantly higher $\chi^2$ than Model 5 when applied to the star-forming galaxy data only, indicating that a non-linear scaling with SFR is required.

These results indicate a more complicated relationship between \LX\ and SFR than initially revealed in Figure~\ref{fig:lx_vs_sfr}. 
For lower stellar mass galaxies ($\log \mstel/\msun \lesssim 10$) with moderate SFRs ($\gtrsim 1$~\Msun~yr$^{-1}$) the X-ray emission appears to be dominated by the HMXB population and thus the observed X-ray luminosity traces the SFR. 
However, in more massive galaxies there can be a significant contribution from LMXBs and the observed X-ray luminosity may be relatively high, even if the current SFR of the galaxy (seen at other wavelengths) is low.
Nevertheless, when considering just the star-forming galaxy samples, a simple scaling between \LX\ and SFR (Model 4) can be applied across all stellar masses and redshifts.
In such star-forming galaxies---pre-selected on the basis of their blue rest-frame optical colours indicating young stellar populations---the observed X-ray luminosities \emph{are} correlated with the SFR estimated at other wavelengths.
We discuss these findings further and compare with previous studies of the X-ray emission from normal galaxies in Section~\ref{sec:discuss_lxsfr} below.

In Figure~\ref{fig:ms_vs_literature} we use the scaling relation between \LX\ and SFR for star-forming galaxies (Model 4) to convert our measurements of the ``X-ray main sequence" to SFRs (black crosses and error bars). 
These estimates of the SFR, based on the observed \Lmode, are also listed in Table~\ref{tab:xrayms}.
In addition, we set a \emph{conservative} lower limit on the SFR (black upward pointing arrows in Figure~\ref{fig:ms_vs_literature}), based on the lower limit on \Lmode\ \emph{and} allowing for a stellar-mass dependent contribution to the X-ray luminosity (i.e. using Model 5). 
We believe these lower limits are conservative as we have pre-selected star-forming galaxies based on their optical colours; we know there must be significant star formation in such galaxies and a substantial population of HMXBs that contribute to the observed \LX.
We also note that the conversion of our \LX\ measurements to SFRs applied here is reliant on the calibration relative to our multiwavelength SFR estimates 
\refresp{and as such are \emph{not} completely independent of the UV-to-IR data. 
Nevertheless, converting our measured X-ray luminosities (and the associated uncertainties) to SFRs provides a new, distinct measurement of the main sequence of star formation} and allows us to compare our results with previous studies \refresp{that use a wide variety of multiwavelength datasets and different methodologies}
(see Section~\ref{sec:discuss_msevol} below).

\begin{figure*}
\includegraphics[width=\textwidth,trim=20 10 20 0]{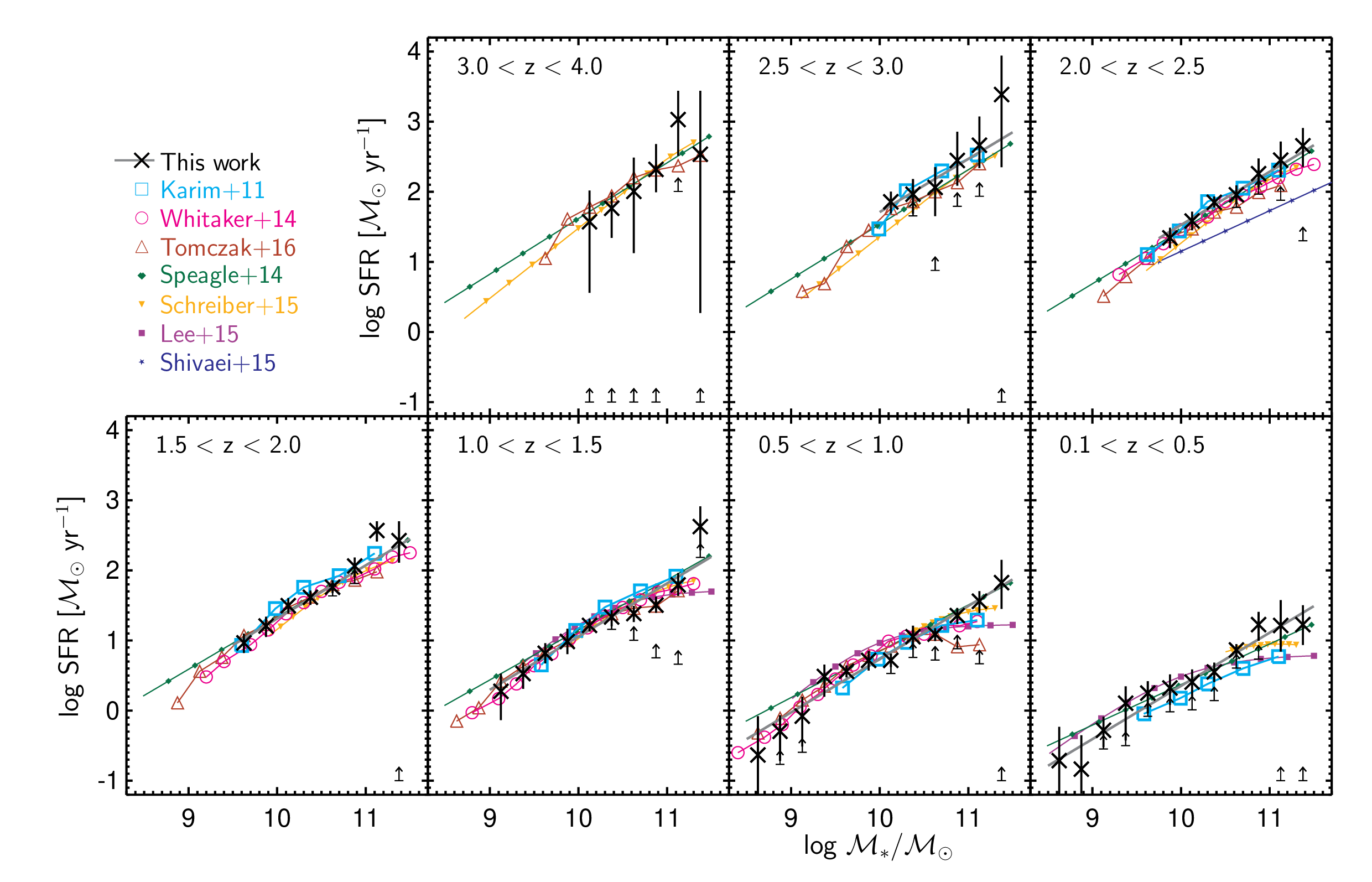}
\caption{
Our estimates of the main sequence of star formation from our X-ray analysis compared to previous measurements in the literature at other wavelengths.
Black crosses indicate the X-ray based measurements from this work, converting \Lmode\ to SFR using the scaling given by Model 4 in Table~\ref{tab:lxsfrmodels}. 
The grey lines indicate our best fit for the evolution of the X-ray main sequence to $z\approx3$ with a constant slope, converted to SFRs (see Equation~\ref{eq:ms_sfr}). 
We do not show this fit in the highest redshift panel where there is poor agreement with our measurements.
Black upward pointing arrows indicate a conservative lower limit from the X-ray data, allowing for a stellar-mass-dependent contribution from LMXBs to the observed X-ray luminosity (converting the lower limit on \Lmode\ to an SFR using Model 5 from Table~\ref{tab:lxsfrmodels}); the arrow is place at $\log$ SFR [\Msun~yr$^{-1}$] $= -1$ when the lower limit on \Lmode\ is consistent with the stellar-mass-dependent contribution only. 
Previous measurements of the main sequence of star formation are shown by the coloured symbols (we omit uncertainties for clarity). 
Large, open symbols indicate direct measurements that are tabulated in the following papers:
\citet[cyan squares]{Karim11} based on stacking of the 1.4~Ghz radio emission in the 2~deg$^2$ COSMOS field; \citet[pink circles]{Whitaker14} based on stacking of the UV+IR(24\micron) emission in the CANDELS-3DHST fields; and \citet[dark red triangles]{Tomczak16} based on stacking of UV+IR(24\micron) in ZFOURGE.
Smaller, solid symbols instead indicate parametrizations of the main sequence from \citet{Speagle14}, \citet{Schreiber15}, \citet{Lee15} and \citet{Shivaei15}. 
Previous results are only shown over the range of stellar masses and redshifts covered by a given study; the exception is \citet{Speagle14}, who compiled multiple earlier results from the literature, covering a wide stellar mass and redshift range, and thus we show their best fit across all stellar masses in all redshift panels. 
For \citet{Karim11}, we show their results at $z=0.2$--0.4, 0.6--0.8, 1.2--1.6 and 1.6--2.0 in our lowest four redshift panels (the redshift bins are identical for $z=$2.0--2.5 and 2.5--3.0). 
For \citet{Tomczak16}, we take the logarithmic mean of their narrower redshift bins for $z=0.5$--1.0 and 1.0--1.5. The \citet{Whitaker14} work matches our redshift bins exactly (over the range considered). For \citet{Lee15}, we show the best fits in their $z=0.25$--0.46, 0.63--0.78, and 1.11--1.30 bins in our three lowest redshift panels. 
For \citet{Shivaei15} we show the best fit to a sample at $z=2.09$--2.61.
For \citet{Speagle14} and \citet{Schreiber15} we show the best fit including redshift evolution, evaluated at the centre of our redshift bins.
}
\label{fig:ms_vs_literature}
\end{figure*}

\section{Discussion}
\label{sec:discuss}

\newt{
In this paper, we measure an ``X-ray main sequence" in star-forming galaxies (Section~\ref{sec:msofsf}) and determine the scaling between \LX\ and SFR (Section~\ref{sec:xraysfr}). 
This section discusses our results and compares with previous studies.
In Section~\ref{sec:discuss_lxsfr} we discuss our measurements of the scaling between X-ray luminosity, SFR, and \Mstel, compare with previous studies, and discuss \more{the constraints our results place on the evolution of X-ray binary populations within galaxies.}
Section~\ref{sec:discuss_agn} discusses whether emission from AGN could be contaminating our measurements and influencing our results.
In Section~\ref{sec:discuss_msevol} we convert our X-ray main sequence to SFRs, compare with previous studies of the main sequence of star formation, and discuss the implications of our X-ray--based measurements.
}

\subsection{X-ray binary populations and the X-ray emission of normal galaxies}
\label{sec:discuss_lxsfr}

\newt{
In Section~\ref{sec:xraysfr} above we used multiwavelength (UV+IR or SED-based) estimates of the SFRs to determine the relationship between \LX\ and the SFR for our star-forming galaxy sample.
We also considered all galaxies (both star-forming and quiescent) and probed the X-ray emission as a function of SFR, \Mstel, and $z$, finding evidence for an additional contribution to \LX\ that scales with stellar mass.}
Here we compare with previous studies of the X-ray luminosities of normal galaxies and discuss the origin of the X-ray emission.
\newt{We first consider our \LX-SFR relation for star-forming galaxies, which primarily traces HMXBs, and 
compare with previous measurements 
(predominantly at low redshifts).
Next, we discuss the redshift evolution of the \LX-SFR relation, found in both our work and previous studies, 
as well as the non-linear scaling between \LX\ and SFR ($\lx\propto\mathrm{SFR}^{0.83}$),
that may reflect changes in the overall stellar populations (and thus the HMXBs) across our stellar-mass-selected samples of star-forming galaxies.
Finally, we discuss the additional contribution to the X-ray luminosity that scales with the stellar mass and is attributed to the LMXB population. 
}

A number of previous studies have investigated the scaling between \LX\ and SFR using samples of individual galaxies with X-ray detections \citep[e.g.][]{Ranalli03,Grimm03,Persic07,Symeonidis11}.
 \citet{Lehmer10} determined the \LX-SFR relation based on a sample of 66 local star-forming galaxies (including luminous and ultraluminous infrared galaxies with high SFRs).
Their best-fitting linear relation ($\lx \approx 1.8\times 10^{39} \times$~SFR) gives a factor $\sim3$ lower \LX\ per unit SFR than our best linear fit for star-forming galaxies ($\lx \approx 5.2 \times 10^{39} \times$~SFR: Model 1, black line in Figure~\ref{fig:lx_vs_sfr}).
However, more recent work using a sample of star-forming galaxies spanning $z\sim0-1.4$ by \citet{Mineo14} found a higher normalisation: $\lx \approx 4.0 \times 10^{39} \times $~SFR (converted to our X-ray band and assumed IMF), which is closer to our result.
These discrepancies could be due to the redshift evolution of \LX-SFR relation. 
Indeed, extrapolating our Model 3 (with a simple redshift-dependence only) gives $\lx \approx 2.9 \times 10^{39} \times$~SFR
at $z=0$, which is in closer agreement with the \citet{Lehmer10} measurement for local galaxies. 

Evidence for redshift evolution in the \LX-SFR relation has been found in a number of prior studies that is consistent with our measured evolution of $\sim (1+z)^{1.3}$  (e.g. \citealt{Basu-Zych13, Lehmer16}, but see also \citealt{Symeonidis14} who find no evidence for an intrinsic evolution in the ratio of X-ray to far-IR luminosities for star-forming galaxies).
\newt{
The evolution most likely reflects changes in the properties of the stellar populations of galaxies (e.g. stellar ages, star formation histories, metallicities) which in turn affect the number and luminosities of HMXBs within a galaxy \citep[e.g.][]{Belczynski08,Fragos13,Tzanavaris13}.
Indeed, \citet{Fragos13} predict a higher \LX\ per unit SFR at higher redshifts, primarily due to the lower metallicities of high redshift galaxies that results in more numerous and more luminous HMXBs  \citep[see also][]{Kaaret11,Basu-Zych13b,Douna15}.
However, the extent of the predicted evolution is somewhat weaker than our observed trend indicating that younger galaxy ages at high redshift or differing star formation histories may also play a role.
}

\newt{
The sub-linear scaling between \LX\ and SFR that we observe for our stellar-mass-selected samples of star-forming galaxies (i.e. $\lx \propto \mathrm{SFR}^{0.83}$, based on Model 4) may also be related to changes in the underlying stellar populations that alter the properties of HMXBs. 
Our highest SFRs at a given redshift correspond to higher stellar mass samples of galaxies (see Figure~\ref{fig:lx_vs_sfr}). 
Thus, the expected changes in the properties of the stellar population across our mass range \citep[e.g. higher metallicities in higher mass galaxies:][]{Tremonti04,Erb06,Kewley08} may alter the X-ray luminosity per unit SFR, leading to our observed non-linear dependence.}
\newt{
Nonetheless, our results constrain the relationship between the \emph{average} X-ray luminosity (traced by the mode of the overall distribution of \LX) and the \emph{average} SFRs for different (complete) stellar mass limits.
To reveal the underlying driver of the non-linear SFR-dependence
we would need to identify and select galaxies at fixed redshift, stellar mass, SFR, age, metallicity etc. and then study their X-ray properties, which is not possible with the current dataset.
Ultimately, combining X-ray luminosity measurements with data at other wavelengths may allow constraints to be placed on a variety of galaxy properties \citep[e.g. both SFR and metallicity, see][]{Brorby16}.
}

\newt{
We also find evidence for a contribution to the X-ray luminosity that \more{appears to scales more directly with the galaxy stellar mass} (independent of the non-linear SFR-dependent contribution), when considering \emph{both} star-forming and quiescent galaxies and thus probing a broader range of SFR for a given stellar mass and redshift (see Figure~\ref{fig:lx_vs_sfr_sfrbins}).
This contribution to the X-ray luminosity is most likely due to LMXBs, which are a long-lived population that
\more{trace the result of star formation within the last $\sim100$~Myr -- 3~Gyr}, and 
thus will scale more closely with the total stellar mass of a galaxy than the current SFR.}
\citet{Lehmer10} also found that a scaling of  \LX\ with both stellar mass and SFR (corresponding to the LMXB and HMXB populations, respectively) reduced the scatter compared to a single scaling between \LX\ and SFR. 
In addition, we find that the stellar-mass-dependent contribution to \LX\ evolves strongly with redshift, $\sim(1+z)^{3.9}$ (Model 5),
indicating that the overall X-ray luminosity from LMXBs per unit stellar mass is greater at higher redshifts.
\citet{Lehmer16} stacked samples of galaxies in the CDFS (using recently acquired data bringing the total exposure to $\sim6$~Ms) and also found evidence for a stellar-mass and redshift dependent contribution to the observed X-ray luminosities of normal galaxies, given by $\alpha(z) \approx 1.6 \times 10^{29}(1+z)^{3.8}$ (based on the 0.5--2~keV band, converted here to our X-ray band and IMF).
This relation has a similar redshift dependence and a slightly higher normalisation than our relation, but is generally consistent with our work considering the differing methodologies and potential systematic issues in \Mstel\ and SFR estimates. 
\citet{Lehmer16} and \citet{Fragos13} both attribute the strong evolution of the total X-ray luminosity from LMXBs to the younger ages of galaxies at higher redshifts \citep[resulting in higher mass donor stars in LMXBs, increasing the X-ray luminosity per unit stellar mass, but see also][]{Zhang12}.
Ultimately, the total observed X-ray luminosity from LMXBs in a galaxy will reflect the star formation history of a galaxy and thus trace the relic of star formation over longer time-scales than the HMXB emission or SFR tracers at other wavelengths. 
\more{The strong evolution of the X-ray luminosity for our quiescent galaxy sample, $\sim(1+z)^{3.8}$, indicates that at high redshift the quiescent population may be dominated by galaxies where the star formation has been quenched relatively recently.}

\more{Our results place important constraints on the formation and evolution of X-ray binary populations out to $z\sim4$ and indicate that large populations of HMXBs are forming at high redshifts.
Some fraction of these short-lived HMXB systems will form neutron star--neutron star, black hole--neutron star or black hole--black hole binaries that will eventually merge and could be detected by Advanced LIGO and future gravitational wave experiments.
Indeed, the initial Advanced LIGO discoveries indicate that black hole binary systems may be fairly common \citep[e.g.][]{Abbott16b,LIGO16}, although we defer a direct comparison of merger rates with our constraints on the high-redshift HMXB population to future studies \citep[see also e.g.][]{Belczynski08}.
}

In conclusion, the overall X-ray luminosity (from both HMXBs and LMXBs) provides a unique and independent probe of star formation in galaxies that will trace the history of star formation on different time-scales 
to estimates at other wavelengths (e.g. UV emission that traces recent, unobscured star formation; or far-IR emission that traces dusty, obscured star formation on slightly different time-scales).
Our estimates of \Lmode\ appear to provide a robust and relatively direct tracer of the SFRs in \emph{star-forming} galaxies over the wide range of stellar mass and redshift probed by our study; 
however, X-ray based estimates of SFR remain subject to uncertainties related to the underlying stellar populations, star-formation histories, and metallicities of the galaxy population.  
SFR estimators at other wavelengths are also affected by these issues (see e.g.~\citealt{Madau14} for an overview and discussion).
An advantage of the X-ray luminosity as an SFR tracer is that it should be relatively immune to the effects of dust, which severely affects UV-based SFR estimates and must be modelled correctly to determine IR-based SFRs. 
The sensitivity of current X-ray observatories limits the use of X-rays as an SFR tracer for individual galaxies, although our advanced analysis provides estimates of the underlying distributions and---most crucially---recovers the overall mode of the population, reducing biases compared to a more basic stacking analysis and allowing X-ray data of varying depths to be combined robustly. 
The existence of an ``X-ray main sequence" with a single slope over a wide range of stellar masses and redshifts thus places important constraints on the evolution of the galaxy population (see Section~\ref{sec:discuss_msevol} below).

\subsection{AGN contamination}
\label{sec:discuss_agn}

The majority of the star-forming galaxies in our sample are not individually detected in our \textit{Chandra} X-ray imaging (at our nominal detection threshold).
For most of the galaxies that \emph{are} detected, the X-ray emission tends to be associated with an AGN with  $\lx \gtrsim 10^{42}$~\ergs, not the galactic processes that we are tracking in this paper. 
A relatively weak AGN can easily overwhelm the X-ray emission associated with star formation in a galaxy \citep[e.g.][]{Brandt15}.
Thus, it is worth considering the potential impact of AGNs on our estimates of \Lmode\ for the star-forming galaxy samples and our ability to trace the main sequence of star formation via the X-ray emission.

In our analysis, we do not separate or exclude AGNs (other than optically luminous QSOs, that are also very X-ray bright), instead estimating the overall distributions of X-ray luminosities.\footnote{It is clear from our results that using a single luminosity limit of e.g.~$\lx >10^{42}$~\ergs\ to distinguish AGN from star-forming galaxies is inappropriate when considering galaxies of various stellar masses and redshifts \citep[see also discussion in][]{Aird15}.}
By modelling the full distribution we are able to account for the broad, roughly power-law distributions of AGN luminosities \citep[see also Paper II]{Aird12,Bongiorno12,Georgakakis14,Bongiorno16,Jones16} and subsequently separate this component from the distinct low-luminosity peaks that we associate with galactic processes. 
Our technique, using the available X-ray information for all galaxies, allows for  
AGN emission from galaxies that are not detected in the X-ray imaging (at the nominal detection threshold). 
Our method should therefore be significantly more robust than previous studies, particularly stacking analyses that may not reliably exclude all AGN and could give a biased indication of the ``average" luminosity from star formation \citep[e.g.][]{Symeonidis14,Lehmer16}.
For the AGN emission to have a significant impact on our results would require a very large population of AGNs that all have luminosities within a very narrow range, close to our observed peaks. 
Furthermore, a very high AGN duty cycle \newt{($\gtrsim 60$~per cent, based on our measured luminosity distributions)} would need to be inferred to make a substantial contribution and explain the peak.
Thus, we expect any systematic effects due to AGN contamination are much smaller than our (relatively conservative) estimates of the uncertainties in \Lmode\ and our ability to trace the main sequence of star formation via the X-ray emission is not significantly affected by the presence of AGNs.

\subsection{The evolution of the main sequence of star formation}
\label{sec:discuss_msevol}

In this section 
we compare our X-ray-based work with previous measurements of the main sequence of star formation.
\refresp{The primary finding of our work is the identification of an ``X-ray main sequence": a linear relationship between $\log \mstel$ and $\log \lmode$, the mode of the distribution of X-ray luminosities, with a slope of $b=0.63\pm0.03$ and a normalization that evolves as $(1+z)^{3.79\pm0.12}$ (see Section~\ref{sec:msofsf}, Equation~\ref{eq:lxms_all} and Figure~\ref{fig:lx_ms}). 
Our X-ray main sequence provides an independent tracer of the main sequence of star formation.
However, to compare with previous measurements, we use Model~4 from Table~\ref{tab:lxsfrmodels}, derived in Section~\ref{sec:xraysfr}, to convert \Lmode\ to SFRs (indicated by the black crosses in Figure~\ref{fig:ms_vs_literature}).
We note that this \LX--SFR relation relies on our multiwavelength measurements of SFRs from our UV-to-IR SFR ladder.
Thus, while our measurement of the main sequence of star formation is distinct from previous studies, it is not completely independent of the UV-to-IR data.}
We allow for the uncertainty in this conversion due to the potential contribution of LMXBs by using Model~5 (with a stellar-mass-dependent contribution to the X-ray luminosity, see Table~\ref{tab:lxsfrmodels}) to convert the lower limit on \Lmode\ to a conservative lower limit on SFR (indicated by the upward pointing arrows in Figure~\ref{fig:ms_vs_literature}). 

Using the conversions described above, our X-ray measurements correspond to a main sequence of star formation, given by
\begin{equation}
\log \mathrm{SFR}\; [\msun\; \mathrm{yr}^{-1}] \approx -7.6 + 0.76 \log \left(\frac{\mstel}{\msun}\right) + 2.95\log(1+z)
\label{eq:ms_sfr}
\end{equation}
i.e. a main sequence with a constant slope $m\approx 0.76\pm0.06$ and a normalization that evolves with redshift as $\sim(1+z)^{2.95\pm0.33}$, shown by the grey lines in Figure~\ref{fig:ms_vs_literature}, where the uncertainties are propagated from our fit to the X-ray main sequence (Equation~\ref{eq:lxms_all}) and the \LX-SFR relation (Model 4 in Table~\ref{tab:lxsfrmodels}).
Based on our X-ray measurements, there is no evidence for a turnover or flattening of the slope of the main sequence at high stellar masses or low redshifts. 
However, we cannot rule out such a turnover at $z\lesssim1$ and $\log \mstel/\msun\gtrsim10.5$ because of the uncertainties in how accurately the X-ray luminosity traces SFR in this regime (due to the potential contribution from LMXBs). 
\refresp{Furthermore, our estimate of the slope and evolution of the main sequence in the SFR--\Mstel\ plane relies on our calibration of the \LX--SFR relation.
Any systematic uncertainties in our UV-to-IR SFR estimates used for the cross-calibration (e.g. a stellar mass dependence for lower metallicty galaxies) could bias our estimate of the slope and evolution of the main sequence of star formation.
However, there is currently no evidence of a significant impact on our results from such systematic biases.
}

In Figure~\ref{fig:ms_vs_literature}, we compare our X-ray based estimates of the main sequence to a number of previous studies based on SFR estimators at other wavelengths.
In general, there is good agreement between our work and the previous studies, revealing a main sequence that rises with stellar mass and evolves strongly with redshift. 
A general agreement with prior measurements 
\refresp{is expected as we have calibrated \LX\ as a SFR tracer using our UV-to-IR SFR ladder.
Nevertheless, our method is distinct from prior studies, relying only on the overall calibration between \LX\ and UV-to-IR based SFRs rather than using such estimates directly for each individual galaxy.}
Indeed, using the UV-to-IR SEDs to estimate \emph{both} the stellar mass and SFR \refresp{directly} for individual galaxies can lead to correlated uncertainties that may bias any measurement of the main sequence \citep[][]{Reddy12,Speagle14}.
Measurements of the main sequence that are based on far-IR data \citep[e.g. from \textit{Herschel}:][]{Schreiber15,Lee15} or radio data \citep[e.g. the 1.4~Ghz stacking analysis of][]{Karim11}
should be less susceptible to these issues, although the limited depth of such data can introduce additional selection effects and biases \citep[see][]{Speagle14,Ilbert15}.

We find that the normalization of the main sequence evolves strongly with redshift as $\sim(1+z)^{2.95}$ to $z\sim3$.
This evolution is consistent with most recent studies \citep[e.g.][]{Whitaker14,Schreiber15,Tasca15}.
\citet{Karim11} found a somewhat stronger evolution $\sim(1+z)^{3.9}$ based on stacking of 1.4~Ghz radio data, although their measurements (cyan squares in Figure~\ref{fig:ms_vs_literature}) are generally consistent with our X-ray estimates, given the uncertainties, falling slightly below our data at low redshifts and slightly above at higher redshifts due to the stronger evolution.
\citet{Speagle14} also noted the discrepancy in radio-based measurements of the main sequence and proposed a redshift-dependent correction factor to bring them in line with the ${\sim(1+z)^{2.8}}$ evolution found by other studies.
The origin of this redshift-dependent systematic offset remains unclear.
Interestingly, we find a very similar evolution to \citet{Karim11}, $\sim(1+z)^{3.8}$, in our original measurements of the X-ray main sequence (i.e. tracking \Lmode\ as a function of redshift, see Section~\ref{sec:msofsf} and Figure~\ref{fig:lx_ms}). The strength of this evolution is counteracted by the redshift dependence of the \LX-SFR relation (derived in Section~\ref{sec:xraysfr}).
Thus, the radio-based measurements may suffer from a systematic dependence on metallicity or a related physical property of star-forming galaxies in a similar way to our X-ray measurements.
Alternatively, the UV-to-IR based SFR estimates could be systematically biased and the true evolution of the main sequence may be closer to $(1+z)^{3.8}$. 
Further study and comparison of X-ray, radio and UV-to-IR SFR estimates out to high redshifts is required to fully reconcile these discrepancies  \citep[see also][]{Basu15,Magnelli15}.

\citet{Shivaei15} measure the main sequence at $z=2.09-2.61$ using dust-corrected H$\alpha$ luminosities to trace the SFR. 
Their best fitting relation (shown by the small dark blue squares in the $2.0<z<2.5$ panel of Figure~\ref{fig:ms_vs_literature}) falls significantly below our X-ray based estimates and the other measurements from the literature, although our main sequence does fall within the range spanned by their data points for individual galaxies, which have a relatively large intrinsic scatter ($\sim0.3$~dex), \newt{and our slope ($m=0.76\pm0.06$) is statistically consistent with their measurement ($m=0.65\pm0.08$).}
Furthermore, Shivaei et al. find that using SED-based estimates of the SFRs yielded a steeper slope and a higher normalization, which is closer to our X-ray-based measurements and the other studies shown in Figure~\ref{fig:ms_vs_literature}.
However, using SED-based measurements of both SFR and \Mstel\ may underestimate the intrinsic scatter due to underlying correlations between the parameters, and bias measurements of the slope of the main sequence, as essentially the same data are used to estimate both \Mstel\ and SFR.
A different fitting method (treating both \Mstel\ and SFR as dependent variables) also led to a steeper slope.
These findings highlight the potential issues in any measurements of the slope of the main sequence based on UV-to-IR data only.
A major advantage of our study is that our X-ray-based measurements provide independent constraints on the average SFRs of galaxies (at a given stellar mass), with realistic uncertainties.

The slope of the main sequence ($m$) and the existence of any turnover or flattening at the highest stellar masses has been the subject of much debate \citep[e.g.][]{Whitaker14,Shivaei15,Johnston15}. 
The slope reflects the efficiency at which stars are forming relative to the total stellar mass.
A sub-linear slope indicates that specific SFRs are lower, on average, in higher mass galaxies and thus the star-formation efficiency is lower. 
Recent studies probing lower mass galaxies have instead suggested a roughly linear slope (i.e. $m\approx1$) at low stellar masses, with a break or turnover at $\mstel \gtrsim10^{10}\msun$. The slope of the main sequence above this break becomes progressively flatter as redshift decreases \citep[e.g.][]{Whitaker14,Schreiber15,Tomczak16}. 
Earlier studies (sensitive to a more limited range of stellar mass and redshift) may have effectively found the same behaviour but parametrized this pattern with a single slope that becomes flatter toward lower redshifts \citep[e.g.][]{Speagle14}. 
A linear slope at low masses, indicating a constant star-formation efficiency per unit stellar mass built up, is in agreement with some theoretical predictions \citep[e.g.][]{Somerville08,Sparre15}, while a break or turnover 
at higher masses could indicate a reduction in star formation efficiency or the onset of quenching in higher mass galaxies.

Our X-ray based measurements are consistent with a main sequence of star formation with a constant slope of $m=0.76\pm0.06$ (see Equation~\ref{eq:ms_sfr} above) across a wide range of stellar masses ($8.5\lesssim \log \mstel/\msun \lesssim11.5$) and out to at least $z\sim3$, ostensibly ruling out a linear ($m=1$) slope at the $\gtrsim 3\sigma$ level. 
However, the uncertainty in this slope is somewhat larger at low masses ($\log \mstel/\msun \lesssim 10.5$) and it is clear in Figure~\ref{fig:ms_vs_literature} that our X-ray based measurements are consistent with previous studies that measure a linear ($m\approx1$) slope in this regime \citep[e.g.][]{Whitaker14,Schreiber15}.

At higher stellar masses ($\log \mstel/\msun \gtrsim10.5$)  and lower redshifts ($z\lesssim1$) our X-ray-based main sequence is significantly steeper than prior estimates and lacks the strong turnover found by some recent studies \citep[e.g.][]{Lee15,Schreiber15,Tomczak16}.
The impact of dust on other SFR estimators is expected to be more severe at higher stellar masses; thus our X-ray based measurements may be more robust and reliable in this regime, revealing a higher level of star formation and thus a steeper main sequence that does not turn over.
Our conservative lower limits (black arrows in Figure~\ref{fig:ms_vs_literature}), however, are consistent with the high-mass turnover seen in prior studies. 
Nevertheless, it is important to note that our measurement of an X-ray main sequence (in terms of \LX) with a constant slope that continues to rise to high stellar masses and does not turn over is \emph{independent} of our calibration relative to UV-to-IR SFRs.
The existence of the X-ray main sequence indicates a fundamental connection between the total stellar mass and the X-ray luminosity of galaxies. 
Interpreting this relation relies on an understanding of how the HMXB and LMXB populations are related to other physical galaxy properties (e.g. SFR, stellar mass, metallicity, star formation history).
In broad terms, however, the X-ray luminosity will provide a tracer of the star formation within a galaxy that is insensitive to dust and may effectively average star formation over longer time-scales than estimators at other wavelengths. 
Based on this probe, there is no evidence for 
a change in the efficiency of star formation at high stellar masses or the onset of quenching within the star-forming galaxy population.

Fully reconciling the various measurements of the main sequence of star formation could provide important insights into the physical nature and evolution of the galaxy population (but is beyond the scope of this paper).
Different SFR estimators will be sensitive to the stellar populations of galaxies in different ways.
Thus, measurements of the main sequence are affected by metallicities, star-formation histories and dust attenuation curves, as well as selection biases inherent to any individual method. 
Our measurements of the X-ray main sequence, providing an additional probe of the SFR with a unique dependence on these key parameters, are thus a vital addition to the current observational landscape.

\section{Summary and conclusions}
\label{sec:summary}

In this paper, we present measurements of the ``X-ray main sequence", tracing the main sequence of star formation out to $z\sim4$ using new, novel methods. Here, we summarize our work and overall conclusions. 

\begin{itemize}[leftmargin=*,itemsep=3pt]
\item
We construct stellar-mass limited samples of star-forming galaxies out to $z=4$, based on deep NIR-selected catalogues from the CANDELS/3DHST and UltraVISTA surveys.
We extract X-ray data for all sources from the deep ($\sim160$~ks -- 4~Ms) \textit{Chandra} imaging. 

\item
We measure the intrinsic probability distribution functions of X-ray luminosities at a given stellar mass and redshift.
We implement a flexible Bayesian modelling technique to fully utilize the X-ray data from all galaxies, correct for incompleteness, probe below the nominal detection limits and  accurately recover the underlying distribution.
Our measured distributions all exhibit a clear peak at low luminosities ($\lx \lesssim 10^{42}$~\ergs), which we associate with star formation processes, and a broad tail to higher luminosities (associated with AGN activity).

\item
We track the position of the low-luminosity peak (\Lmode) as a function of stellar mass and redshift, revealing an ``X-ray main sequence" \refresp{with a constant slope ($b=0.63\pm0.03$) and a normalization that evolves strongly with redshift as $(1+z)^{3.79\pm0.12}$.}
There is no evidence for a break or turnover in the X-ray main sequence at high stellar masses. 

\item
To relate our X-ray main sequence to the main sequence of star formation, we compare our measurements of  \Lmode\ with UV-to-IR based estimates of the SFR. 
For our star-forming galaxy sample, we find a sub-linear scaling between the X-ray luminosity and SFR, $\lx \propto$~SFR$^{0.83\pm0.05}$, that evolves with redshift as $\sim(1+z)^{1.3\pm0.2}$ and 
\newt{
may be related to changes in the stellar populations of galaxies with stellar mass and redshift (that affect the number and luminosities of HMXBs).}
\newt{We also consider all galaxies (both star-forming and quiescent) as a function of SFR, stellar mass and redshift and find evidence for a contribution to the X-ray luminosity that \more{appears to scale more directly with the total stellar mass and is likely due to the contribution from longer-lived LMXBs.}
We use these relations to convert our X-ray luminosities to SFRs to provide an X-ray based measurement of the main sequence of star formation.}

\item
Our X-ray based measurements are consistent with an SFR--\Mstel\ relation 
(i.e. a main sequence of star formation) with a constant slope of $m\approx0.76\pm0.06$ and a normalization that evolves with redshift as $\sim(1+z)^{2.95\pm0.33}$.
However, the stellar-mass-dependent contribution to the X-ray luminosity introduces an uncertainty such that we are unable to rule out a turnover in the main sequence at high stellar masses and low redshifts.
\end{itemize}

Our X-ray based measurements provide a robust, independent tracer of the main sequence that is not significantly affected by dust and may trace star formation over longer time-scales than probes at other wavelengths. 
Our work illustrates the potential of X-ray surveys to probe star formation over cosmic time.
The deepest \textit{Chandra} surveys are now reaching flux limits (over small areas) that enable direct detection of normal star-forming galaxies out to at least $z\sim1.5$. 
In the future, the \textit{Athena} X-ray observatory \citep{Nandra13} will reach comparable X-ray flux limits over areas of several square degrees \citep{Aird13b}, providing large, statistical samples of X-ray detected star-forming galaxies that will enable vital constraints on both the slope and scatter of the main sequence of star formation over $\sim$half of cosmic time.  
The all-sky X-ray surveys carried out with \textit{eROSITA} \citep{Merloni12} will also directly detect large numbers of star-forming galaxies in the local Universe, enabling detailed studies of X-ray binary populations as a function of galaxy properties and providing a critical baseline for higher redshift studies. 
In the meantime, combining the full range of deep and wide \textit{Chandra} surveys and using advanced statistical techniques to push to the limits of these data, as in our work, will allow further studies of the relation between the X-ray emission and the physical properties of galaxies, placing unique constraints on the star formation history of the universe.

\section*{acknowledgements}
We thank the referee for helpful comments that have improved this paper.
We acknowledge helpful discussions with Kirpal Nandra.
JA acknowledges support from ERC Advanced Grant FEEDBACK 340442. 
ALC acknowledges support from NSF CAREER award AST-1055081.
AG acknowledges the {\sc thales} project 383549 that is jointly funded by the European Union  and the  Greek Government  in  the framework  of the  programme ``Education and lifelong learning''. 
This work is based in part on observations taken by the 3D-HST Treasury Program (GO 12177 and 12328) with the NASA/ESA HST, which is operated by the Association of Universities for Research in Astronomy, Inc., under NASA contract NAS5-26555.
Based in part on data obtained with the European Southern Observatory Very Large Telescope, Paranal, Chile, under Large Program 185.A-0791, and made available by the VUDS team at the CESAM data center, Laboratoire d'Astrophysique de Marseille, France.
The scientific results reported in this article are based to a significant degree on observations made by the \textit{Chandra} X-ray Observatory.

\appendix

\section{Multiwavelength estimates of star formation rates and spectral energy distribution fitting for galaxies}
\label{app:sedfits}

In this appendix, we describe our construction of an ``SFR ladder" \citep[e.g.][]{Wuyts11} to provide estimates of the SFR for every galaxy in our sample based on either the UV+IR emission or (when a 24\micron\ detection is unavailable) from fitting the UV-to-MIR spectral energy distributions (SEDs) with stellar population synthesis (SPS) models.
Our SED fitting is performed using the FAST code \citep{Kriek09}, subject to a number of modifications that are described in this appendix. 

A widely used and reliable method of estimating the SFR of a galaxy, that we adopt when possible, is to sum the unobscured star formation traced by the UV light and the obscured star formation traced by the total IR luminosity \citep[e.g.][]{Kennicutt98,Gordon00,Wuyts08,DominguezSanchez14,Ilbert15}.
The total SFR, assuming a \citet{Chabrier03} initial mass function (IMF), is given by
\begin{equation}
\mathrm{SFR}_\mathrm{UV+IR} [\msun\;\mathrm{yr}^{-1}] = 
 1.09\times 10^{-10} ( L_\mathrm{IR} + 2.2 L_\mathrm{UV})\; [L_\odot]
 \label{eq:sedsfr}
 \end{equation}
where $L_\mathrm{IR}$ is the total rest-frame 8--1000\micron\ IR luminosity and $L_\mathrm{UV}$ is the total rest-frame 1216--3000~\AA\ UV luminosity \citep{Bell05,Whitaker14}.
The additional factor 2.2 accounts for UV emission outside the 1216--3000~\AA\ range. 
We estimate $L_\mathrm{UV}$ from the monochromatic UV luminosity at rest-frame 2800~\AA, 
$L_\mathrm{UV}\approx1.5 \nu L_{\nu 2800}$, where $L_{\nu 2800}$ is interpolated from the observed photometry of a galaxy using our SED fits (described below, although this quantity is insensitive to the details of the SED modelling). 
To estimate $L_\mathrm{IR}$, we require a $>3\sigma$ detection in the \textit{Spitzer}/MIPS 24\micron\ imaging\footnote{We note that \textit{Herschel} imaging at longer wavelengths with PACS and SPIRE is available in our fields \citep{Elbaz11,Lutz11,Magnelli13} that could be used to probe the peak of the IR emission or fit the far-IR SED. However, given the limited depths, only a very small fraction of the galaxies in our sample are actually detected by \textit{Herschel} ($\lesssim7$ per cent are detected in the PACS 100\micron\ imaging) and thus for this work we only consider the deeper \textit{Spitzer} 24\micron\ data.}
and scale the observed 24~\micron\ flux density to $L_\mathrm{IR}$ assuming a single template, the log average of the \citet{Dale02} templates \citep[see][]{Wuyts08,Elbaz10,Muzzin13,Whitaker14}.
\refresp{These estimates should be relatively insensitive to changes in the level of dust.
Decreased levels of dust (e.g. due to lower metallicities) result in less obscured star formation, traced at 24\micron, which is counteracted by an increase in the unobscured star formation, traced in the UV.
However, a recent study \citep{Shivaei16b} proposed that the extrapolation of the 24\micron\ emission to a total IR luminosity is dependent on metallicity at higher redshifts ($z\sim2$), when poly-aromatic hydrocarbon (PAH) features at rest-frame $\sim 7.7$\micron\ start to enter the observed 24\micron\ band. 
As there is currently no consensus on this effect, however, we retain the standard approach of adopting a single, luminosity-independent IR template to extrapolate from the observed 24\micron\ flux to the total IR luminosity at all stellar masses and redshifts \citep[see][]{Wuyts08,Tomczak16,Straatman16}.}

Combining the observed UV and IR(24\micron) emission should provide an accurate estimate of the total SFR.
However, 24\micron\ detections are only available for $\sim 40$ per cent of our star-forming galaxy sample. 
For the remainder, we must estimate the SFR based on the UV-to-MIR SED and apply a correction for any obscuration due to dust. 
This is achieved by fitting the SEDs with SPS models using FAST. 
The SED fits are also used to estimate the stellar masses for all of the galaxies in our sample.

FAST determines galaxy physical properties by generating SPS model templates over a grid of stellar population parameters (age, star formation time-scale $\tau$, dust content $A_V$, metallicity, and redshift), comparing the predicted broadband fluxes to the observed photometry, and adopting the template from the grid that gives the lowest $\chi^2$. 
We choose to use Flexible Stellar Population Synthesis (FSPS) models \citep{Conroy09,Conroy10} with fixed solar metallicity\footnote{
\refresp{We tested the effect of allowing a broader range of metallicities in our SED fitting ($\sim$0.1--1.5 times solar metallicity) for a representative subset of our galaxy sample. We found that our estimates of \Mstel\ changed by $<$0.02~dex and that our estimates of \SFRSED\ changed by $<$0.05~dex, on average, with no strong systematics with \Mstel\ or redshift. 
Thus, our SED fitting is relatively unaffected by the assumed metallicity, at least when combining large samples of galaxies, and we retain fixed solar metallicity in our final SED fitting for the full galaxy sample.
}}
and a \citet{Chabrier03} Initial Mass Function (IMF).
We allow for dust reddening of $A_V=0-4$~magnitudes, assuming the dust attenuation curves of \citet{Kriek13}.
We adopt ``delayed-$\tau$" models with $\tau$ in the range 0.1--10~Gyr, which allows for both exponentially declining star formation histories and linearly rising star formation histories that are more appropriate for modelling the SEDs of high-redshift ($z\gtrsim2$) galaxies \citep[e.g.][]{Maraston10,Behroozi13}.
Initially, we allow for galaxy ages in the range $\sim$100~Myr -- 13~Gyr (but see further discussion of a redshift-dependent minimum age below). 
FAST automatically excludes any templates where the age is \emph{older} than the observable universe at a given $z$.

We identified a number of issues when running the standard FAST code with the parameters described above.
First, we found that the recovered SFRs were restricted to a range of relatively discrete values.
This discreteness is due to the coarse grid of stellar population parameters.
Of more concern, we found that a non-negligible fraction of galaxies with very blue optical-to-NIR colours (classified as star-forming galaxies in the UVJ diagram) were assigned very low SFRs (in poor agreement with $\mathrm{SFR_{UV+IR}}$, where available).
Conversely, red (quiescent) galaxies could be assigned relatively large SFRs that appear at odds with their overall SEDs. 
Further investigation revealed that this issue was also related to the coarse parameter grid.
A single stellar population template can provide the best $\chi^2$, but many other templates---corresponding to more realistic SFRs---give a comparable (but higher) $\chi^2$, especially when the photometric uncertainties are large. 
Further refinement of the grid for each individual galaxy is computationally prohibitive for the large number of galaxies in our sample.
Instead, FAST includes a template error function, which introduces an additional uncertainty to account for the coarse parameter grid, intended to partly mitigate this issue.
We include the template error function but still find that a single, unrealistic template can produce the best $\chi^2$.
We therefore take a ``semi-Bayesian" approach \citep[see also][]{Moustakas13}, where we weight each template according to the $\chi_i^2$ value (for a given galaxy) and calculate a weighted average of the SFR over the entire parameter space, effectively applying a prior that prefers well-occupied regions of the parameter grid. Thus,
\begin{equation}
\mathrm{SFR}_\mathrm{SED} = \frac{\sum_i \left[ \exp(-\chi_i^2/2) \times \psi_i \right]}{\sum_i \left[\exp(-\chi_i^2/2)\right]}
\end{equation}
where $\psi_i$ denotes the SFR associated with template $i$ and the sum is taken over the entire grid of stellar population parameters. 
Applying this weighting scheme not only solves the issue of a single, outlier template being erroneously assigned to a galaxy but also eliminates the discreteness in the distribution of SFRs for our overall galaxy sample, producing a continuous range of values.
For consistency, we also apply this $\chi^2$-weighted averaging when determining the stellar mass, although this has less of an impact (changing \Mstel\ by less than 0.1~dex for the majority of galaxies and with no overall systematic shift) as the stellar masses are primarily determined by the NIR flux and are much less sensitive to the details of the SPS models. 

\begin{figure}
\includegraphics[width=\columnwidth,trim=0 48 0 0]{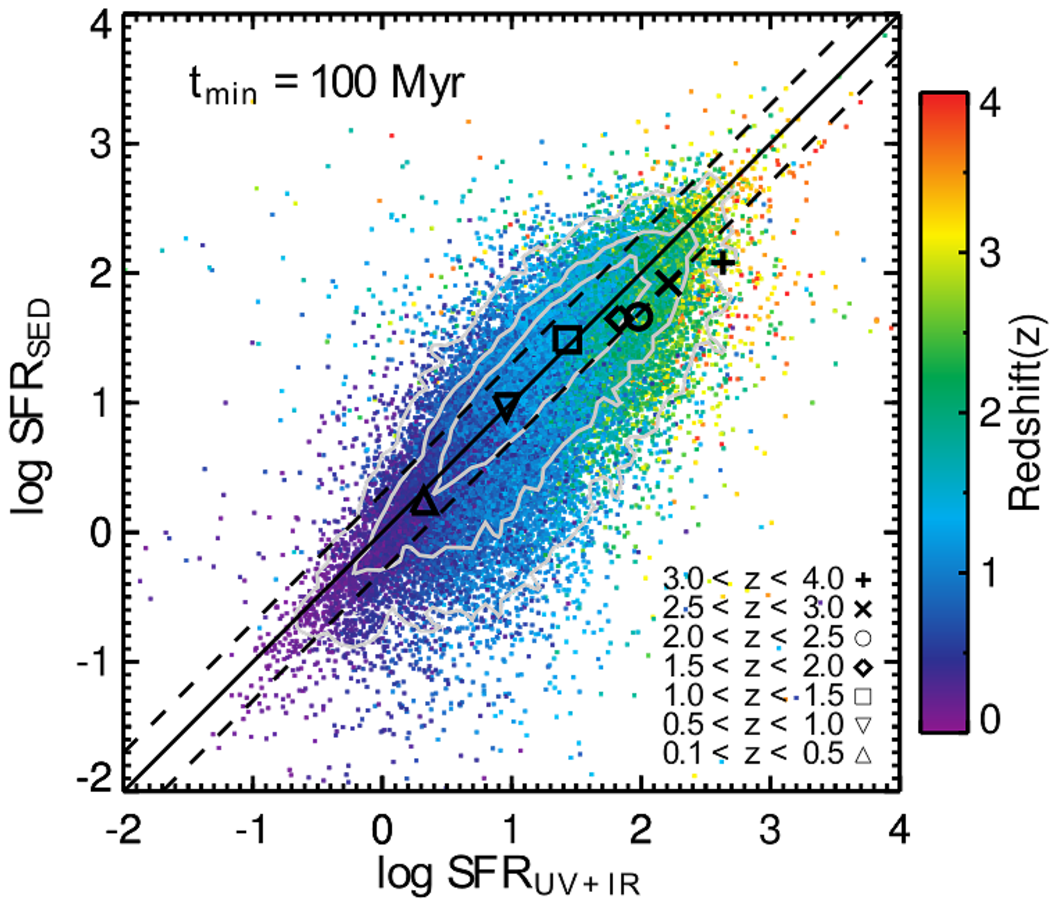}
\includegraphics[width=\columnwidth,trim=0 48 0 0]{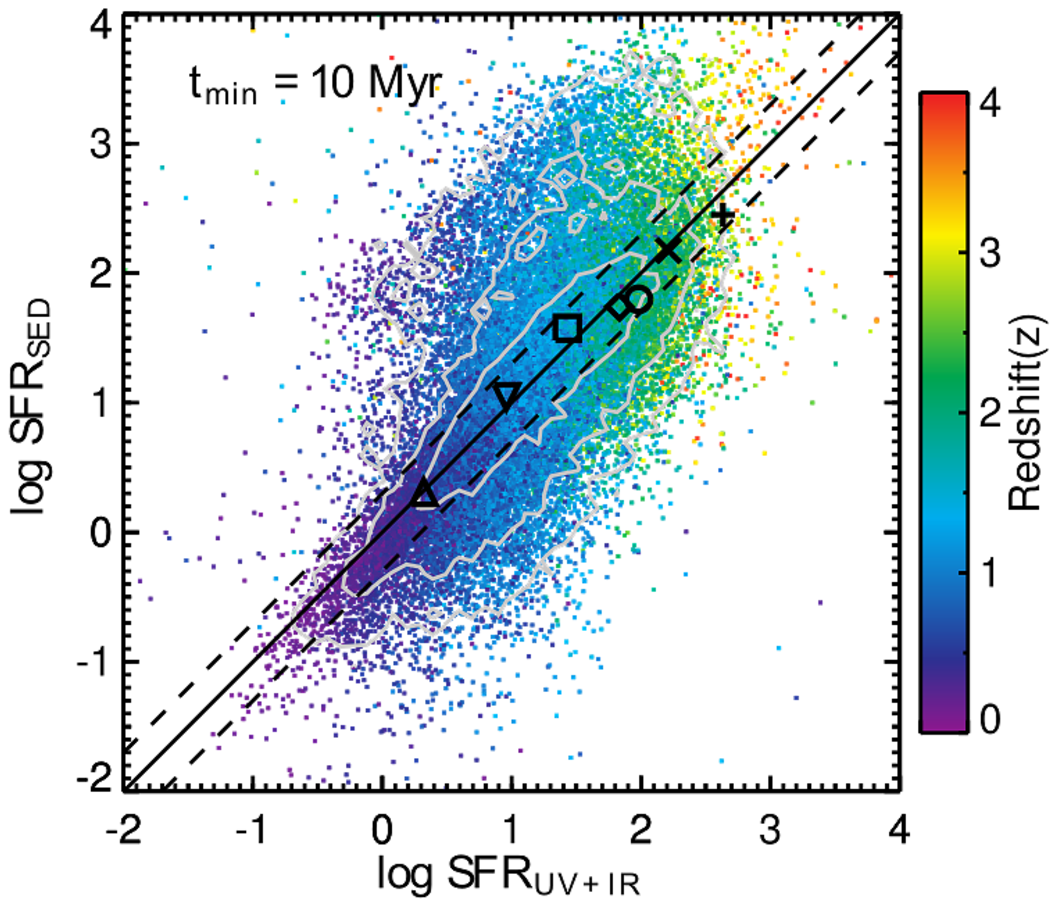}
\includegraphics[width=\columnwidth,trim=0 20 0 0]{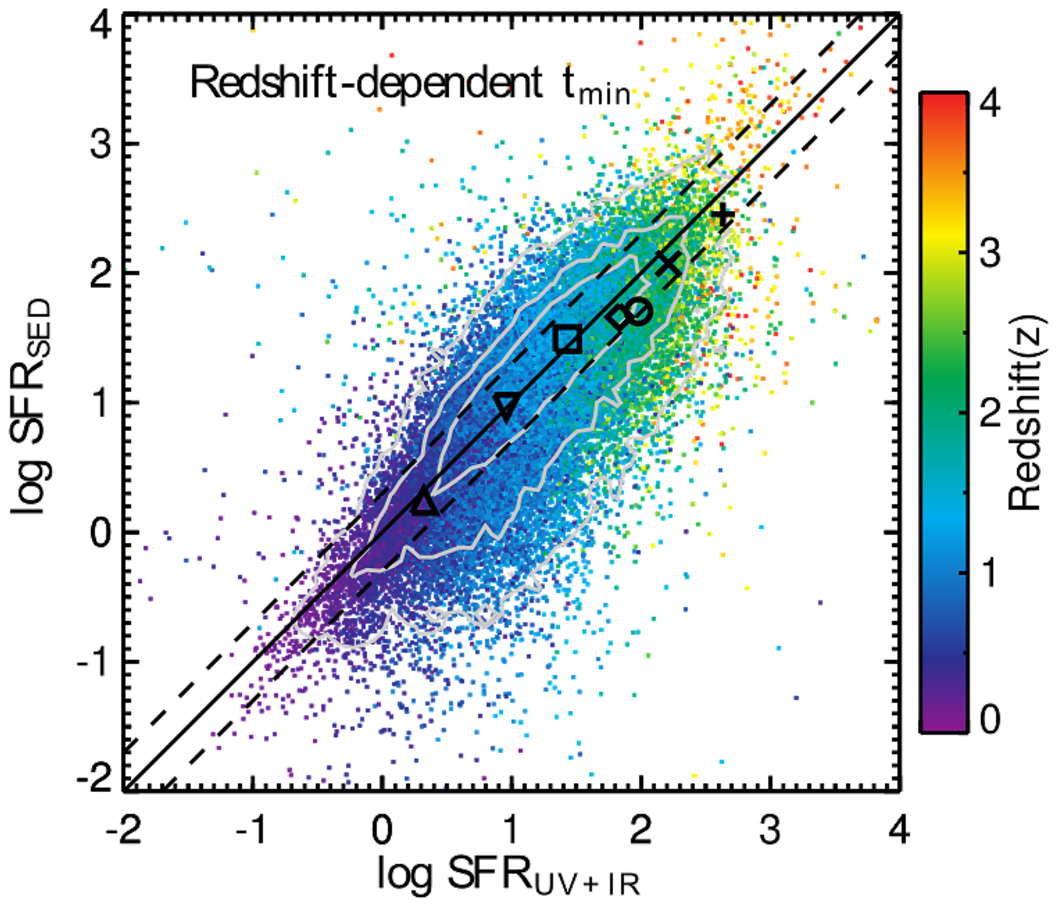}
\caption{Comparison of UV+IR estimates of the SFR (\SFRUVIR) to SED-based estimates (\SFRSED) for our star-forming galaxy sample with different assumptions on the minimum allowed age of the stellar population, $t_\mathrm{min}$, in the SED fitting. 
Small coloured points indicate individual galaxies (with an estimate of \SFRUVIR\ i.e. detected at 24\micron), colour-coded by redshift. 
The large black points show the median SFR in a redshift bin, as indicated. 
Grey contours enclose 50, 80, and 95 per cent of the galaxies. 
The solid black line indicates a 1:1 relation whereas the dashed black lines indicate $\pm0.3$~dex (i.e. a factor $\sim2$). 
With a standard $t_\mathrm{min}=100$~Myr (\emph{top panel}) the SED fitting tends to under-estimate the SFR (compared to UV+IR) at the highest redshifts ($z\gtrsim2$). 
Allowing for younger ages (e.g. $t_\mathrm{min}=10$~Myr, \emph{middle panel}) improves the SFR estimates for high-$z$ galaxies but leads to over-estimates at lower $z$, resulting in an increased scatter. 
Adopting a redshift-dependent $t_\mathrm{min}$ (\emph{bottom panel}, see Equation~\ref{eq:sedtmin}), whereby very young galaxies are not allowed at lower redshifts, provides a reasonable compromise. 
}
\label{fig:sfr_vs_sfr}
\end{figure}

To test the accuracy and reliability of our SED-based SFR estimates, in Figure~\ref{fig:sfr_vs_sfr} we compare with the UV+IR estimates, when available (i.e. for sources with 24\micron\ detections). 
The top panel compares our initial \SFRSED\ estimates with the parameters described above, setting the minimum galaxy age to $t_\mathrm{min}=100$~Myr. 
In general, there is a reasonable agreement between the two estimates. 
However, at higher redshifts ($z \gtrsim 2$) the SED fitting tends to underestimate the SFR 
(compared to \SFRUVIR) by up to $\sim0.5$~dex. 
Our limited range of SPS templates are unable to produce the young, high SFR galaxies observed at these early times \citep[see also][]{Reddy12,Reddy15,Shivaei16}. 
Reducing the minimum allowed galaxy age to 10~Myr (middle panel of Figure~\ref{fig:sfr_vs_sfr}) solves this issue and improves the agreement between \SFRSED\ and \SFRUVIR\ at $z\gtrsim2$. 
However, allowing for younger templates results in an increased scatter at lower redshift and a tendency to over-estimate the SFR from the SED. 
To solve this issue, we implement a redshift-dependent minimum age, 
\begin{equation}
\log t_\mathrm{min}(z) [\mathrm{Myr}] = 2.2 - 0.4z
\label{eq:sedtmin}
\end{equation}
which corresponds to setting a minimum allowed age of 100~Myr at $z=0.5$ and 10~Myr at $z=3$. 
We exclude any templates with ages less than the limit given by Equation \ref{eq:sedtmin} when marginalizing over the parameter grid to estimate \SFRSED\ (i.e. applying Equation~\ref{eq:sedsfr}).
The bottom panel of Figure~\ref{fig:sfr_vs_sfr} shows that our redshift-dependent $t_\mathrm{min}$ ensures good agreement (to within a factor $\sim2$, on average) between \SFRSED\ and \SFRUVIR\ over the full redshift range without the large scatter introduced when allowing for young galaxy ages at all redshifts. 

Comparing \SFRSED\ and \SFRUVIR\ allows us to confirm the reliability of our SED fitting and calibrate our SFR ladder. 
In practice, we only adopt \SFRSED\ as our best estimate of the SFR when \SFRUVIR\ is not available, i.e. for sources that lack 24\micron\ detections. 
\newt{
The good agreement between the two SFR estimates indicates that \SFRSED\ is reliable, even at high redshifts and for the highest SFRs.}
Sources with high levels of dust (where the reliability of SED-based estimates may break down) will be detected at 24\micron, ensuring our SFR ladder provides a reliable estimate.

\section{Bayesian mixture modelling of the distribution of X-ray luminosities}
\label{app:bayesmix}

In this appendix we describe the Bayesian mixture modelling approach that we use to estimate the intrinsic distribution of X-ray luminosities for galaxies in a given stellar mass--redshift bin, $p(\log \lx \giv \mstel,z)$. 

For a single galaxy in our sample (with a given redshift, $z_i$, and stellar mass, $\mathcal{M}_i$), the observed X-ray data consist of the total observed counts, $N_i$, the estimated background count rate, $b_i$, and the effective exposure at the position of the source, $t_i$. 
In a Bayesian framework, our knowledge of the X-ray luminosity of the galaxy, $L_\mathrm{X}$,\footnote{Following \citet{Aird15}, $L_\mathrm{X}$ refers to the rest-frame 2--10~keV luminosity, regardless of the observed energy band (here, the 0.5--2~keV band).}
is described by a probability distribution,
\begin{equation}
p(\lx \giv D_i) d\lx = \mathcal{L}(N_i \giv b_i, t_i, z_i) \; \pi(\lx \giv \mathcal{M}_i, z_i) d\lx
\label{eq:pi}
\end{equation}
where $D_i$ indicates the observed data from source $i$, $\mathcal{L}(N_i \giv \lx, b_i, t_i, z_i)$ is the likelihood of observing $N_i$ counts from a source with luminosity \LX, and $\pi(\lx \giv \mathcal{M}_i, z_i)$ acts as a prior, describing the true underlying distribution of luminosities of star-forming galaxies with mass $\mathcal{M}_i$ and redshift $z_i$. 
The likelihood of observing $N_i$ X-ray counts can be described by a Poisson process, thus,
\begin{equation}
\mathcal{L}(N_i \giv b_i, t_i, z_i) = \frac{(k_i \lx  + b_i)^{N_i}}{N_i!} \;e ^{-(k_i \lx  + b_i)}
\label{eq:poisslik}
\end{equation}
where
\begin{equation}
k_i = \eta(z_i) t_i
\end{equation}
and $\eta(z_i)$ is a factor that converts between \LX\ and the observed count rate in the 0.5--2~keV band. 
In this paper, we assume the X-ray emission has a fixed spectral shape that is described by a power-law with photon index $\Gamma=1.9$ (subject to absorption due to Galactic $N_\mathrm{H}$ only) and thus the conversion factor depends on redshift only.
A correction for the size of the aperture used to extract X-ray counts (which corresponds to a 70 per cent EEF, see Section \ref{sec:xraydata} above) is included in this term.
We note that Equations~\ref{eq:pi} and \ref{eq:poisslik} are valid for sources that are associated with significant X-ray detections (above our detection threshold) and for galaxies that are ``non-detections" in the X-ray imaging where we have extracted X-ray information at the galaxy position. 

As a mathematical convenience, we re-write Equation~\ref{eq:poisslik} as
\begin{equation}
\mathcal{L}(N_i \giv b_i, t_i, z_i) = \sum_{S=0}^{N_i} 
			\left(\frac{(k_i\lx)^{S}}{S!}e^{-k_i \lx} \; 
			\frac{b_i^{N_i-S}}{(N_i-S)!} e^{-b_i}\right).
\label{eq:poisslik2}			
\end{equation}
This equation can be interpretted as describing two independent Poisson processes that produce the observed counts: 
the source of luminosity \LX\ which produces an integer number of observed counts, $S$;
and a background component with expected rate $b_i$ that also produces an integer number of observed counts, $B=N_i-S_i$. 
The summation can then be seen as integrating over the possible values of the unknown nuisance parameter, $S$. 
Equation~\ref{eq:poisslik2} is also mathematically equivalent to the binomial expansion of Equation~\ref{eq:poisslik}.

The likelihood of observing all sources within a stellar mass--redshift bin is given by
integrating Equation~\ref{eq:pi} over all possible \LX\ for an individual source and then taking the product of the probabilities for all sources in a bin. Thus,
\begin{align}
\mathcal{L}(\mathbf{D}_\mathrm{bin}) &=   
\prod_{i=1}^{n_\mathrm{source}} \int_0^\infty p(\lx \giv D_i) \; d\lx \label{eq:likall}\\
	 &=  \prod_{i=1}^{n_\mathrm{source}}
		\int_0^\infty \mathcal{L}(N_i \giv b_i, t_i, z_i) \;\pi(\lx \giv \mathcal{M}_\mathrm{bin}, z_\mathrm{bin}) \; d\lx \nonumber
\end{align}
where $\mathbf{D}_\mathrm{bin}$ indicates the data from all sources within a bin and $n_\mathrm{source}$ is the number of sources in the bin. 
Here, we assume that the intrinsic distribution of \LX\ is the same throughout the stellar mass--redshift bin and described by $\pi(\lx \giv \mathcal{M}_\mathrm{bin}, z_\mathrm{bin})$.

Our next task is to define a model for the intrinsic probability distribution function of luminosities for galaxies of a given stellar mass and redshift,
\begin{equation}
\pi(\lx \giv \mathcal{M}_\mathrm{bin},z_\mathrm{bin}) \;d\lx \equiv p(\log \lx \giv \mathcal{M}_*,z) \;d\log\lx
\label{eq:ploglx}
\end{equation}
where we have re-written our target function as $p(\log \lx \giv \mathcal{M}_*,z)$ to indicate a probability density per unit $\log \lx$ at a given stellar mass and redshift.

We choose to adopt a flexible, non-parametric description of $p(\log \lx \giv \mathcal{M}_*,z)$ based on a Bayesian mixture modelling approach. 
The overall probability function is described by the sum of $K$ mixture components,
\begin{equation}
p(\log \lx \giv \mathcal{M}_*,z) \;d\log\lx = \sum_{j=1}^{K} f(\log \lx \giv \theta_j) \; d\log\lx
\end{equation}
where $f(\log \lx \giv \theta_j)$ is a single function that describes each mixture component with a different set of parameters, $\theta_j$.
We choose to model our distribution by a series of Gamma distributions as this function is the conjugate of our Poisson likelihood and thus significantly simplifies our computation compared to using e.g. a sum of Gaussian distributions that is more generally adopted in Bayesian mixture modelling \citep[e.g.][]{Kelly08}.
A single Gamma distribution component is given by 
\begin{equation}
f(\log \lx \giv \theta_j) d\log \lx = A_j \frac{\ln 10}{\Gamma(\alpha_j)} \left(\frac{\lx}{L_j}\right)^{\alpha_j} e^{-\lx/L_j} d\log\lx
\label{eq:gammacomp}
\end{equation}
where $\alpha_j$ controls the shape of the distribution, $L_j$ is a scale parameter (which determines the position of the peak of the distribution), and $\Gamma(\alpha_j)$ is the Gamma function. 
Each component is normalised such that
\begin{equation}
\int_{-\infty}^{+\infty} f(\log \lx \giv \theta_j) d\log \lx = A_j
\end{equation}
and we required that
\begin{equation}
\sum_{j=1}^K A_j = 1
\end{equation}
and thus each $A_j$ represents the fraction of the overall probability distribution function that is associated with component $j$ of the Bayesian mixture model.
We note that a single component of our Gamma distribution is analogous to the Schechter distribution, although we require $\alpha_j>0$ to ensure that the integral of a component does not diverge; a rising ``faint-end slope" is instead described by the combination of multiple components. 

With the form of our mixture components defined by Equation~\ref{eq:gammacomp} and the Poisson likelihood for each source given by Equation~\ref{eq:poisslik2}, our overall likelihood (Equation~\ref{eq:likall}) can be written as
\resizebox{1.05\columnwidth}{!}{
\begin{minipage}{1.05\columnwidth}
\begin{align}
\mathcal{L}(\mathbf{D}_\mathrm{bin}) &=& 
	\prod_{i=1}^{n_\mathrm{source}} & \left[\int_{-\infty}^{+\infty}
		\sum_{S=0}^{N_i} 
			\left( \frac{(k_i \lx)^{S}}{S!}e^{-k_i \lx} \; 
			\frac{b_i^{N_i-S}}{(N_i-S)!} e^{-b_i} \right) \right. \nonumber \\
			&  & &
			\left. \times   \sum_{j=1}^{K} \left(A_j \frac{\ln 10}{\Gamma(\alpha_j)} 
      \left( \frac{\lx}{L_j}\right)^{\alpha_j} e^{-\frac{\lx}{L_j}} \right) d\log\lx \right]
      \nonumber\\    
           &=& \prod_{i=1}^{n_\mathrm{source}} &
      	   \left[ \sum_{j=1}^{K} A_j w_{ij} \right] \label{eq:likall2}
\end{align}
\end{minipage}
}
where
\begin{align}
w_{ij} &=&  \sum_{S=0}^{N_i} & \left[ 
                  \frac{b_i^{N_i-S}}{(N_i-S)!} e^{-b_i} \;
		          \frac{\ln 10}{\Gamma(\alpha_j)} \; \times \right. \nonumber\\
	& & &\left.    \int_{-\infty}^{+\infty}  
		     				\frac{(k_i \lx)^{S}}{S!}e^{-k_i \lx} 
		 	                \left(\frac{\lx}{L_j}\right)^{\alpha_j} e^{-\frac{\lx}{L_j}}
		 	        \;d\log\lx 
		 	   \right].		\label{eq:wij}
\end{align}			  
The integral in Equation~\ref{eq:wij} has an analytic solution for a given $L_j$, $\alpha_j$ and $S$. 
To ease our computation (while retaining sufficient flexibility), we thus choose to fix all $\alpha_j=3.0$ and adopt a fixed, logaritmically spaced grid for $L_j$ spanning $38 \le \log L_j \;(\mathrm{erg\;s^{-1}}) \le 45$ in steps of 0.2~dex.
\footnote{An additional mixture component with $\log L_j \;(\mathrm{erg\;s^{-1}})=36$ is also included to account for the possibility that some fraction of our sources produce essentially no X-ray luminosity (in practice, modelled as a luminosity $\sim$2 orders of magnitude lower than the rest of the population and significantly below the limits probed by our data).}
Evaluation of our overall likelihood (Equation~\ref{eq:likall2}) then reduces to the summation of a series of pre-computed weights, $w_{ij}$, multiplied by each $A_j$ that constitute the free parameters to be determined. 

The final step in our Bayesian inference is to combine our overall likelihood (Equation \ref{eq:likall2}) with a prior on the free parameters (the set of $A_j$) to recover the posterior distribution, which in turn describes our estimate of the intrinsic distribution of luminosities, $p(\log \lx \giv \mstel,z)$. 
We apply an additional ``smoothness" prior \citep[see also][]{Buchner15} which requires that the values of $A_j$ do not vary rapidly between adjacent mixture components in our grid of $L_j$, as well as ensuring that our recovered distribution is not dominated by a small number of the mixture components. 
We adopt a log-normal prior that penalises large deviations in logarithmic space between adjacent components. 
Thus, 
\begin{equation}
\pi(\log A_{j+1}) \sim \mathrm{Normal}(\log A_j, \sigma)
\end{equation}
where we have assumed an ordering of the mixture components in increasing $L_j$.
The level of smoothness is controlled by $\sigma$, which we fix at 0.3, providing a reasonable balance between a smoothness requirement and ensuring we are able to recover distinct features. 
No smoothness prior is applied for $A_1$, corresponding to the lowest luminosity component, which is instead allowed to take any value between 0 and 1. 
The remaining components are subject to the additional constraint that 
\begin{equation}
\sum_{j=2}^K A_j = 1 - A_1.
\end{equation}
We adopt the \textsc{MultiNest} algorithm \citep{Feroz08,Feroz09,Feroz13} to efficiently explore the multi-modal parameter space and produce posterior draws of the set of parameters, $\{A_j\}$. 
Our final estimates in Figure~\ref{fig:plxm_soft} are derived by evaluating $p(\log \lx \giv \mstel,z)$ for each posterior draw of $\{A_j\}$ and determining the median (solid lines) and 90 per cent confidence intervals (hatched regions) at a given $\log \lx$.

\section{Simulations to verify our Bayesian methodology}
\label{app:sims}

\begin{figure*}
\includegraphics[width=0.9\textwidth,trim=0 1.5cm 0 0]{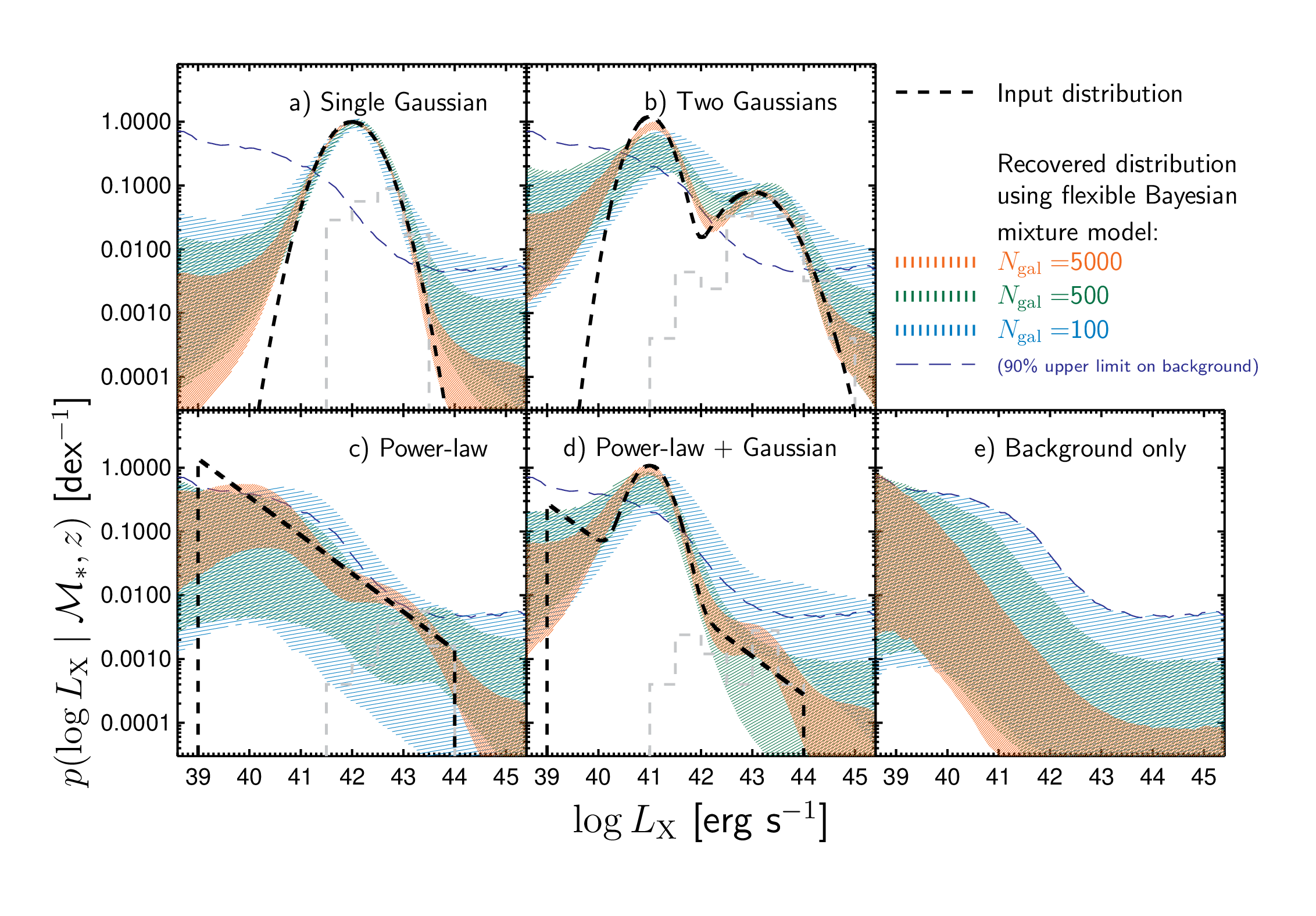}
\caption{
Demonstration of the reconstruction of $p(\log \lx \giv \mstel,z)$ with our Bayesian mixture modelling \refresp{(coloured regions, corresponding to different galaxy sample sizes, as indicated)} for simulating various underlying distributions (black dashed lines):
a) a single Gaussian (log-normal) function; b) two Gaussian (log-normal) distributions centred at different $\log \lx$; c) a power-law distribution over $39<\log \lx$~(\ergs)~$<44$; d) the combination of a Gaussian and a power-law; and e) assuming the observed counts are purely from the background for all galaxies in a sample. 
The grey dashed histogram shows the distribution of sources in our $N_\mathrm{gal}-=5000$ simulated sample that would satisfy the nominal X-ray detection limits.
Our flexible Bayesian method is able to recover distributions with a range of underlying shapes, correcting for incompleteness and probing below the nominal detection limits.
\refresp{However, with smaller samples of galaxies the constraints become progressively weaker and our prior has a stronger impact, smoothing out features in the underlying distribution. 
For the smallest sample size considered ($N_\mathrm{gal}=100$), it is difficult to distinguish between our recovered distribution and the upper limits based on the background (shown by the blue long-dashed line in every panel).
}
}
\label{fig:plx_sims}
\end{figure*}

\refresp{
We have performed a number of simulations to assess the ability of our Bayesian methodology (described above) to recover different underlying distributions and identify a low-luminosity peak (as seen in Figure~\ref{fig:plxm_soft}).
}

\refresp{
Our initial simulations are shown in Figure~\ref{fig:plx_sims}.
For these simulations, we start with the sample of 6937 galaxies in our $10.0<\log \mstel/\msun<10.5$ and $1.0<z<1.5$ bin and select a random subset of $N_\mathrm{gal}= 5000$, 500 and 100 galaxies. 
We adopt the observed values of the redshift, background rate, and effective exposure ($z_i,b_i,t_i$) for the randomly selected galaxies, ensuring our simulations adopt a realistic distribution for depths of the X-ray data.
We then simulate the observed total counts, $N_i$, from each galaxy, assuming different underlying distributions of $p(\log \lx \giv \mstel,z)$, as shown by the black dashed lines in Figure~\ref{fig:plx_sims}. 
Finally, we attempt to recover the underlying distribution using our Bayesian mixture modelling algorithm for each of the different $N_\mathrm{gal}$ samples.}

\refresp{ 
Our recovered constraints on \Plx\ for each of the $N_\mathrm{gal}$ samples are shown by the coloured regions in Figure~\ref{fig:plx_sims}. 
We are able to recover a wide range of distributions.
With a large sample of galaxies ($N_\mathrm{gal}=5000$), 
our modelling only deviates significantly from the input distributions at the faintest luminosities, where the data provide little constraint and thus our prior for a smoothly varying distribution tends to dominate.
Such discrepancies are reasonable and should have a minimal impact on our results based on real data.
As expected, with fewer galaxies in a sample (e.g. $N_\mathrm{gal}=500$ or 100) the recovered constraints on \Plx\ become progressively weaker. 
Furthermore, our prior, which prefers a smoothly varying distribution, has a stronger impact for the smaller sample sizes and thus smooths out the more acute features of the input distribution.
This effect is most apparent in Panel b) for the ``Two Gaussians" input distribution. 
We note that the identification of a distinct, narrow peak at lower luminosities in \Plx, as seen in the real data, tends to go \emph{against} our prior for a smoothly varying distribution, indicating that these features are real and driven by the observed data, rather than the prior.} 

\refresp{
The final panel (e) of Figure~\ref{fig:plx_sims} assumes that there is no X-ray emission from any of the galaxies in \refresp{a given sample} and that the data are purely due to Poisson fluctuations of the background.
Our recovered posterior distributions for \Plx\ are consistent with a minimal population of X-ray emitting sources. 
The posterior distributions set upper limits on the fraction of galaxies at a given luminosity. 
With fewer galaxies in a sample, the posterior upper limits on \Plx\ shift to higher probability densities. 
For the smallest sample considered ($N_\mathrm{gal}=100$) the upper limit from the ``Background only" simulation (shown by the thin, long-dashed blue line, which is replicated in every panel) is roughly consistent with the posterior distributions in panels c) and d), indicating that with such a small sample the posterior constraints are consistent with the data coming from background fluctuations and we are unable to place strong constraints on the underlying distribution.}

\refresp{
The ability of our Bayesian modelling to recover a distinct, low-luminosity peak in \Plx\ that can be attributed to the bulk of the population (and thus to star formation processes) in a given sample of galaxies will depend on a number of different factors: 
i) the size of the galaxy sample, ii) the depth of the X-ray data for such a sample, iii) the extent of AGN activity within the sample, and iv) the true luminosity of any such star-formation peak. 
To explore these effects in more detail, and thus verify the robustness of the results presented in this paper, we have carried out a large suite of additional simulations.
We start with the observed sample of galaxies in a given stellar mass--redshift bin (as shown in Figure~\ref{fig:plxm_soft}). 
We retain the observed X-ray data (total counts, background counts and effective exposure) for any galaxies with significant X-ray detections \emph{and} with an observed $\lx>10^{41}$~\ergs.
This ensures we retain a realistic AGN contribution to the observed \Plx\ that will vary with stellar mass and redshift. 
For the remaining galaxies, we simulate new X-ray data assuming the X-ray luminosity is drawn from a log-Gaussian distribution centred at $\lmode=10^{39}$~\ergs\ with a width of 0.25~dex.}

\begin{figure*}
\includegraphics[width=\textwidth,trim=0cm 1cm 0cm 0cm]{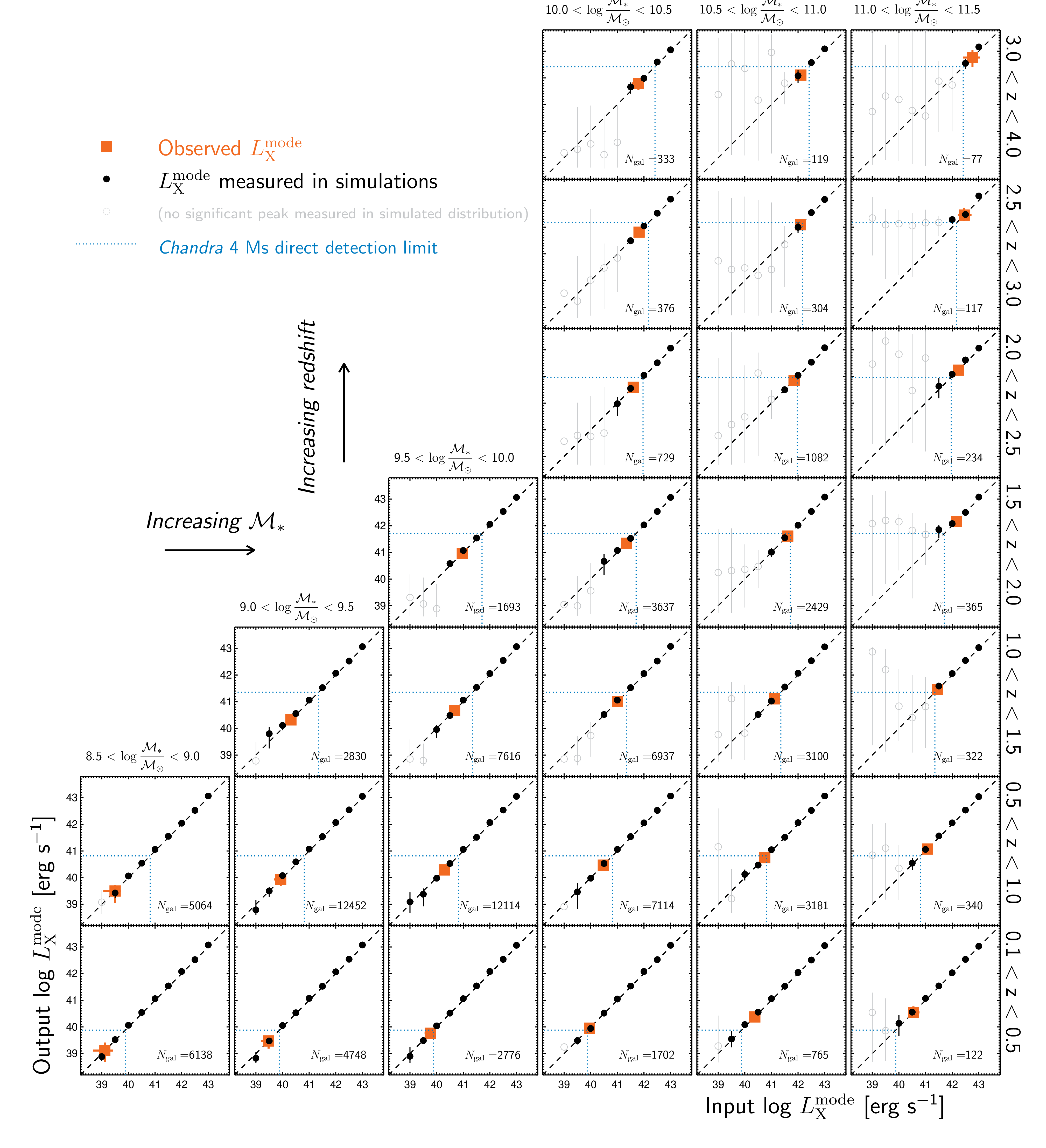}
\caption{
\refresp{
Comparison of the input $L_\mathrm{X}^\mathrm{mode}$ in simulations ($x$-axis) to that recovered using our Bayesian modelling ($y$-axis), running simulations based on the observed galaxy samples in each stellar mass--redshift bin for a range of input luminosities (black circles).  
Grey open symbols indicate bins where the mode of the recovered distribution either lies below the background constraints or only accounts for a small fraction of higher luminosity sources (i.e. AGN) and thus we do not have a significant detection of a low-luminosity peak of X-ray emission from galaxies in our simulated data. 
Large orange squares show the observed $L_\mathrm{X}^\mathrm{mode}$ in the real data. 
The blue dotted lines indicate the luminosity corresponding to the flux limit for direct detection of point sources in the field with our deepest \textit{Chandra} data ($\sim$4~Ms in the GOODS-S field). With sufficiently large galaxy samples we can reliably probe below these limits.    
}}
\label{fig:sim_lx_vs_lx}
\end{figure*}

\refresp{
We then run our Bayesian analysis on the simulated data and attempt to identify the mode of the recovered luminosity distribution. 
The process is repeated for a grid of input \Lmode\ (increasing in steps of 0.5~dex) and for every stellar mass--redshift bin. 
The results are shown in Figure~\ref{fig:sim_lx_vs_lx}, which compares the input $L_\mathrm{X}^\mathrm{mode}$ in the simulation with that recovered by our Bayesian modelling. 
We set two conditions for the identification of a significant low-luminosity peak given the posterior constraints on \Plx\ in our simulations:
\begin{enumerate}[leftmargin=*,itemsep=3pt]
\item
The probability density at the mode of the measured \Plx\ lies above the 90 per cent upper limit from the random sampling of the background.
\item
Integrating \Plx\ down to 1~dex below the peak accounts for $>50$ per cent of galaxies.
\end{enumerate}
The first condition corresponds to requiring that any observed peak cannot be accounted for purely by background fluctuations. 
The second condition is required for bins with substantial AGN contributions at higher $L_\mathrm{X}$; in such bins the mode of the recovered distribution can correspond to a broad, high-luminosity peak that is associated with AGN activity, but only accounts for at most $\sim$20 per cent of the galaxies.
These conditions correspond to our original identification of significant, low-luminosity peaks that account for the bulk of galaxies in the real data.
Simulations where the recovered $L_\mathrm{X}^\mathrm{mode}$ does not satisfy these conditions are indicated by grey open symbols in Figure~\ref{fig:sim_lx_vs_lx}.}

\refresp{The results from the simulations shown in Figure~\ref{fig:sim_lx_vs_lx} illustrate that our Bayesian modelling is able to accurately recover the position of $L_\mathrm{X}^\mathrm{mode}$ over a wide range of luminosities. 
The simulations reveal a complex relationship between the number of galaxies in a bin, the level of AGN activity in such galaxies, and our ability to identify a distinct peak in the underlying distribution at different luminosities. 
In all stellar mass--redshift bins there is a lower limit on the luminosity of the peak that can be recovered (i.e. the lowest luminosity black point) that depends in a fairly complex manner on the number of galaxies in a bin and the level of AGN activity in such galaxies.
The errors on \Lmode\ below this limit (grey points) become very large as the mode either tracks a broad, poorly defined peak at higher luminosities (associated with AGN in a small fraction of the galaxies) or the recovered distribution is consistent with background noise. 
The position of the observed $L_\mathrm{X}^\mathrm{mode}$ in the real data (indicated by the large orange squares) varies relative to this effective limit but always lies within the range where a reliable measurement is possible. 
In many stellar mass--redshift bins, it would be possible to measure an \Lmode\
at $\sim$0.5--1~dex fainter in luminosity than the observed value, providing further evidence that our measurements are robust and not an artefact of the data. 
These simulations thus verify the reliability of our Bayesian method and the results presented in the paper.}

\refresp{
In stellar mass--redshift bins with sufficiently large number of galaxies (i.e. for $\log \mathcal{M}_*/\mathcal{M}_\odot < 10.5$),
the simulations show that our method can robustly probe a
factor $\sim$10--100 below the nominal point source detection
limits of our deepest \textit{Chandra} data, indicated by the dotted blue lines 
in Figure~\ref{fig:sim_lx_vs_lx} which correspond to the on-axis flux limit for the $\sim4$~Ms data in the GOODS-S field. 
Such sensitivity limits are only achieved over $<1$ per cent of our total area coverage.
Our method is thus a powerful way of fully exploiting \textit{Chandra} survey data. 
}

\section{The distribution of X-ray luminosities for quiescent galaxies and the relation to stellar mass}
\label{app:qugals}

\begin{figure*}
\includegraphics[width=0.57\textwidth,trim=25 5 0 0]{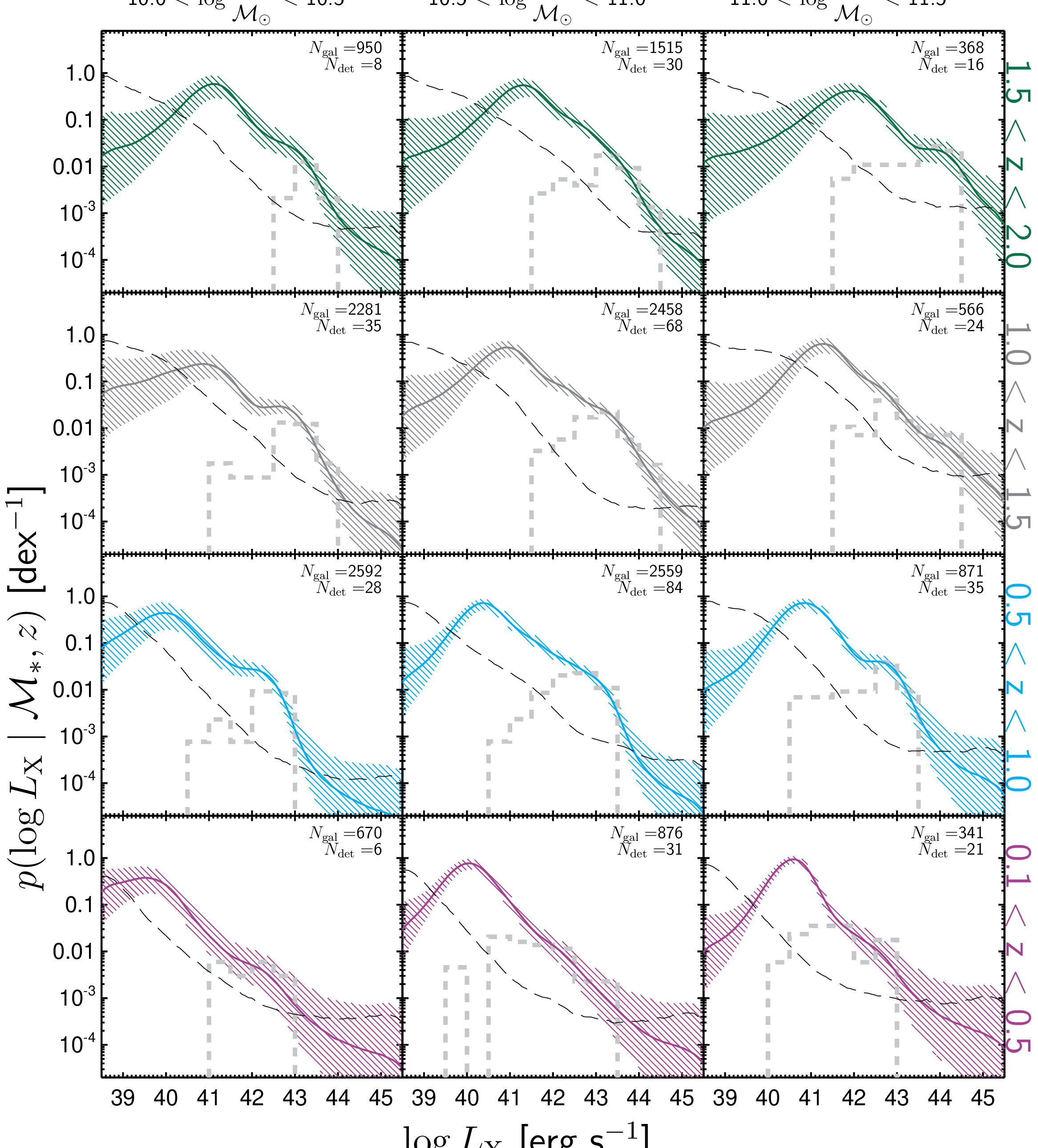}
\caption{Measurements of the intrinsic probability distributions of X-ray luminosities for samples of quiescent galaxies (selected using the UVJ diagram), as a function of stellar mass and redshift (cf. Figure~\ref{fig:plxm_soft} for star-forming galaxies). 
We only show results at lower redshifts and higher stellar masses where we are able to place robust constraints on the distribution and identify a peak at low luminosities. 
As in Figure~\ref{fig:plxm_soft}, the thick coloured lines show our best estimates of $p(\lx \giv \mstel,z)$ and shaded regions indicate the 90 per cent confidence interval. 
The grey histograms correspond to significant detections in the 0.5--2~keV X-ray band (without any corrections for incompleteness). The thin dashed black curve is the 90 per cent upper limit based on shifting the galaxy positions to randomly sample the background.
}
\label{fig:plxm_quiescent}
\end{figure*}

The main results of this paper focus on the X-ray emission from star-forming galaxies, revealing the ``X-ray main sequence". 
In this appendix, we apply the same analysis to sources that lie in the quiescent\footnote{
We note that this definition of  ``quiescent" selects galaxies on the basis of red optical-to-NIR colours. 
Such sources may not be completely quiescent and have low-levels of ongoing star formation.
Furthermore, a non-negligible fraction have relatively high levels of star formation that is identified in the IR (and traced by our SFR estimates). However, such star formation does not have a significant impact on the observed UVJ colours, which are dominated by the light from old stellar populations, especially in the most massive galaxies.}
region of the UVJ diagram.
Figure~\ref{fig:plxm_quiescent} presents measurements of $p(\log \lx \giv \mstel,z)$ at moderate-to-high stellar masses ($10<\log \mstel/\msun<11.5$) and lower redshifts ($z<2$) where we are able to place good constraints on the distributions.
A low-luminosity peak can be identified in all of these distributions that we associate with galactic processes, as well as a tail to higher luminosities associated with AGN accretion activity. 
This low-luminosity peak is found despite the apparently quiescent nature of these galaxies (based on their UVJ colours). 
The X-ray emission is likely to come from a combination of HMXBs (tracing low levels of ongoing star formation  that may be obscured at optical wavelengths), LMXBs (a longer lived source population that provide a delayed tracer of star formation activity in the galaxy), and any hot gas component.
At lower stellar masses ($\log \mstel/\msun<10$) or higher redshifts ($z>2$), not shown in Figure~\ref{fig:plxm_quiescent}, we are unable to identify a clear peak (significantly above the constraints from a sampling of the \textit{Chandra} background) in our measurements of $p(\log \lx \giv \mstel,z)$, most likely due to the relatively small numbers of quiescent galaxies in our samples and the limited depths probed by our X-ray data.

\begin{figure*}
\begin{center}
\includegraphics[width=0.45\textwidth,trim=30 30 10 35]{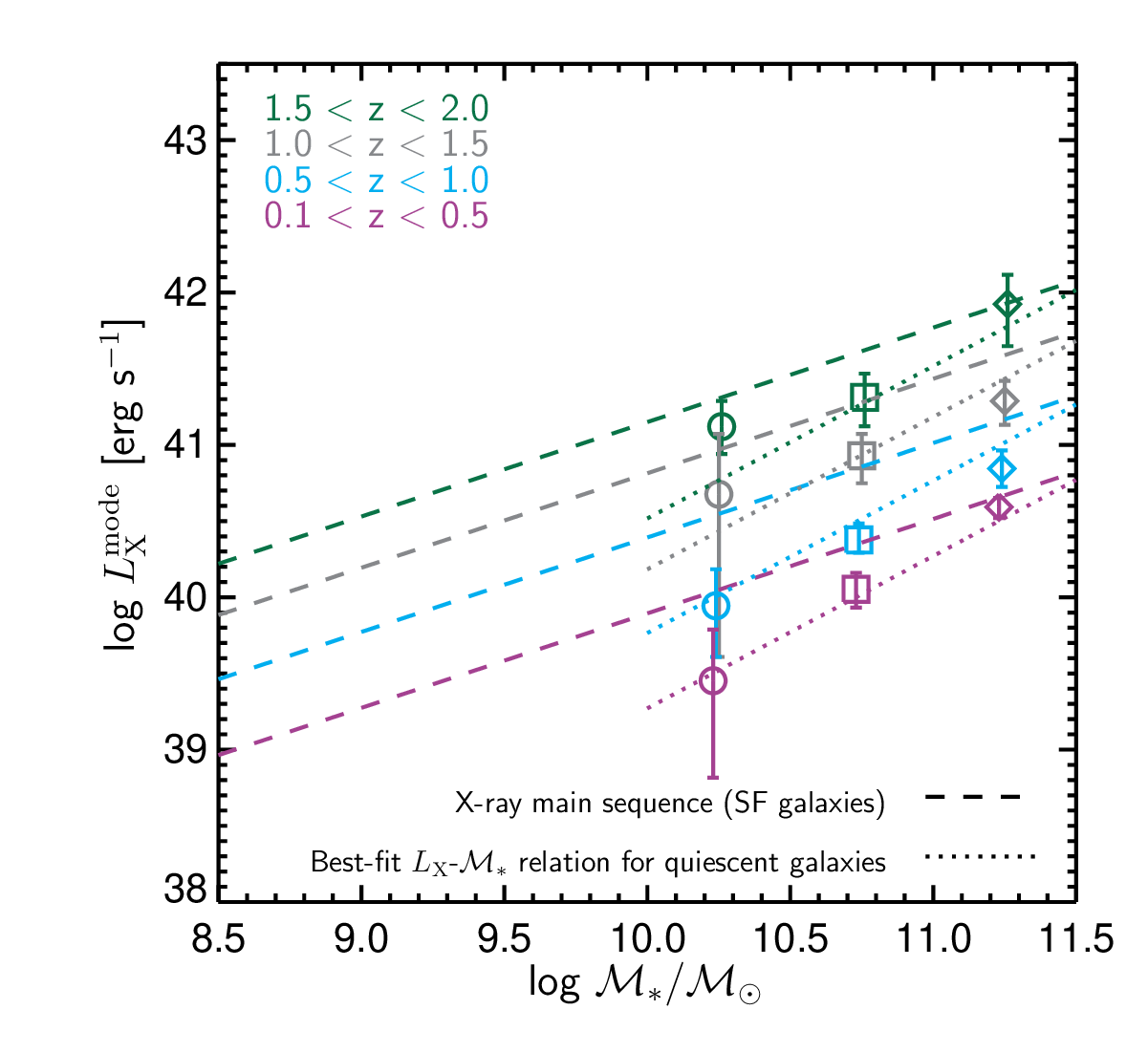}
\end{center}
\caption{The peak of the distribution of X-ray luminosities from quiescent galaxies, $\lmode$, as a function of stellar mass and redshift (data points, dotted lines indicated the best fit to these data using Equation~\ref{eq:lx_vs_mstel_qu}), compared to the ``X-ray main sequence" for star-forming galaxies (dashed lines) measured in Section~\ref{sec:msofsf}.
At high masses and lower redshifts the average X-ray luminosity from quiescent galaxies is comparable to that observed from star-forming galaxies, indicating longer time-scale processes (e.g. LMXBs) that track the overall stellar mass (rather than the current SFR seen at other wavelengths) can make a significant contribution to the observed X-ray luminosities.
}
\label{fig:lx_vs_mass_qu}
\label{lastpage}
\end{figure*}

In Figure~\ref{fig:lx_vs_mass_qu} we plot the luminosity of the peak (the mode of the distribution) as a function of stellar mass for the quiescent galaxy samples (cf. Figure~\ref{fig:lx_ms} for star-forming galaxies). 
The dashed lines show the ``X-ray main sequence" for star-forming galaxies identified in Section~\ref{sec:msofsf}. 
The dotted lines show the best fit to a linear relationship between \Lmode\ and \Mstel\ for quiescent galaxies that evolves with redshift, of the form
\begin{equation}
\log \lmode [\mathrm{erg\;s^{-1}}]= a + \left(\log \frac{\mstel}{\msun}-10.2\right) + b \log \frac{1+z}{1+z_0}
\label{eq:lx_vs_mstel_qu}
\end{equation}
where we find $a=40.19\pm0.05$ and $b=3.83\pm0.33$. 
Allowing for a non-linear scaling between \Lmode\ and \Mstel\ (i.e. a non-unity mass-dependent slope in Equation~\ref{eq:lx_vs_mstel_qu}) does not significantly improve the fit. 
Thus, the relationship between \Lmode\ and \Mstel\ for quiescent galaxies has a steeper slope than the ``X-ray main sequence" for star-forming galaxies. At high stellar masses our measured $\lx^\mathrm{mode}$ in quiescent galaxies is comparable to the luminosity observed from star-forming galaxies, indicating that for galaxies in this regime there may be a significant contribution from longer-lived LMXBs to the observed X-ray luminosities. This issue is discussed further in Section~\ref{sec:xraysfr} above where we explore the scaling between SFR and \LX.

{\footnotesize

\begin{thebibliography}{147}
\expandafter\ifx\csname natexlab\endcsname\relax\def\natexlab#1{#1}\fi

\bibitem[{{Abbott} {et~al.}(2016){Abbott}, {Abbott}, {Abbott}, {Abernathy},
  {Acernese}, {Ackley}, {Adams}, {Adams}, {Addesso}, {Adhikari}, {Adya},
  {Affeldt}, {Agathos}, {Agatsuma}, {Aggarwal}, {Aguiar}, {Aiello}, {Ain},
  {Ajith}, {Allen}, {Allocca}, {Altin}, {Anderson}, {Anderson}, {Arai},
  {Araya}, {Arceneaux}, {Areeda}, {Arnaud}, {Arun}, {Ascenzi}, {Ashton}, {Ast},
  {Aston}, {Astone}, {Aufmuth}, {Aulbert}, {Babak}, {Bacon}, {Bader}, {Baker},
  {Baldaccini}, {Ballardin}, {Ballmer}, {Barayoga}, {Barclay}, {Barish},
  {Barker}, {Barone}, {Barr}, {Barsotti}, {Barsuglia}, {Barta}, {Bartlett},
  {Bartos}, {Bassiri}, {Basti}, {Batch}, {Baune}, {Bavigadda}, {Bazzan},
  {Behnke}, {Bejger}, {Belczynski}, {Bell}, {Bell}, {Berger}, {Bergman},
  {Bergmann}, {Berry}, {Bersanetti}, {Bertolini}, {Betzwieser}, {Bhagwat},
  {Bhandare}, {Bilenko}, {Billingsley}, {Birch}, {Birney}, {Biscans}, {Bisht},
  {Bitossi}, {Biwer}, {Bizouard}, {Blackburn}, {Blair}, {Blair}, {Blair},
  {Bloemen}, {Bock}, {Bodiya}, {Boer}, {Bogaert}, {Bogan}, {Bohe}, {Bojtos},
  {Bond}, {Bondu}, {Bonnand}, {Boom}, {Bork}, {Boschi}, {Bose}, {Bouffanais},
  {Bozzi}, {Bradaschia}, {Brady}, {Braginsky}, {Branchesi}, {Brau}, {Briant},
  {Brillet}, {Brinkmann}, {Brisson}, {Brockill}, {Brooks}, {Brown}, {Brown},
  {Brown}, {Buchanan}, {Buikema}, {Bulik}, {Bulten}, {Buonanno}, {Buskulic},
  {Buy}, {Byer}, {Cadonati}, {Cagnoli}, {Cahillane}, {Calder\'{o}n Bustillo},
  {Callister}, {Calloni}, {Camp}, {Cannon}, {Cao}, {Capano}, {Capocasa},
  {Carbognani}, {Caride}, {Casanueva Diaz}, {Casentini}, {Caudill},
  {Cavagli\`{a}}, {Cavalier}, {Cavalieri}, {Cella}, {Cepeda}, {Cerboni
  Baiardi}, {Cerretani}, {Cesarini}, {Chakraborty}, {Chalermsongsak},
  {Chamberlin}, {Chan}, {Chao}, {Charlton}, {Chassande-Mottin}, {Chen}, {Chen},
  {Cheng}, {Chincarini}, {Chiummo}, {Cho}, {Cho}, {Chow}, {Christensen}, {Chu},
  {Chua}, {Chung}, {Ciani}, {Clara}, {Clark}, {Cleva}, {Coccia}, {Cohadon},
  {Colla}, {Collette}, {Cominsky}, {Constancio}, {Conte}, {Conti}, {Cook},
  {Corbitt}, {Cornish}, {Corsi}, {Cortese}, {Costa}, {Coughlin}, {Coughlin},
  {Coulon}, {Countryman}, {Couvares}, {Cowan}, {Coward}, {Cowart}, {Coyne},
  {Coyne}, {Craig}, {Creighton}, {Cripe}, {Crowder}, {Cumming}, {Cunningham},
  {Cuoco}, {Dal Canton}, {Danilishin}, {D'Antonio}, {Danzmann}, {Darman},
  {Dattilo}, {Dave}, {Daveloza}, {Davier}, {Davies}, {Daw}, {Day}, {DeBra},
  {Debreczeni}, {Degallaix}, {De Laurentis}, {Del\'{e}glise}, {Del Pozzo},
  {Denker}, {Dent}, {Dereli}, {Dergachev}, {DeRosa}, {DeRosa}, {DeSalvo},
  {Dhurandhar}, {D\'{\i}az}, {Di Fiore}, {Di Giovanni}, {Di Lieto}, {Di Pace},
  {Di Palma}, {Di Virgilio}, {Dojcinoski}, {Dolique}, {Donovan}, {Dooley},
  {Doravari}, {Douglas}, {Downes}, {Drago}, {Drever}, {Driggers}, {Du},
  {Ducrot}, {Dwyer}, {Edo}, {Edwards}, {Effler}, {Eggenstein}, {Ehrens},
  {Eichholz}, {Eikenberry}, {Engels}, {Essick}, {Etzel}, {Evans}, {Evans},
  {Everett}, {Factourovich}, {Fafone}, {Fair}, {Fairhurst}, {Fan}, {Fang},
  {Farinon}, {Farr}, {Farr}, {Favata}, {Fays}, {Fehrmann}, {Fejer}, {Ferrante},
  {Ferreira}, {Ferrini}, {Fidecaro}, {Fiori}, {Fiorucci}, {Fisher}, {Flaminio},
  {Fletcher}, {Fournier}, {Franco}, {Frasca}, {Frasconi}, {Frei}, {Freise},
  {Frey}, {Frey}, {Fricke}, {Fritschel}, {Frolov}, {Fulda}, {Fyffe}, {Gabbard},
  {Gair}, {Gammaitoni}, {Gaonkar}, {Garufi}, {Gatto}, {Gaur}, {Gehrels},
  {Gemme}, {Gendre}, {Genin}, {Gennai}, {George}, {Gergely}, {Germain},
  {Ghosh}, {Ghosh}, {Giaime}, {Giardina}, {Giazotto}, {Gill}, {Glaefke},
  {Goetz}, {Goetz}, {Gondan}, {Gonz\'{a}lez}, {Gonzalez Castro}, {Gopakumar},
  {Gordon}, {Gorodetsky}, {Gossan}, {Gosselin}, {Gouaty}, {Graef}, {Graff},
  {Granata}, {Grant}, {Gras}, {Gray}, {Greco}, {Green}, {Groot}, {Grote},
  {Grunewald}, {Guidi}, {Guo}, {Gupta}, {Gupta}, {Gushwa}, {Gustafson},
  {Gustafson}, {Hacker}, {Hall}, {Hall}, {Hammond}, {Haney}, {Hanke}, {Hanks},
  {Hanna}, {Hannam}, {Hanson}, {Hardwick}, {Harms}, {Harry}, {Harry}, {Hart},
  {Hartman}, {Haster}, {Haughian}, {Heidmann}, {Heintze}, {Heitmann}, {Hello},
  {Hemming}, {Hendry}, {Heng}, {Hennig}, {Heptonstall}, {Heurs}, {Hild},
  {Hoak}, {Hodge}, {Hofman}, {Hollitt}, {Holt}, {Holz}, {Hopkins}, {Hosken},
  {Hough}, {Houston}, {Howell}, {Hu}, {Huang}, {Huerta}, {Huet}, {Hughey},
  {Husa}, {Huttner}, {Huynh-Dinh}, {Idrisy}, {Indik}, {Ingram}, {Inta}, {Isa},
  {Isac}, {Isi}, {Islas}, {Isogai}, {Iyer}, {Izumi}, {Jacqmin}, {Jang}, {Jani},
  {Jaranowski}, {Jawahar}, {Jim\'{e}nez-Forteza}, {Johnson}, {Jones}, {Jones},
  {Jonker}, {Ju}, {K}, {Kalaghatgi}, {Kalogera}, {Kandhasamy}, {Kang},
  {Kanner}, {Karki}, {Kasprzack}, {Katsavounidis}, {Katzman}, {Kaufer}, {Kaur},
  {Kawabe}, {Kawazoe}, {K\'{e}f\'{e}lian}, {Kehl}, {Keitel}, {Kelley}, {Kells},
  {Kennedy}, {Key}, {Khalaidovski}, {Khalili}, {Khan}, {Khan}, {Khan},
  {Khazanov}, {Kijbunchoo}, {Kim}, {Kim}, {Kim}, {Kim}, {Kim}, {Kim}, {King},
  {King}, {Kinzel}, {Kissel}, {Kleybolte}, {Klimenko}, {Koehlenbeck},
  {Kokeyama}, {Koley}, {Kondrashov}, {Kontos}, {Korobko}, {Korth}, {Kowalska},
  {Kozak}, {Kringel}, {Krishnan}, {Kr\'{o}lak}, {Krueger}, {Kuehn}, {Kumar},
  {Kuo}, {Kutynia}, {Lackey}, {Landry}, {Lange}, {Lantz}, {Lasky}, {Lazzarini},
  {Lazzaro}, {Leaci}, {Leavey}, {Lebigot}, {Lee}, {Lee}, {Lee}, {Lee}, {Lenon},
  {Leonardi}, {Leong}, {Leroy}, {Letendre}, {Levin}, {Levine}, {Li}, {Libson},
  {Littenberg}, {Lockerbie}, {Logue}, {Lombardi}, {Lord}, {Lorenzini},
  {Loriette}, {Lormand}, {Losurdo}, {Lough}, {L\"{u}ck}, {Lundgren}, {Luo},
  {Lynch}, {Ma}, {MacDonald}, {Machenschalk}, {MacInnis}, {Macleod},
  {Maga\~{n}a-Sandoval}, {Magee}, {Mageswaran}, {Majorana}, {Maksimovic},
  {Malvezzi}, {Man}, {Mandel}, {Mandic}, {Mangano}, {Mansell}, {Manske},
  {Mantovani}, {Marchesoni}, {Marion}, {M\'{a}rka}, {M\'{a}rka}, {Markosyan},
  {Maros}, {Martelli}, {Martellini}, {Martin}, {Martin}, {Martynov}, {Marx},
  {Mason}, {Masserot}, {Massinger}, {Masso-Reid}, {Matichard}, {Matone},
  {Mavalvala}, {Mazumder}, {Mazzolo}, {McCarthy}, {McClelland}, {McCormick},
  {McGuire}, {McIntyre}, {McIver}, {McManus}, {McWilliams}, {Meacher},
  {Meadors}, {Meidam}, {Melatos}, {Mendell}, {Mendoza-Gandara}, {Mercer},
  {Merilh}, {Merzougui}, {Meshkov}, {Messenger}, {Messick}, {Meyers},
  {Mezzani}, {Miao}, {Michel}, {Middleton}, {Mikhailov}, {Milano}, {Miller},
  {Millhouse}, {Minenkov}, {Ming}, {Mirshekari}, {Mishra}, {Mitra},
  {Mitrofanov}, {Mitselmakher}, {Mittleman}, {Moggi}, {Mohan}, {Mohapatra},
  {Montani}, {Moore}, {Moore}, {Moraru}, {Moreno}, {Morriss}, {Mossavi},
  {Mours}, {Mow-Lowry}, {Mueller}, {Mueller}, {Muir}, {Mukherjee}, {Mukherjee},
  {Mukherjee}, {Mukund}, {Mullavey}, {Munch}, {Murphy}, {Murray}, {Mytidis},
  {Nardecchia}, {Naticchioni}, {Nayak}, {Necula}, {Nedkova}, {Nelemans},
  {Neri}, {Neunzert}, {Newton}, {Nguyen}, {Nielsen}, {Nissanke}, {Nitz},
  {Nocera}, {Nolting}, {Normandin}, {Nuttall}, {Oberling}, {Ochsner}, {O'Dell},
  {Oelker}, {Ogin}, {Oh}, {Oh}, {Ohme}, {Oliver}, {Oppermann}, {Oram},
  {O'Reilly}, {O'Shaughnessy}, {Ottaway}, {Ottens}, {Overmier}, {Owen}, {Pai},
  {Pai}, {Palamos}, {Palashov}, {Palomba}, {Pal-Singh}, {Pan}, {Pankow},
  {Pannarale}, {Pant}, {Paoletti}, {Paoli}, {Papa}, {Paris}, {Parker},
  {Pascucci}, {Pasqualetti}, {Passaquieti}, {Passuello}, {Patricelli},
  {Patrick}, {Pearlstone}, {Pedraza}, {Pedurand}, {Pekowsky}, {Pele}, {Penn},
  {Perreca}, {Phelps}, {Piccinni}, {Pichot}, {Piergiovanni}, {Pierro},
  {Pillant}, {Pinard}, {Pinto}, {Pitkin}, {Poggiani}, {Popolizio}, {Post},
  {Powell}, {Prasad}, {Predoi}, {Premachandra}, {Prestegard}, {Price},
  {Prijatelj}, {Principe}, {Privitera}, {Prix}, {Prodi}, {Prokhorov},
  {Puncken}, {Punturo}, {Puppo}, {P\"{u}rrer}, {Qi}, {Qin}, {Quetschke},
  {Quintero}, {Quitzow-James}, {Raab}, {Rabeling}, {Radkins}, {Raffai}, {Raja},
  {Rakhmanov}, {Rapagnani}, {Raymond}, {Razzano}, {Re}, {Read}, {Reed},
  {Regimbau}, {Rei}, {Reid}, {Reitze}, {Rew}, {Reyes}, {Ricci}, {Riles},
  {Robertson}, {Robie}, {Robinet}, {Rocchi}, {Rolland}, {Rollins}, {Roma},
  {Romano}, {Romano}, {Romanov}, {Romie}, {Rosi\'{n}ska}, {Rowan},
  {R\"{u}diger}, {Ruggi}, {Ryan}, {Sachdev}, {Sadecki}, {Sadeghian}, {Salconi},
  {Saleem}, {Salemi}, {Samajdar}, {Sammut}, {Sanchez}, {Sandberg}, {Sandeen},
  {Sanders}, {Sassolas}, {Sathyaprakash}, {Saulson}, {Sauter}, {Savage},
  {Sawadsky}, {Schale}, {Schilling}, {Schmidt}, {Schmidt}, {Schnabel},
  {Schofield}, {Sch\"{o}nbeck}, {Schreiber}, {Schuette}, {Schutz}, {Scott},
  {Scott}, {Sellers}, {Sentenac}, {Sequino}, {Sergeev}, {Serna}, {Setyawati},
  {Sevigny}, {Shaddock}, {Shah}, {Shahriar}, {Shaltev}, {Shao}, {Shapiro},
  {Shawhan}, {Sheperd}, {Shoemaker}, {Shoemaker}, {Siellez}, {Siemens}, {Sigg},
  {Silva}, {Simakov}, {Singer}, {Singer}, {Singh}, {Singh}, {Singhal},
  {Sintes}, {Slagmolen}, {Smith}, {Smith}, {Smith}, {Son}, {Sorazu},
  {Sorrentino}, {Souradeep}, {Srivastava}, {Staley}, {Steinke}, {Steinlechner},
  {Steinlechner}, {Steinmeyer}, {Stephens}, {Stevenson}, {Stone}, {Strain},
  {Straniero}, {Stratta}, {Strauss}, {Strigin}, {Sturani}, {Stuver},
  {Summerscales}, {Sun}, {Sutton}, {Swinkels}, {Szczepa\'{n}czyk}, {Tacca},
  {Talukder}, {Tanner}, {T\'{a}pai}, {Tarabrin}, {Taracchini}, {Taylor},
  {Theeg}, {Thirugnanasambandam}, {Thomas}, {Thomas}, {Thomas}, {Thorne},
  {Thorne}, {Thrane}, {Tiwari}, {Tiwari}, {Tokmakov}, {Tomlinson}, {Tonelli},
  {Torres}, {Torrie}, {T\"{o}yr\"{a}}, {Travasso}, {Traylor}, {Trifir\`{o}},
  {Tringali}, {Trozzo}, {Tse}, {Turconi}, {Tuyenbayev}, {Ugolini},
  {Unnikrishnan}, {Urban}, {Usman}, {Vahlbruch}, {Vajente}, {Valdes}, {van
  Bakel}, {van Beuzekom}, {van den Brand}, {van den Broeck}, {Vander-Hyde},
  {van der Schaaf}, {van Heijningen}, {van Veggel}, {Vardaro}, {Vass},
  {Vas\'{u}th}, {Vaulin}, {Vecchio}, {Vedovato}, {Veitch}, {Veitch},
  {Venkateswara}, {Verkindt}, {Vetrano}, {Vicer\'{e}}, {Vinciguerra}, {Vine},
  {Vinet}, {Vitale}, {Vo}, {Vocca}, {Vorvick}, {Voss}, {Vousden}, {Vyatchanin},
  {Wade}, {Wade}, {Wade}, {Walker}, {Wallace}, {Walsh}, {Wang}, {Wang}, {Wang},
  {Wang}, {Wang}, {Ward}, {Warner}, {Was}, {Weaver}, {Wei}, {Weinert},
  {Weinstein}, {Weiss}, {Welborn}, {Wen}, {We{\ss}els}, {Westphal}, {Wette},
  {Whelan}, {White}, {Whiting}, {Williams}, {Williamson}, {Willis}, {Willke},
  {Wimmer}, {Winkler}, {Wipf}, {Wittel}, {Woan}, {Worden}, {Wright}, {Wu},
  {Yablon}, {Yam}, {Yamamoto}, {Yancey}, {Yap}, {Yu}, {Yvert}, {Zadro{\.z}ny},
  {Zangrando}, {Zanolin}, {Zendri}, {Zevin}, {Zhang}, {Zhang}, {Zhang},
  {Zhang}, {Zhao}, {Zhou}, {Zhou}, {Zhu}, {Zucker}, {Zuraw}, {and}, {Zweizig},
  {LIGO Scientific Collaboration}, \& {Virgo Collaboration}}]{Abbott16b}
{Abbott}, B.~P., {et~al.} 2016, \apjl, 818, L22

\bibitem[{{Aird} {et~al.}(2016){Aird}, {Coil}, \& {Georgakakis}}]{Aird16b}
{Aird}, J., {Coil}, A.~L., \& {Georgakakis}, A. 2016, in prep.

\bibitem[{{Aird} {et~al.}(2015){Aird}, {Coil}, {Georgakakis}, {Nandra},
  {Barro}, \& {P\'{e}rez-Gonz\'{a}lez}}]{Aird15}
{Aird}, J., {Coil}, A.~L., {Georgakakis}, A., {Nandra}, K., {Barro}, G., \&
  {P\'{e}rez-Gonz\'{a}lez}, P.~G. 2015, \mnras, 451, 1892

\bibitem[{{Aird} {et~al.}(2012){Aird}, {Coil}, {Moustakas}, {Blanton},
  {Burles}, {Cool}, {Eisenstein}, {Smith}, {Wong}, \& {Zhu}}]{Aird12}
{Aird}, J., {et~al.} 2012, \apj, 746, 90

\bibitem[{{Aird} {et~al.}(2013){Aird}, {Comastri}, {Brusa}, {Cappelluti},
  {Moretti}, {Vanzella}, {Volonteri}, {Alexander}, {Afonso}, {Fiore},
  {Georgantopoulos}, {Iwasawa}, {Merloni}, {Nandra}, {Salvaterra}, {Salvato},
  {Severgnini}, {Schawinski}, {Shankar}, {Vignali}, \& {Vito}}]{Aird13b}
---. 2013, An Athena+ Supporting Paper (arXiv:1306.2325)

\bibitem[{{Aird} {et~al.}(2008){Aird}, {Nandra}, {Georgakakis}, {Laird},
  {Steidel}, \& {Sharon}}]{Aird08}
{Aird}, J., {Nandra}, K., {Georgakakis}, A., {Laird}, E.~S., {Steidel}, C.~C.,
  \& {Sharon}, C. 2008, \mnras, 387, 883

\bibitem[{{Aird} {et~al.}(2010){Aird}, {Nandra}, {Laird}, {Georgakakis},
  {Ashby}, {Barmby}, {Coil}, {Huang}, {Koekemoer}, {Steidel}, \&
  {Willmer}}]{Aird10}
{Aird}, J., {et~al.} 2010, \mnras, 401, 2531

\bibitem[{{Alexander} {et~al.}(2003){Alexander}, {Bauer}, {Brandt},
  {Schneider}, {Hornschemeier}, {Vignali}, {Barger}, {Broos}, {Cowie},
  {Garmire}, {Townsley}, {Bautz}, {Chartas}, \& {Sargent}}]{Alexander03}
{Alexander}, D.~M., {et~al.} 2003, \aj, 126, 539

\bibitem[{{Basu} {et~al.}(2015){Basu}, {Wadadekar}, {Beelen}, {Singh},
  {Archana}, {Sirothia}, \& {Ishwara-Chandra}}]{Basu15}
{Basu}, A., {Wadadekar}, Y., {Beelen}, A., {Singh}, V., {Archana}, K.~N.,
  {Sirothia}, S., \& {Ishwara-Chandra}, C.~H. 2015, \apj, 803, 51

\bibitem[{{Basu-Zych} {et~al.}(2013{\natexlab{a}}){Basu-Zych}, {Lehmer},
  {Hornschemeier}, {Bouwens}, {Fragos}, {Oesch}, {Belczynski}, {Brandt},
  {Kalogera}, {Luo}, {Miller}, {Mullaney}, {Tzanavaris}, {Xue}, \&
  {Zezas}}]{Basu-Zych13}
{Basu-Zych}, A.~R., {et~al.} 2013{\natexlab{a}}, \apj, 762, 45

\bibitem[{{Basu-Zych} {et~al.}(2013{\natexlab{b}}){Basu-Zych}, {Lehmer},
  {Hornschemeier}, {Gon{\c c}alves}, {Fragos}, {Heckman}, {Overzier}, {Ptak},
  \& {Schiminovich}}]{Basu-Zych13b}
---. 2013{\natexlab{b}}, \apj, 774, 152

\bibitem[{{Basu-Zych} {et~al.}(2007){Basu-Zych}, {Schiminovich}, {Johnson},
  {Hoopes}, {Overzier}, {Treyer}, {Heckman}, {Barlow}, {Bianchi}, {Conrow},
  {Donas}, {Forster}, {Friedman}, {Lee}, {Madore}, {Martin}, {Milliard},
  {Morrissey}, {Neff}, {Rich}, {Salim}, {Seibert}, {Small}, {Szalay}, {Wyder},
  \& {Yi}}]{Basu-Zych07}
---. 2007, \apjs, 173, 457

\bibitem[{{Behroozi} {et~al.}(2013){Behroozi}, {Wechsler}, \&
  {Conroy}}]{Behroozi13}
{Behroozi}, P.~S., {Wechsler}, R.~H., \& {Conroy}, C. 2013, \apj, 770, 57

\bibitem[{{Belczynski} {et~al.}(2008){Belczynski}, {Kalogera}, {Rasio}, {Taam},
  {Zezas}, {Bulik}, {Maccarone}, \& {Ivanova}}]{Belczynski08}
{Belczynski}, K., {Kalogera}, V., {Rasio}, F.~A., {Taam}, R.~E., {Zezas}, A.,
  {Bulik}, T., {Maccarone}, T.~J., \& {Ivanova}, N. 2008, \apjs, 174, 223

\bibitem[{{Bell} {et~al.}(2005){Bell}, {Papovich}, {Wolf}, {Le Floc'h},
  {Caldwell}, {Barden}, {Egami}, {McIntosh}, {Meisenheimer},
  {P\'{e}rez-Gonz\'{a}lez}, {Rieke}, {Rieke}, {Rigby}, \& {Rix}}]{Bell05}
{Bell}, E.~F., {et~al.} 2005, \apj, 625, 23

\bibitem[{{Bell} {et~al.}(2004){Bell}, {Wolf}, {Meisenheimer}, {Rix}, {Borch},
  {Dye}, {Kleinheinrich}, {Wisotzki}, \& {McIntosh}}]{Bell04}
---. 2004, \apj, 608, 752

\bibitem[{{Blanton} \& {Moustakas}(2009)}]{Blanton09}
{Blanton}, M.~R., \& {Moustakas}, J. 2009, \araa, 47, 159

\bibitem[{{Bongiorno} {et~al.}(2012){Bongiorno}, {Merloni}, {Brusa},
  {Magnelli}, {Salvato}, {Mignoli}, {Zamorani}, {Fiore}, {Rosario}, {Mainieri},
  {Hao}, {Comastri}, {Vignali}, {Balestra}, {Bardelli}, {Berta}, {Civano},
  {Kampczyk}, {Le Floc'h}, {Lusso}, {Lutz}, {Pozzetti}, {Pozzi}, {Riguccini},
  {Shankar}, \& {Silverman}}]{Bongiorno12}
{Bongiorno}, A., {et~al.} 2012, \mnras, 427, 3103

\bibitem[{{Bongiorno} {et~al.}(2016){Bongiorno}, {Schulze}, {Merloni},
  {Zamorani}, {Ilbert}, {La Franca}, {Peng}, {Piconcelli}, {Mainieri},
  {Silverman}, {Brusa}, {Fiore}, {Salvato}, \& {Scoville}}]{Bongiorno16}
---. 2016, \aap, 588, A78

\bibitem[{{Brammer} {et~al.}(2008){Brammer}, {van Dokkum}, \&
  {Coppi}}]{Brammer08}
{Brammer}, G.~B., {van Dokkum}, P.~G., \& {Coppi}, P. 2008, \apj, 686, 1503

\bibitem[{{Brammer} {et~al.}(2012){Brammer}, {van Dokkum}, {Franx},
  {Fumagalli}, {Patel}, {Rix}, {Skelton}, {Kriek}, {Nelson}, {Schmidt},
  {Bezanson}, {da Cunha}, {Erb}, {Fan}, {F\"{o}rster Schreiber}, {Illingworth},
  {Labb\'{e}}, {Leja}, {Lundgren}, {Magee}, {Marchesini}, {McCarthy},
  {Momcheva}, {Muzzin}, {Quadri}, {Steidel}, {Tal}, {Wake}, {Whitaker}, \&
  {Williams}}]{Brammer12}
{Brammer}, G.~B., {et~al.} 2012, \apjs, 200, 13

\bibitem[{{Brammer} {et~al.}(2011){Brammer}, {Whitaker}, {van Dokkum},
  {Marchesini}, {Franx}, {Kriek}, {Labb\'{e}}, {Lee}, {Muzzin}, {Quadri},
  {Rudnick}, \& {Williams}}]{Brammer11}
---. 2011, \apj, 739, 24

\bibitem[{{Brandt} \& {Alexander}(2015)}]{Brandt15}
{Brandt}, W.~N., \& {Alexander}, D.~M. 2015, \aapr, 23, 1

\bibitem[{{Brorby} {et~al.}(2016){Brorby}, {Kaaret}, {Prestwich}, \&
  {Mirabel}}]{Brorby16}
{Brorby}, M., {Kaaret}, P., {Prestwich}, A., \& {Mirabel}, I.~F. 2016, \mnras,
  457, 4081

\bibitem[{{Brown} {et~al.}(2007){Brown}, {Dey}, {Jannuzi}, {Brand}, {Benson},
  {Brodwin}, {Croton}, \& {Eisenhardt}}]{Brown07}
{Brown}, M.~J.~I., {Dey}, A., {Jannuzi}, B.~T., {Brand}, K., {Benson}, A.~J.,
  {Brodwin}, M., {Croton}, D.~J., \& {Eisenhardt}, P.~R. 2007, \apj, 654, 858

\bibitem[{{Brusa} {et~al.}(2007){Brusa}, {Zamorani}, {Comastri}, {Hasinger},
  {Cappelluti}, {Civano}, {Finoguenov}, {Mainieri}, {Salvato}, {Vignali},
  {Elvis}, {Fiore}, {Gilli}, {Impey}, {Lilly}, {Mignoli}, {Silverman}, {Trump},
  {Urry}, {Bender}, {Capak}, {Huchra}, {Kneib}, {Koekemoer}, {Leauthaud},
  {Lehmann}, {Massey}, {Matute}, {McCarthy}, {McCracken}, {Rhodes}, {Scoville},
  {Taniguchi}, \& {Thompson}}]{Brusa07}
{Brusa}, M., {et~al.} 2007, \apjs, 172, 353

\bibitem[{{Buchner} {et~al.}(2015){Buchner}, {Georgakakis}, {Nandra},
  {Brightman}, {Menzel}, {Liu}, {Hsu}, {Salvato}, {Rangel}, {Aird}, {Merloni},
  \& {Ross}}]{Buchner15}
{Buchner}, J., {et~al.} 2015, \apj, 802, 89

\bibitem[{{Capak} {et~al.}(2007){Capak}, {Aussel}, {Ajiki}, {McCracken},
  {Mobasher}, {Scoville}, {Shopbell}, {Taniguchi}, {Thompson}, {Tribiano},
  {Sasaki}, {Blain}, {Brusa}, {Carilli}, {Comastri}, {Carollo}, {Cassata},
  {Colbert}, {Ellis}, {Elvis}, {Giavalisco}, {Green}, {Guzzo}, {Hasinger},
  {Ilbert}, {Impey}, {Jahnke}, {Kartaltepe}, {Kneib}, {Koda}, {Koekemoer},
  {Komiyama}, {Leauthaud}, {Le Fevre}, {Lilly}, {Liu}, {Massey}, {Miyazaki},
  {Murayama}, {Nagao}, {Peacock}, {Pickles}, {Porciani}, {Renzini}, {Rhodes},
  {Rich}, {Salvato}, {Sanders}, {Scarlata}, {Schiminovich}, {Schinnerer},
  {Scodeggio}, {Sheth}, {Shioya}, {Tasca}, {Taylor}, {Yan}, \&
  {Zamorani}}]{Capak07}
{Capak}, P., {et~al.} 2007, \apjs, 172, 99

\bibitem[{{Chabrier}(2003)}]{Chabrier03}
{Chabrier}, G. 2003, \pasp, 115, 763

\bibitem[{{Ciliegi} {et~al.}(2003){Ciliegi}, {Zamorani}, {Hasinger}, {Lehmann},
  {Szokoly}, \& {Wilson}}]{Ciliegi03}
{Ciliegi}, P., {Zamorani}, G., {Hasinger}, G., {Lehmann}, I., {Szokoly}, G., \&
  {Wilson}, G. 2003, \aap, 398, 901

\bibitem[{{Civano} {et~al.}(2016){Civano}, {Marchesi}, {Comastri}, {Urry},
  {Elvis}, {Cappelluti}, {Puccetti}, {Brusa}, {Zamorani}, {Hasinger},
  {Aldcroft}, {Alexander}, {Allevato}, {Brunner}, {Capak}, {Finoguenov},
  {Fiore}, {Fruscione}, {Gilli}, {Glotfelty}, {Griffiths}, {Hao}, {Harrison},
  {Jahnke}, {Kartaltepe}, {Karim}, {LaMassa}, {Lanzuisi}, {Miyaji}, {Ranalli},
  {Salvato}, {Sargent}, {Scoville}, {Schawinski}, {Schinnerer}, {Silverman},
  {Smolcic}, {Stern}, {Toft}, {Trakhtenbrot}, {Treister}, \&
  {Vignali}}]{Civano16}
{Civano}, F., {et~al.} 2016, \apj, 819, 62

\bibitem[{{Coil} {et~al.}(2011){Coil}, {Blanton}, {Burles}, {Cool},
  {Eisenstein}, {Moustakas}, {Wong}, {Zhu}, {Aird}, {Bernstein}, {Bolton}, \&
  {Hogg}}]{Coil11}
{Coil}, A.~L., {et~al.} 2011, \apj, 741, 8

\bibitem[{{Conroy}(2013)}]{Conroy13b}
{Conroy}, C. 2013, \araa, 51, 393

\bibitem[{{Conroy} \& {Gunn}(2010)}]{Conroy10}
{Conroy}, C., \& {Gunn}, J.~E. 2010, \apj, 712, 833

\bibitem[{{Conroy} {et~al.}(2009){Conroy}, {Gunn}, \& {White}}]{Conroy09}
{Conroy}, C., {Gunn}, J.~E., \& {White}, M. 2009, \apj, 699, 486

\bibitem[{{Cool} {et~al.}(2013){Cool}, {Moustakas}, {Blanton}, {Burles},
  {Coil}, {Eisenstein}, {Wong}, {Zhu}, {Aird}, {Bernstein}, {Bolton}, {Hogg},
  \& {Mendez}}]{Cool13}
{Cool}, R.~J., {et~al.} 2013, \apj, 767, 118

\bibitem[{{Cowley} {et~al.}(2016){Cowley}, {Spitler}, {Tran}, {Rees},
  {Labb\'{e}}, {Allen}, {Brammer}, {Glazebrook}, {Hopkins}, {Juneau},
  {Kacprzak}, {Mullaney}, {Nanayakkara}, {Papovich}, {Quadri}, {Straatman},
  {Tomczak}, \& {van Dokkum}}]{Cowley16}
{Cowley}, M.~J., {et~al.} 2016, \mnras, 457, 629

\bibitem[{{Daddi} {et~al.}(2007){Daddi}, {Dickinson}, {Morrison}, {Chary},
  {Cimatti}, {Elbaz}, {Frayer}, {Renzini}, {Pope}, {Alexander}, {Bauer},
  {Giavalisco}, {Huynh}, {Kurk}, \& {Mignoli}}]{Daddi07b}
{Daddi}, E., {et~al.} 2007, \apj, 670, 156

\bibitem[{{Dale} \& {Helou}(2002)}]{Dale02}
{Dale}, D.~A., \& {Helou}, G. 2002, \apj, 576, 159

\bibitem[{{David} {et~al.}(1992){David}, {Jones}, \& {Forman}}]{David92}
{David}, L.~P., {Jones}, C., \& {Forman}, W. 1992, \apj, 388, 82

\bibitem[{{Dom\'{\i}nguez S\'{a}nchez} {et~al.}(2014){Dom\'{\i}nguez
  S\'{a}nchez}, {Bongiovanni}, {Lara-L\'{o}pez}, {Oteo}, {Cepa}, {P\'{e}rez
  Garc\'{\i}a}, {S\'{a}nchez-Portal}, {Ederoclite}, {Lutz}, {Cresci},
  {Delvecchio}, {Berta}, {Magnelli}, {Popesso}, {Pozzi}, \&
  {Riguccini}}]{DominguezSanchez14}
{Dom\'{\i}nguez S\'{a}nchez}, H., {et~al.} 2014, \mnras, 441, 2

\bibitem[{{Douna} {et~al.}(2015){Douna}, {Pellizza}, {Mirabel}, \&
  {Pedrosa}}]{Douna15}
{Douna}, V.~M., {Pellizza}, L.~J., {Mirabel}, I.~F., \& {Pedrosa}, S.~E. 2015,
  \aap, 579, A44

\bibitem[{{Elbaz} {et~al.}(2007){Elbaz}, {Daddi}, {Le Borgne}, {Dickinson},
  {Alexander}, {Chary}, {Starck}, {Brandt}, {Kitzbichler}, {MacDonald},
  {Nonino}, {Popesso}, {Stern}, \& {Vanzella}}]{Elbaz07}
{Elbaz}, D., {et~al.} 2007, \aap, 468, 33

\bibitem[{{Elbaz} {et~al.}(2011){Elbaz}, {Dickinson}, {Hwang},
  {D\'{\i}az-Santos}, {Magdis}, {Magnelli}, {Le Borgne}, {Galliano},
  {Pannella}, {Chanial}, {Armus}, {Charmandaris}, {Daddi}, {Aussel}, {Popesso},
  {Kartaltepe}, {Altieri}, {Valtchanov}, {Coia}, {Dannerbauer}, {Dasyra},
  {Leiton}, {Mazzarella}, {Alexander}, {Buat}, {Burgarella}, {Chary}, {Gilli},
  {Ivison}, {Juneau}, {Le Floc'h}, {Lutz}, {Morrison}, {Mullaney}, {Murphy},
  {Pope}, {Scott}, {Brodwin}, {Calzetti}, {Cesarsky}, {Charlot}, {Dole},
  {Eisenhardt}, {Ferguson}, {F\"{o}rster Schreiber}, {Frayer}, {Giavalisco},
  {Huynh}, {Koekemoer}, {Papovich}, {Reddy}, {Surace}, {Teplitz}, {Yun}, \&
  {Wilson}}]{Elbaz11}
---. 2011, \aap, 533, A119

\bibitem[{{Elbaz} {et~al.}(2010){Elbaz}, {Hwang}, {Magnelli}, {Daddi},
  {Aussel}, {Altieri}, {Amblard}, {Andreani}, {Arumugam}, {Auld}, {Babbedge},
  {Berta}, {Blain}, {Bock}, {Bongiovanni}, {Boselli}, {Buat}, {Burgarella},
  {Castro-Rodriguez}, {Cava}, {Cepa}, {Chanial}, {Chary}, {Cimatti},
  {Clements}, {Conley}, {Conversi}, {Cooray}, {Dickinson}, {Dominguez},
  {Dowell}, {Dunlop}, {Dwek}, {Eales}, {Farrah}, {F\"{o}rster Schreiber},
  {Fox}, {Franceschini}, {Gear}, {Genzel}, {Glenn}, {Griffin}, {Gruppioni},
  {Halpern}, {Hatziminaoglou}, {Ibar}, {Isaak}, {Ivison}, {Lagache}, {Le
  Borgne}, {Le Floc'h}, {Levenson}, {Lu}, {Lutz}, {Madden}, {Maffei}, {Magdis},
  {Mainetti}, {Maiolino}, {Marchetti}, {Mortier}, {Nguyen}, {Nordon},
  {O'Halloran}, {Okumura}, {Oliver}, {Omont}, {Page}, {Panuzzo},
  {Papageorgiou}, {Pearson}, {Perez Fournon}, {P\'{e}rez Garc\'{\i}a},
  {Poglitsch}, {Pohlen}, {Popesso}, {Pozzi}, {Rawlings}, {Rigopoulou},
  {Riguccini}, {Rizzo}, {Rodighiero}, {Roseboom}, {Rowan-Robinson},
  {Saintonge}, {Sanchez Portal}, {Santini}, {Sauvage}, {Schulz}, {Scott},
  {Seymour}, {Shao}, {Shupe}, {Smith}, {Stevens}, {Sturm}, {Symeonidis},
  {Tacconi}, {Trichas}, {Tugwell}, {Vaccari}, {Valtchanov}, {Vieira},
  {Vigroux}, {Wang}, {Ward}, {Wright}, {Xu}, \& {Zemcov}}]{Elbaz10}
---. 2010, \aap, 518, L29

\bibitem[{{Elvis} {et~al.}(2009){Elvis}, {Civano}, {Vignali}, {Puccetti},
  {Fiore}, {Cappelluti}, {Aldcroft}, {Fruscione}, {Zamorani}, {Comastri},
  {Brusa}, {Gilli}, {Miyaji}, {Damiani}, {Koekemoer}, {Finoguenov}, {Brunner},
  {Urry}, {Silverman}, {Mainieri}, {Hasinger}, {Griffiths}, {Carollo}, {Hao},
  {Guzzo}, {Blain}, {Calzetti}, {Carilli}, {Capak}, {Ettori}, {Fabbiano},
  {Impey}, {Lilly}, {Mobasher}, {Rich}, {Salvato}, {Sanders}, {Schinnerer},
  {Scoville}, {Shopbell}, {Taylor}, {Taniguchi}, \& {Volonteri}}]{Elvis09}
{Elvis}, M., {et~al.} 2009, \apjs, 184, 158

\bibitem[{{Erb} {et~al.}(2006){Erb}, {Shapley}, {Pettini}, {Steidel}, {Reddy},
  \& {Adelberger}}]{Erb06}
{Erb}, D.~K., {Shapley}, A.~E., {Pettini}, M., {Steidel}, C.~C., {Reddy},
  N.~A., \& {Adelberger}, K.~L. 2006, \apj, 644, 813

\bibitem[{{Fabbiano}(1989)}]{Fabbiano89}
{Fabbiano}, G. 1989, \araa, 27, 87

\bibitem[{{Faber} {et~al.}(2007){Faber}, {Willmer}, {Wolf}, {Koo}, {Weiner},
  {Newman}, {Im}, {Coil}, {Conroy}, {Cooper}, {Davis}, {Finkbeiner}, {Gerke},
  {Gebhardt}, {Groth}, {Guhathakurta}, {Harker}, {Kaiser}, {Kassin},
  {Kleinheinrich}, {Konidaris}, {Kron}, {Lin}, {Luppino}, {Madgwick},
  {Meisenheimer}, {Noeske}, {Phillips}, {Sarajedini}, {Schiavon}, {Simard},
  {Szalay}, {Vogt}, \& {Yan}}]{Faber07}
{Faber}, S.~M., {et~al.} 2007, \apj, 665, 265

\bibitem[{{Feroz} \& {Hobson}(2008)}]{Feroz08}
{Feroz}, F., \& {Hobson}, M.~P. 2008, \mnras, 384, 449

\bibitem[{{Feroz} {et~al.}(2009){Feroz}, {Hobson}, \& {Bridges}}]{Feroz09}
{Feroz}, F., {Hobson}, M.~P., \& {Bridges}, M. 2009, \mnras, 398, 1601

\bibitem[{{Feroz} {et~al.}(2013){Feroz}, {Hobson}, {Cameron}, \&
  {Pettitt}}]{Feroz13}
{Feroz}, F., {Hobson}, M.~P., {Cameron}, E., \& {Pettitt}, A.~N. 2013,
  arXiv:1306.2144

\bibitem[{{Fragos} {et~al.}(2013){Fragos}, {Lehmer}, {Tremmel}, {Tzanavaris},
  {Basu-Zych}, {Belczynski}, {Hornschemeier}, {Jenkins}, {Kalogera}, {Ptak}, \&
  {Zezas}}]{Fragos13}
{Fragos}, T., {et~al.} 2013, \apj, 764, 41

\bibitem[{{Georgakakis} {et~al.}(2015){Georgakakis}, {Aird}, {Buchner},
  {Salvato}, {Menzel}, {Brandt}, {McGreer}, {Dwelly}, {Mountrichas}, {Koki},
  {Georgantopoulos}, {Hsu}, {Merloni}, {Liu}, {Nandra}, \&
  {Ross}}]{Georgakakis15}
{Georgakakis}, A., {et~al.} 2015, \mnras, 453, 1946

\bibitem[{{Georgakakis} {et~al.}(2008){Georgakakis}, {Nandra}, {Laird}, {Aird},
  \& {Trichas}}]{Georgakakis08}
{Georgakakis}, A., {Nandra}, K., {Laird}, E.~S., {Aird}, J., \& {Trichas}, M.
  2008, \mnras, 388, 1205

\bibitem[{{Georgakakis} {et~al.}(2014){Georgakakis}, {P\'{e}rez-Gonz\'{a}lez},
  {Fanidakis}, {Salvato}, {Aird}, {Messias}, {Lotz}, {Barro}, {Hsu}, {Nandra},
  {Rosario}, {Cooper}, {Kocevski}, \& {Newman}}]{Georgakakis14}
{Georgakakis}, A., {et~al.} 2014, \mnras, 440, 339

\bibitem[{{Georgakakis} {et~al.}(2006){Georgakakis}, {Chavushyan}, {Plionis},
  {Georgantopoulos}, {Koulouridis}, {Leonidaki}, \& {Mercado}}]{Georgakakis06b}
{Georgakakis}, A.~E., {Chavushyan}, V., {Plionis}, M., {Georgantopoulos}, I.,
  {Koulouridis}, E., {Leonidaki}, I., \& {Mercado}, A. 2006, \mnras, 367, 1017

\bibitem[{{Ghosh} \& {White}(2001)}]{Ghosh01}
{Ghosh}, P., \& {White}, N.~E. 2001, \apjl, 559, L97

\bibitem[{{Gordon} {et~al.}(2000){Gordon}, {Clayton}, {Witt}, \&
  {Misselt}}]{Gordon00}
{Gordon}, K.~D., {Clayton}, G.~C., {Witt}, A.~N., \& {Misselt}, K.~A. 2000,
  \apj, 533, 236

\bibitem[{{Grimm} {et~al.}(2003){Grimm}, {Gilfanov}, \& {Sunyaev}}]{Grimm03}
{Grimm}, H.-J., {Gilfanov}, M., \& {Sunyaev}, R. 2003, Astronomische
  Nachrichten, 324, 171

\bibitem[{{Grogin} {et~al.}(2011){Grogin}, {Kocevski}, {Faber}, {Ferguson},
  {Koekemoer}, {Riess}, {Acquaviva}, {Alexander}, {Almaini}, {Ashby}, {Barden},
  {Bell}, {Bournaud}, {Brown}, {Caputi}, {Casertano}, {Cassata}, {Castellano},
  {Challis}, {Chary}, {Cheung}, {Cirasuolo}, {Conselice}, {Roshan Cooray},
  {Croton}, {Daddi}, {Dahlen}, {Dav\'{e}}, {de Mello}, {Dekel}, {Dickinson},
  {Dolch}, {Donley}, {Dunlop}, {Dutton}, {Elbaz}, {Fazio}, {Filippenko},
  {Finkelstein}, {Fontana}, {Gardner}, {Garnavich}, {Gawiser}, {Giavalisco},
  {Grazian}, {Guo}, {Hathi}, {H\"{a}ussler}, {Hopkins}, {Huang}, {Huang},
  {Jha}, {Kartaltepe}, {Kirshner}, {Koo}, {Lai}, {Lee}, {Li}, {Lotz}, {Lucas},
  {Madau}, {McCarthy}, {McGrath}, {McIntosh}, {McLure}, {Mobasher},
  {Moustakas}, {Mozena}, {Nandra}, {Newman}, {Niemi}, {Noeske}, {Papovich},
  {Pentericci}, {Pope}, {Primack}, {Rajan}, {Ravindranath}, {Reddy}, {Renzini},
  {Rix}, {Robaina}, {Rodney}, {Rosario}, {Rosati}, {Salimbeni}, {Scarlata},
  {Siana}, {Simard}, {Smidt}, {Somerville}, {Spinrad}, {Straughn}, {Strolger},
  {Telford}, {Teplitz}, {Trump}, {van der Wel}, {Villforth}, {Wechsler},
  {Weiner}, {Wiklind}, {Wild}, {Wilson}, {Wuyts}, {Yan}, \& {Yun}}]{Grogin11}
{Grogin}, N.~A., {et~al.} 2011, \apjs, 197, 35

\bibitem[{{Hao} {et~al.}(2011){Hao}, {Kennicutt}, {Johnson}, {Calzetti},
  {Dale}, \& {Moustakas}}]{Hao11}
{Hao}, C.-N., {Kennicutt}, R.~C., {Johnson}, B.~D., {Calzetti}, D., {Dale},
  D.~A., \& {Moustakas}, J. 2011, \apj, 741, 124

\bibitem[{{Hsu} {et~al.}(2014){Hsu}, {Salvato}, {Nandra}, {Brusa}, {Bender},
  {Buchner}, {Donley}, {Kocevski}, {Guo}, {Hathi}, {Rangel}, {Willner},
  {Brightman}, {Georgakakis}, {Budav\'{a}ri}, {Szalay}, {Ashby}, {Barro},
  {Dahlen}, {Faber}, {Ferguson}, {Galametz}, {Grazian}, {Grogin}, {Huang},
  {Koekemoer}, {Lucas}, {McGrath}, {Mobasher}, {Peth}, {Rosario}, \&
  {Trump}}]{Hsu14}
{Hsu}, L.-T., {et~al.} 2014, \apj, 796, 60

\bibitem[{{Ilbert} {et~al.}(2015){Ilbert}, {Arnouts}, {Le Floc'h}, {Aussel},
  {Bethermin}, {Capak}, {Hsieh}, {Kajisawa}, {Karim}, {Le F\`{e}vre}, {Lee},
  {Lilly}, {McCracken}, {Michel-Dansac}, {Moutard}, {Renzini}, {Salvato},
  {Sanders}, {Scoville}, {Sheth}, {Silverman}, {Smol{\v c}i\'{c}}, {Taniguchi},
  \& {Tresse}}]{Ilbert15}
{Ilbert}, O., {et~al.} 2015, \aap, 579, A2

\bibitem[{{Johnston} {et~al.}(2015){Johnston}, {Vaccari}, {Jarvis}, {Smith},
  {Giovannoli}, {H\"{a}u{\ss}ler}, \& {Prescott}}]{Johnston15}
{Johnston}, R., {Vaccari}, M., {Jarvis}, M., {Smith}, M., {Giovannoli}, E.,
  {H\"{a}u{\ss}ler}, B., \& {Prescott}, M. 2015, \mnras, 453, 2540

\bibitem[{{Jones} {et~al.}(2016){Jones}, {Hickox}, {Black}, {Hainline},
  {DiPompeo}, \& {Goulding}}]{Jones16}
{Jones}, M.~L., {Hickox}, R.~C., {Black}, C.~S., {Hainline}, K.~N., {DiPompeo},
  M.~A., \& {Goulding}, A.~D. 2016, \apj, 826, 12

\bibitem[{{Kaaret} {et~al.}(2011){Kaaret}, {Schmitt}, \& {Gorski}}]{Kaaret11}
{Kaaret}, P., {Schmitt}, J., \& {Gorski}, M. 2011, \apj, 741, 10

\bibitem[{{Karim} {et~al.}(2011){Karim}, {Schinnerer},
  {Mart\'{\i}nez-Sansigre}, {Sargent}, {van der Wel}, {Rix}, {Ilbert}, {Smol{\v
  c}i\'{c}}, {Carilli}, {Pannella}, {Koekemoer}, {Bell}, \&
  {Salvato}}]{Karim11}
{Karim}, A., {et~al.} 2011, \apj, 730, 61

\bibitem[{{Kelly} {et~al.}(2008){Kelly}, {Fan}, \& {Vestergaard}}]{Kelly08}
{Kelly}, B.~C., {Fan}, X., \& {Vestergaard}, M. 2008, \apj, 682, 874

\bibitem[{{Kennicutt} \& {Evans}(2012)}]{Kennicutt12}
{Kennicutt}, R.~C., \& {Evans}, N.~J. 2012, \araa, 50, 531

\bibitem[{{Kennicutt}(1998)}]{Kennicutt98}
{Kennicutt}, Jr., R.~C. 1998, \araa, 36, 189

\bibitem[{{Kewley} \& {Ellison}(2008)}]{Kewley08}
{Kewley}, L.~J., \& {Ellison}, S.~L. 2008, \apj, 681, 1183

\bibitem[{{Koekemoer} {et~al.}(2011){Koekemoer}, {Faber}, {Ferguson}, {Grogin},
  {Kocevski}, {Koo}, {Lai}, {Lotz}, {Lucas}, {McGrath}, {Ogaz}, {Rajan},
  {Riess}, {Rodney}, {Strolger}, {Casertano}, {Castellano}, {Dahlen},
  {Dickinson}, {Dolch}, {Fontana}, {Giavalisco}, {Grazian}, {Guo}, {Hathi},
  {Huang}, {van der Wel}, {Yan}, {Acquaviva}, {Alexander}, {Almaini}, {Ashby},
  {Barden}, {Bell}, {Bournaud}, {Brown}, {Caputi}, {Cassata}, {Challis},
  {Chary}, {Cheung}, {Cirasuolo}, {Conselice}, {Roshan Cooray}, {Croton},
  {Daddi}, {Dav\'{e}}, {de Mello}, {de Ravel}, {Dekel}, {Donley}, {Dunlop},
  {Dutton}, {Elbaz}, {Fazio}, {Filippenko}, {Finkelstein}, {Frazer}, {Gardner},
  {Garnavich}, {Gawiser}, {Gruetzbauch}, {Hartley}, {H\"{a}ussler},
  {Herrington}, {Hopkins}, {Huang}, {Jha}, {Johnson}, {Kartaltepe},
  {Khostovan}, {Kirshner}, {Lani}, {Lee}, {Li}, {Madau}, {McCarthy},
  {McIntosh}, {McLure}, {McPartland}, {Mobasher}, {Moreira}, {Mortlock},
  {Moustakas}, {Mozena}, {Nandra}, {Newman}, {Nielsen}, {Niemi}, {Noeske},
  {Papovich}, {Pentericci}, {Pope}, {Primack}, {Ravindranath}, {Reddy},
  {Renzini}, {Rix}, {Robaina}, {Rosario}, {Rosati}, {Salimbeni}, {Scarlata},
  {Siana}, {Simard}, {Smidt}, {Snyder}, {Somerville}, {Spinrad}, {Straughn},
  {Telford}, {Teplitz}, {Trump}, {Vargas}, {Villforth}, {Wagner}, {Wandro},
  {Wechsler}, {Weiner}, {Wiklind}, {Wild}, {Wilson}, {Wuyts}, \&
  {Yun}}]{Koekemoer11}
{Koekemoer}, A.~M., {et~al.} 2011, \apjs, 197, 36

\bibitem[{{Kriek} \& {Conroy}(2013)}]{Kriek13}
{Kriek}, M., \& {Conroy}, C. 2013, \apjl, 775, L16

\bibitem[{{Kriek} {et~al.}(2015){Kriek}, {Shapley}, {Reddy}, {Siana}, {Coil},
  {Mobasher}, {Freeman}, {de Groot}, {Price}, {Sanders}, {Shivaei}, {Brammer},
  {Momcheva}, {Skelton}, {van Dokkum}, {Whitaker}, {Aird}, {Azadi}, {Kassis},
  {Bullock}, {Conroy}, {Dav{\'e}}, {Kere{\v s}}, \& {Krumholz}}]{Kriek15}
{Kriek}, M., {et~al.} 2015, \apjs, 218, 15

\bibitem[{{Kriek} {et~al.}(2009){Kriek}, {van Dokkum}, {Labb\'{e}}, {Franx},
  {Illingworth}, {Marchesini}, \& {Quadri}}]{Kriek09}
{Kriek}, M., {van Dokkum}, P.~G., {Labb\'{e}}, I., {Franx}, M., {Illingworth},
  G.~D., {Marchesini}, D., \& {Quadri}, R.~F. 2009, \apj, 700, 221

\bibitem[{{Labb\'{e}} {et~al.}(2005){Labb\'{e}}, {Huang}, {Franx}, {Rudnick},
  {Barmby}, {Daddi}, {van Dokkum}, {Fazio}, {Schreiber}, {Moorwood}, {Rix},
  {R\"{o}ttgering}, {Trujillo}, \& {van der Werf}}]{Labbe05}
{Labb\'{e}}, I., {et~al.} 2005, \apjl, 624, L81

\bibitem[{{Laird} {et~al.}(2009){Laird}, {Nandra}, {Georgakakis}, {Aird},
  {Barmby}, {Conselice}, {Coil}, {Davis}, {Faber}, {Fazio}, {Guhathakurta},
  {Koo}, {Sarajedini}, \& {Willmer}}]{Laird09}
{Laird}, E.~S., {et~al.} 2009, \apjs, 180, 102

\bibitem[{{Laird} {et~al.}(2006){Laird}, {Nandra}, {Hobbs}, \&
  {Steidel}}]{Laird06}
{Laird}, E.~S., {Nandra}, K., {Hobbs}, A., \& {Steidel}, C.~C. 2006, \mnras,
  373, 217

\bibitem[{{Le F\`{e}vre} {et~al.}(2015){Le F\`{e}vre}, {Tasca}, {Cassata},
  {Garilli}, {Le Brun}, {Maccagni}, {Pentericci}, {Thomas}, {Vanzella},
  {Zamorani}, {Zucca}, {Amorin}, {Bardelli}, {Capak}, {Cassar\`{a}},
  {Castellano}, {Cimatti}, {Cuby}, {Cucciati}, {de la Torre}, {Durkalec},
  {Fontana}, {Giavalisco}, {Grazian}, {Hathi}, {Ilbert}, {Lemaux}, {Moreau},
  {Paltani}, {Ribeiro}, {Salvato}, {Schaerer}, {Scodeggio}, {Sommariva},
  {Talia}, {Taniguchi}, {Tresse}, {Vergani}, {Wang}, {Charlot}, {Contini},
  {Fotopoulou}, {L\'{o}pez-Sanjuan}, {Mellier}, \& {Scoville}}]{LeFevre15}
{Le F\`{e}vre}, O., {et~al.} 2015, \aap, 576, A79

\bibitem[{{Lee} {et~al.}(2015){Lee}, {Sanders}, {Casey}, {Toft}, {Scoville},
  {Hung}, {Le Floc'h}, {Ilbert}, {Zahid}, {Aussel}, {Capak}, {Kartaltepe},
  {Kewley}, {Li}, {Schawinski}, {Sheth}, \& {Xiao}}]{Lee15}
{Lee}, N., {et~al.} 2015, \apj, 801, 80

\bibitem[{{Lehmer} {et~al.}(2010){Lehmer}, {Alexander}, {Bauer}, {Brandt},
  {Goulding}, {Jenkins}, {Ptak}, \& {Roberts}}]{Lehmer10}
{Lehmer}, B.~D., {Alexander}, D.~M., {Bauer}, F.~E., {Brandt}, W.~N.,
  {Goulding}, A.~D., {Jenkins}, L.~P., {Ptak}, A., \& {Roberts}, T.~P. 2010,
  \apj, 724, 559

\bibitem[{{Lehmer} {et~al.}(2016){Lehmer}, {Basu-Zych}, {Mineo}, {Brandt},
  {Eufrasio}, {Fragos}, {Hornschemeier}, {Luo}, {Xue}, {Bauer}, {Gilfanov},
  {Ranalli}, {Schneider}, {Shemmer}, {Tozzi}, {Trump}, {Vignali}, {Wang},
  {Yukita}, \& {Zezas}}]{Lehmer16}
{Lehmer}, B.~D., {et~al.} 2016, \apj, 825, 7

\bibitem[{{Lehmer} {et~al.}(2008){Lehmer}, {Brandt}, {Alexander}, {Bell},
  {Hornschemeier}, {McIntosh}, {Bauer}, {Gilli}, {Mainieri}, {Schneider},
  {Silverman}, {Steffen}, {Tozzi}, \& {Wolf}}]{Lehmer08}
---. 2008, \apj, 681, 1163

\bibitem[{{Lehmer} {et~al.}(2012){Lehmer}, {Xue}, {Brandt}, {Alexander},
  {Bauer}, {Brusa}, {Comastri}, {Gilli}, {Hornschemeier}, {Luo}, {Paolillo},
  {Ptak}, {Shemmer}, {Schneider}, {Tozzi}, \& {Vignali}}]{Lehmer12}
---. 2012, \apj, 752, 46

\bibitem[{{Luo} {et~al.}(2010){Luo}, {Brandt}, {Xue}, {Brusa}, {Alexander},
  {Bauer}, {Comastri}, {Koekemoer}, {Lehmer}, {Mainieri}, {Rafferty},
  {Schneider}, {Silverman}, \& {Vignali}}]{Luo10}
{Luo}, B., {et~al.} 2010, \apjs, 187, 560

\bibitem[{{Lupton}(1993)}]{Lupton93}
{Lupton}, R. 1993, {Statistics in theory and practice}

\bibitem[{{Lutz} {et~al.}(2011){Lutz}, {Poglitsch}, {Altieri}, {Andreani},
  {Aussel}, {Berta}, {Bongiovanni}, {Brisbin}, {Cava}, {Cepa}, {Cimatti},
  {Daddi}, {Dominguez-Sanchez}, {Elbaz}, {F\"{o}rster Schreiber}, {Genzel},
  {Grazian}, {Gruppioni}, {Harwit}, {Le Floc'h}, {Magdis}, {Magnelli},
  {Maiolino}, {Nordon}, {P\'{e}rez Garc\'{\i}a}, {Popesso}, {Pozzi},
  {Riguccini}, {Rodighiero}, {Saintonge}, {Sanchez Portal}, {Santini}, {Shao},
  {Sturm}, {Tacconi}, {Valtchanov}, {Wetzstein}, \& {Wieprecht}}]{Lutz11}
{Lutz}, D., {et~al.} 2011, \aap, 532, A90

\bibitem[{{Madau} \& {Dickinson}(2014)}]{Madau14}
{Madau}, P., \& {Dickinson}, M. 2014, \araa, 52, 415

\bibitem[{{Magnelli} {et~al.}(2015){Magnelli}, {Ivison}, {Lutz}, {Valtchanov},
  {Farrah}, {Berta}, {Bertoldi}, {Bock}, {Cooray}, {Ibar}, {Karim}, {Le
  Floc'h}, {Nordon}, {Oliver}, {Page}, {Popesso}, {Pozzi}, {Rigopoulou},
  {Riguccini}, {Rodighiero}, {Rosario}, {Roseboom}, {Wang}, \&
  {Wuyts}}]{Magnelli15}
{Magnelli}, B., {et~al.} 2015, \aap, 573, A45

\bibitem[{{Magnelli} {et~al.}(2013){Magnelli}, {Popesso}, {Berta}, {Pozzi},
  {Elbaz}, {Lutz}, {Dickinson}, {Altieri}, {Andreani}, {Aussel},
  {B\'{e}thermin}, {Bongiovanni}, {Cepa}, {Charmandaris}, {Chary}, {Cimatti},
  {Daddi}, {F\"{o}rster Schreiber}, {Genzel}, {Gruppioni}, {Harwit}, {Hwang},
  {Ivison}, {Magdis}, {Maiolino}, {Murphy}, {Nordon}, {Pannella}, {P\'{e}rez
  Garc\'{\i}a}, {Poglitsch}, {Rosario}, {Sanchez-Portal}, {Santini}, {Scott},
  {Sturm}, {Tacconi}, \& {Valtchanov}}]{Magnelli13}
---. 2013, \aap, 553, A132

\bibitem[{{Maraston} {et~al.}(2010){Maraston}, {Pforr}, {Renzini}, {Daddi},
  {Dickinson}, {Cimatti}, \& {Tonini}}]{Maraston10}
{Maraston}, C., {Pforr}, J., {Renzini}, A., {Daddi}, E., {Dickinson}, M.,
  {Cimatti}, A., \& {Tonini}, C. 2010, \mnras, 407, 830

\bibitem[{{Marchesi} {et~al.}(2016){Marchesi}, {Civano}, {Elvis}, {Salvato},
  {Brusa}, {Comastri}, {Gilli}, {Hasinger}, {Lanzuisi}, {Miyaji}, {Treister},
  {Urry}, {Vignali}, {Zamorani}, {Allevato}, {Cappelluti}, {Cardamone},
  {Finoguenov}, {Griffiths}, {Karim}, {Laigle}, {LaMassa}, {Jahnke}, {Ranalli},
  {Schawinski}, {Schinnerer}, {Silverman}, {Smolcic}, {Suh}, \&
  {Trakhtenbrot}}]{Marchesi16}
{Marchesi}, S., {et~al.} 2016, \apj, 817, 34

\bibitem[{Martin {et~al.}(2005)Martin, Fanson, Schiminovich, Morrissey,
  Friedman, Barlow, Conrow, Grange, Jelinsky, Milliard, Siegmund, Bianchi,
  Byun, Donas, Forster, Heckman, Lee, Madore, Malina, Neff, Rich, Small,
  Szalay, \& Wyder}]{Martin05}
Martin, D.~C., {et~al.} 2005, Astrophys.J., 619, L1

\bibitem[{{McCracken} {et~al.}(2012){McCracken}, {Milvang-Jensen}, {Dunlop},
  {Franx}, {Fynbo}, {Le F\`{e}vre}, {Holt}, {Caputi}, {Goranova}, {Buitrago},
  {Emerson}, {Freudling}, {Hudelot}, {L\'{o}pez-Sanjuan}, {Magnard}, {Mellier},
  {M{\o}ller}, {Nilsson}, {Sutherland}, {Tasca}, \& {Zabl}}]{McCracken12}
{McCracken}, H.~J., {et~al.} 2012, \aap, 544, A156

\bibitem[{{Merloni} {et~al.}(2012){Merloni}, {Predehl}, {Becker},
  {B\"{o}hringer}, {Boller}, {Brunner}, {Brusa}, {Dennerl}, {Freyberg},
  {Friedrich}, {Georgakakis}, {Haberl}, {Hasinger}, {Meidinger}, {Mohr},
  {Nandra}, {Rau}, {Reiprich}, {Robrade}, {Salvato}, {Santangelo}, {Sasaki},
  {Schwope}, {Wilms}, \& {German eROSITA Consortium}}]{Merloni12}
{Merloni}, A., {et~al.} 2012, (arXiv:1209.3114)

\bibitem[{{Mineo} {et~al.}(2014){Mineo}, {Gilfanov}, {Lehmer}, {Morrison}, \&
  {Sunyaev}}]{Mineo14}
{Mineo}, S., {Gilfanov}, M., {Lehmer}, B.~D., {Morrison}, G.~E., \& {Sunyaev},
  R. 2014, \mnras, 437, 1698

\bibitem[{{Mineo} {et~al.}(2012){Mineo}, {Gilfanov}, \& {Sunyaev}}]{Mineo12b}
{Mineo}, S., {Gilfanov}, M., \& {Sunyaev}, R. 2012, \mnras, 426, 1870

\bibitem[{{Momcheva} {et~al.}(2016){Momcheva}, {Brammer}, {van Dokkum},
  {Skelton}, {Whitaker}, {Nelson}, {Fumagalli}, {Maseda}, {Leja}, {Franx},
  {Rix}, {Bezanson}, {Da Cunha}, {Dickey}, {F{\"o}rster Schreiber},
  {Illingworth}, {Kriek}, {Labb{\'e}}, {Ulf Lange}, {Lundgren}, {Magee},
  {Marchesini}, {Oesch}, {Pacifici}, {Patel}, {Price}, {Tal}, {Wake}, {van der
  Wel}, \& {Wuyts}}]{Momcheva16}
{Momcheva}, I.~G., {et~al.} 2016, \apjs, 225, 27

\bibitem[{{Moustakas} {et~al.}(2013){Moustakas}, {Coil}, {Aird}, {Blanton},
  {Cool}, {Eisenstein}, {Mendez}, {Wong}, {Zhu}, \& {Arnouts}}]{Moustakas13}
{Moustakas}, J., {et~al.} 2013, \apj, 767, 50

\bibitem[{{Moustakas} {et~al.}(2006){Moustakas}, {Kennicutt}, \&
  {Tremonti}}]{Moustakas06}
{Moustakas}, J., {Kennicutt}, Jr., R.~C., \& {Tremonti}, C.~A. 2006, \apj, 642,
  775

\bibitem[{{Muzzin} {et~al.}(2013{\natexlab{a}}){Muzzin}, {Marchesini},
  {Stefanon}, {Franx}, {McCracken}, {Milvang-Jensen}, {Dunlop}, {Fynbo},
  {Brammer}, {Labb\'{e}}, \& {van Dokkum}}]{Muzzin13b}
{Muzzin}, A., {et~al.} 2013{\natexlab{a}}, \apj, 777, 18

\bibitem[{{Muzzin} {et~al.}(2013{\natexlab{b}}){Muzzin}, {Marchesini},
  {Stefanon}, {Franx}, {Milvang-Jensen}, {Dunlop}, {Fynbo}, {Brammer},
  {Labb\'{e}}, \& {van Dokkum}}]{Muzzin13}
---. 2013{\natexlab{b}}, \apjs, 206, 8

\bibitem[{{Nandra} {et~al.}(2013){Nandra}, {Barret}, {Barcons}, {Fabian}, {den
  Herder}, {Piro}, {Watson}, {Adami}, {Aird}, {Afonso}, {et~al.}}]{Nandra13}
{Nandra}, K., {et~al.} 2013, (arXiv:1306.2307)

\bibitem[{{Nandra} {et~al.}(2015){Nandra}, {Laird}, {Aird}, {Salvato},
  {Georgakakis}, {Barro}, {Perez-Gonzalez}, {Barmby}, {Chary}, {Coil},
  {Cooper}, {Davis}, {Dickinson}, {Faber}, {Fazio}, {Guhathakurta}, {Gwyn},
  {Hsu}, {Huang}, {Ivison}, {Koo}, {Newman}, {Rangel}, {Yamada}, \&
  {Willmer}}]{Nandra15}
---. 2015, \apjs, 220, 10

\bibitem[{{Nandra} {et~al.}(2002){Nandra}, {Mushotzky}, {Arnaud}, {Steidel},
  {Adelberger}, {Gardner}, {Teplitz}, \& {Windhorst}}]{Nandra02}
{Nandra}, K., {Mushotzky}, R.~F., {Arnaud}, K., {Steidel}, C.~C., {Adelberger},
  K.~L., {Gardner}, J.~P., {Teplitz}, H.~I., \& {Windhorst}, R.~A. 2002, \apj,
  576, 625

\bibitem[{{Noeske} {et~al.}(2007){Noeske}, {Weiner}, {Faber}, {Papovich},
  {Koo}, {Somerville}, {Bundy}, {Conselice}, {Newman}, {Schiminovich}, {Le
  Floc'h}, {Coil}, {Rieke}, {Lotz}, {Primack}, {Barmby}, {Cooper}, {Davis},
  {Ellis}, {Fazio}, {Guhathakurta}, {Huang}, {Kassin}, {Martin}, {Phillips},
  {Rich}, {Small}, {Willmer}, \& {Wilson}}]{Noeske07}
{Noeske}, K.~G., {et~al.} 2007, \apjl, 660, L43

\bibitem[{{Persic} \& {Rephaeli}(2007)}]{Persic07}
{Persic}, M., \& {Rephaeli}, Y. 2007, \aap, 463, 481

\bibitem[{{Ranalli} {et~al.}(2003){Ranalli}, {Comastri}, \&
  {Setti}}]{Ranalli03}
{Ranalli}, P., {Comastri}, A., \& {Setti}, G. 2003, \aap, 399, 39

\bibitem[{{Rangel} {et~al.}(2013){Rangel}, {Nandra}, {Laird}, \&
  {Orange}}]{Rangel13}
{Rangel}, C., {Nandra}, K., {Laird}, E.~S., \& {Orange}, P. 2013, \mnras, 428,
  3089

\bibitem[{{Reddy} {et~al.}(2015){Reddy}, {Kriek}, {Shapley}, {Freeman},
  {Siana}, {Coil}, {Mobasher}, {Price}, {Sanders}, \& {Shivaei}}]{Reddy15}
{Reddy}, N.~A., {et~al.} 2015, \apj, 806, 259

\bibitem[{{Reddy} {et~al.}(2012){Reddy}, {Pettini}, {Steidel}, {Shapley},
  {Erb}, \& {Law}}]{Reddy12}
{Reddy}, N.~A., {Pettini}, M., {Steidel}, C.~C., {Shapley}, A.~E., {Erb},
  D.~K., \& {Law}, D.~R. 2012, \apj, 754, 25

\bibitem[{{Rodighiero} {et~al.}(2011){Rodighiero}, {Daddi}, {Baronchelli},
  {Cimatti}, {Renzini}, {Aussel}, {Popesso}, {Lutz}, {Andreani}, {Berta},
  {Cava}, {Elbaz}, {Feltre}, {Fontana}, {F\"{o}rster Schreiber},
  {Franceschini}, {Genzel}, {Grazian}, {Gruppioni}, {Ilbert}, {Le Floch},
  {Magdis}, {Magliocchetti}, {Magnelli}, {Maiolino}, {McCracken}, {Nordon},
  {Poglitsch}, {Santini}, {Pozzi}, {Riguccini}, {Tacconi}, {Wuyts}, \&
  {Zamorani}}]{Rodighiero11}
{Rodighiero}, G., {et~al.} 2011, \apjl, 739, L40

\bibitem[{{Rovilos} {et~al.}(2009){Rovilos}, {Georgantopoulos}, {Tzanavaris},
  {Pracy}, {Whiting}, {Woods}, \& {Goudis}}]{Rovilos09}
{Rovilos}, E., {Georgantopoulos}, I., {Tzanavaris}, P., {Pracy}, M., {Whiting},
  M., {Woods}, D., \& {Goudis}, C. 2009, \aap, 502, 85

\bibitem[{{Salim} {et~al.}(2007){Salim}, {Rich}, {Charlot}, {Brinchmann},
  {Johnson}, {Schiminovich}, {Seibert}, {Mallery}, {Heckman}, {Forster},
  {Friedman}, {Martin}, {Morrissey}, {Neff}, {Small}, {Wyder}, {Bianchi},
  {Donas}, {Lee}, {Madore}, {Milliard}, {Szalay}, {Welsh}, \& {Yi}}]{Salim07}
{Salim}, S., {et~al.} 2007, \apjs, 173, 267

\bibitem[{{Salvato} {et~al.}(2009){Salvato}, {Hasinger}, {Ilbert}, {Zamorani},
  {Brusa}, {Scoville}, {Rau}, {Capak}, {Arnouts}, {Aussel}, {Bolzonella},
  {Buongiorno}, {Cappelluti}, {Caputi}, {Civano}, {Cook}, {Elvis}, {Gilli},
  {Jahnke}, {Kartaltepe}, {Impey}, {Lamareille}, {Le Floc'h}, {Lilly},
  {Mainieri}, {McCarthy}, {McCracken}, {Mignoli}, {Mobasher}, {Murayama},
  {Sasaki}, {Sanders}, {Schiminovich}, {Shioya}, {Shopbell}, {Silverman},
  {Smol{\v c}i\'{c}}, {Surace}, {Taniguchi}, {Thompson}, {Trump}, {Urry}, \&
  {Zamojski}}]{Salvato09}
{Salvato}, M., {et~al.} 2009, \apj, 690, 1250

\bibitem[{{Salvato} {et~al.}(2011){Salvato}, {Ilbert}, {Hasinger}, {Rau},
  {Civano}, {Zamorani}, {Brusa}, {Elvis}, {Vignali}, {Aussel}, {Comastri},
  {Fiore}, {Le Floc'h}, {Mainieri}, {Bardelli}, {Bolzonella}, {Bongiorno},
  {Capak}, {Caputi}, {Cappelluti}, {Carollo}, {Contini}, {Garilli}, {Iovino},
  {Fotopoulou}, {Fruscione}, {Gilli}, {Halliday}, {Kneib}, {Kakazu},
  {Kartaltepe}, {Koekemoer}, {Kovac}, {Ideue}, {Ikeda}, {Impey}, {Le Fevre},
  {Lamareille}, {Lanzuisi}, {Le Borgne}, {Le Brun}, {Lilly}, {Maier},
  {Manohar}, {Masters}, {McCracken}, {Messias}, {Mignoli}, {Mobasher}, {Nagao},
  {Pello}, {Puccetti}, {Perez-Montero}, {Renzini}, {Sargent}, {Sanders},
  {Scodeggio}, {Scoville}, {Shopbell}, {Silvermann}, {Taniguchi}, {Tasca},
  {Tresse}, {Trump}, \& {Zucca}}]{Salvato11}
---. 2011, \apj, 742, 61

\bibitem[{{Sanders} {et~al.}(2007){Sanders}, {Salvato}, {Aussel}, {Ilbert},
  {Scoville}, {Surace}, {Frayer}, {Sheth}, {Helou}, {Brooke}, {Bhattacharya},
  {Yan}, {Kartaltepe}, {Barnes}, {Blain}, {Calzetti}, {Capak}, {Carilli},
  {Carollo}, {Comastri}, {Daddi}, {Ellis}, {Elvis}, {Fall}, {Franceschini},
  {Giavalisco}, {Hasinger}, {Impey}, {Koekemoer}, {Le F\`{e}vre}, {Lilly},
  {Liu}, {McCracken}, {Mobasher}, {Renzini}, {Rich}, {Schinnerer}, {Shopbell},
  {Taniguchi}, {Thompson}, {Urry}, \& {Williams}}]{Sanders07}
{Sanders}, D.~B., {et~al.} 2007, \apjs, 172, 86

\bibitem[{{Santini} {et~al.}(2014){Santini}, {Maiolino}, {Magnelli}, {Lutz},
  {Lamastra}, {Li Causi}, {Eales}, {Andreani}, {Berta}, {Buat}, {Cooray},
  {Cresci}, {Daddi}, {Farrah}, {Fontana}, {Franceschini}, {Genzel}, {Granato},
  {Grazian}, {Le Floc'h}, {Magdis}, {Magliocchetti}, {Mannucci}, {Menci},
  {Nordon}, {Oliver}, {Popesso}, {Pozzi}, {Riguccini}, {Rodighiero}, {Rosario},
  {Salvato}, {Scott}, {Silva}, {Tacconi}, {Viero}, {Wang}, {Wuyts}, \&
  {Xu}}]{Santini14}
{Santini}, P., {et~al.} 2014, \aap, 562, A30

\bibitem[{{Schreiber} {et~al.}(2015){Schreiber}, {Pannella}, {Elbaz},
  {B\'{e}thermin}, {Inami}, {Dickinson}, {Magnelli}, {Wang}, {Aussel}, {Daddi},
  {Juneau}, {Shu}, {Sargent}, {Buat}, {Faber}, {Ferguson}, {Giavalisco},
  {Koekemoer}, {Magdis}, {Morrison}, {Papovich}, {Santini}, \&
  {Scott}}]{Schreiber15}
{Schreiber}, C., {et~al.} 2015, \aap, 575, A74

\bibitem[{{Scoville} {et~al.}(2007){Scoville}, {Aussel}, {Brusa}, {Capak},
  {Carollo}, {Elvis}, {Giavalisco}, {Guzzo}, {Hasinger}, {Impey}, {Kneib},
  {LeFevre}, {Lilly}, {Mobasher}, {Renzini}, {Rich}, {Sanders}, {Schinnerer},
  {Schminovich}, {Shopbell}, {Taniguchi}, \& {Tyson}}]{Scoville07}
{Scoville}, N., {et~al.} 2007, \apjs, 172, 1

\bibitem[{{Shivaei} {et~al.}(2016){Shivaei}, {Kriek}, {Reddy}, {Shapley},
  {Barro}, {Conroy}, {Coil}, {Freeman}, {Mobasher}, {Siana}, {Sanders},
  {Price}, {Azadi}, {Pasha}, \& {Inami}}]{Shivaei16}
{Shivaei}, I., {et~al.} 2016, \apjl, 820, L23

\bibitem[{Shivaei {et~al.}(2016)Shivaei, Reddy, Shapley, Siana, Kriek,
  Mobasher, Coil, Freeman, Sanders, Price, \& Azadi}]{Shivaei16b}
Shivaei, I., {et~al.} 2016, (arXiv:1609.04814)

\bibitem[{{Shivaei} {et~al.}(2015){Shivaei}, {Reddy}, {Shapley}, {Kriek},
  {Siana}, {Mobasher}, {Coil}, {Freeman}, {Sanders}, {Price}, {de Groot}, \&
  {Azadi}}]{Shivaei15}
{Shivaei}, I., {et~al.} 2015, \apj, 815, 98

\bibitem[{{Skelton} {et~al.}(2014){Skelton}, {Whitaker}, {Momcheva}, {Brammer},
  {van Dokkum}, {Labb\'{e}}, {Franx}, {van der Wel}, {Bezanson}, {Da Cunha},
  {Fumagalli}, {F\"{o}rster Schreiber}, {Kriek}, {Leja}, {Lundgren}, {Magee},
  {Marchesini}, {Maseda}, {Nelson}, {Oesch}, {Pacifici}, {Patel}, {Price},
  {Rix}, {Tal}, {Wake}, \& {Wuyts}}]{Skelton14}
{Skelton}, R.~E., {et~al.} 2014, \apjs, 214, 24

\bibitem[{{Somerville} {et~al.}(2008){Somerville}, {Hopkins}, {Cox},
  {Robertson}, \& {Hernquist}}]{Somerville08}
{Somerville}, R.~S., {Hopkins}, P.~F., {Cox}, T.~J., {Robertson}, B.~E., \&
  {Hernquist}, L. 2008, \mnras, 391, 481

\bibitem[{{Sparre} {et~al.}(2015){Sparre}, {Hayward}, {Springel},
  {Vogelsberger}, {Genel}, {Torrey}, {Nelson}, {Sijacki}, \&
  {Hernquist}}]{Sparre15}
{Sparre}, M., {et~al.} 2015, \mnras, 447, 3548

\bibitem[{{Speagle} {et~al.}(2014){Speagle}, {Steinhardt}, {Capak}, \&
  {Silverman}}]{Speagle14}
{Speagle}, J.~S., {Steinhardt}, C.~L., {Capak}, P.~L., \& {Silverman}, J.~D.
  2014, \apjs, 214, 15

\bibitem[{{Straatman} {et~al.}(2016){Straatman}, {Spitler}, {Quadri},
  {Labb{\'e}}, {Glazebrook}, {Persson}, {Papovich}, {Tran}, {Brammer},
  {Cowley}, {Tomczak}, {Nanayakkara}, {Alcorn}, {Allen}, {Broussard}, {van
  Dokkum}, {Forrest}, {van Houdt}, {Kacprzak}, {Kawinwanichakij}, {Kelson},
  {Lee}, {McCarthy}, {Mehrtens}, {Monson}, {Murphy}, {Rees}, {Tilvi}, \&
  {Whitaker}}]{Straatman16}
{Straatman}, C.~M.~S., {et~al.} 2016, \apj, 830, 51

\bibitem[{{Symeonidis} {et~al.}(2014){Symeonidis}, {Georgakakis}, {Page},
  {Bock}, {Bonzini}, {Buat}, {Farrah}, {Franceschini}, {Ibar}, {Lutz},
  {Magnelli}, {Magdis}, {Oliver}, {Pannella}, {Paolillo}, {Rosario},
  {Roseboom}, {Vaccari}, \& {Villforth}}]{Symeonidis14}
{Symeonidis}, M., {et~al.} 2014, \mnras, 443, 3728

\bibitem[{{Symeonidis} {et~al.}(2011){Symeonidis}, {Georgakakis}, {Seymour},
  {Auld}, {Bock}, {Brisbin}, {Buat}, {Burgarella}, {Chanial}, {Clements},
  {Cooray}, {Eales}, {Farrah}, {Franceschini}, {Glenn}, {Griffin},
  {Hatziminaoglou}, {Ibar}, {Ivison}, {Mortier}, {Oliver}, {Page},
  {Papageorgiou}, {Pearson}, {P\'{e}rez-Fournon}, {Pohlen}, {Rawlings},
  {Raymond}, {Rodighiero}, {Roseboom}, {Rowan-Robinson}, {Scott}, {Smith},
  {Tugwell}, {Vaccari}, {Vieira}, {Vigroux}, {Wang}, \&
  {Wright}}]{Symeonidis11}
---. 2011, \mnras, 417, 2239

\bibitem[{{Tal} {et~al.}(2014){Tal}, {Dekel}, {Oesch}, {Muzzin}, {Brammer},
  {van Dokkum}, {Franx}, {Illingworth}, {Leja}, {Magee}, {Marchesini},
  {Momcheva}, {Nelson}, {Patel}, {Quadri}, {Rix}, {Skelton}, {Wake}, \&
  {Whitaker}}]{Tal14}
{Tal}, T., {et~al.} 2014, \apj, 789, 164

\bibitem[{{Tasca} {et~al.}(2015){Tasca}, {Le F\`{e}vre}, {Hathi}, {Schaerer},
  {Ilbert}, {Zamorani}, {Lemaux}, {Cassata}, {Garilli}, {Le Brun}, {Maccagni},
  {Pentericci}, {Thomas}, {Vanzella}, {Zucca}, {Amorin}, {Bardelli},
  {Cassar\`{a}}, {Castellano}, {Cimatti}, {Cucciati}, {Durkalec}, {Fontana},
  {Giavalisco}, {Grazian}, {Paltani}, {Ribeiro}, {Scodeggio}, {Sommariva},
  {Talia}, {Tresse}, {Vergani}, {Capak}, {Charlot}, {Contini}, {de la Torre},
  {Dunlop}, {Fotopoulou}, {Koekemoer}, {L\'{o}pez-Sanjuan}, {Mellier}, {Pforr},
  {Salvato}, {Scoville}, {Taniguchi}, \& {Wang}}]{Tasca15}
{Tasca}, L.~A.~M., {et~al.} 2015, \aap, 581, A54

\bibitem[{{The LIGO Scientific Collaboration} {et~al.}(2016){The LIGO
  Scientific Collaboration}, {the Virgo Collaboration}, {Abbott}, {Abbott},
  {Abbott}, {Abernathy}, {Acernese}, {Ackley}, {Adams}, {Adams}, \&
  et~al.}]{LIGO16}
{The LIGO Scientific Collaboration} {et~al.} 2016, (arXiv:1606.04856)

\bibitem[{{Tomczak} {et~al.}(2016){Tomczak}, {Quadri}, {Tran}, {Labb\'{e}},
  {Straatman}, {Papovich}, {Glazebrook}, {Allen}, {Brammer}, {Cowley},
  {Dickinson}, {Elbaz}, {Inami}, {Kacprzak}, {Morrison}, {Nanayakkara},
  {Persson}, {Rees}, {Salmon}, {Schreiber}, {Spitler}, \&
  {Whitaker}}]{Tomczak16}
{Tomczak}, A.~R., {et~al.} 2016, \apj, 817, 118

\bibitem[{{Tremonti} {et~al.}(2004){Tremonti}, {Heckman}, {Kauffmann},
  {Brinchmann}, {Charlot}, {White}, {Seibert}, {Peng}, {Schlegel}, {Uomoto},
  {Fukugita}, \& {Brinkmann}}]{Tremonti04}
{Tremonti}, C.~A., {et~al.} 2004, \apj, 613, 898

\bibitem[{{Tzanavaris} {et~al.}(2013){Tzanavaris}, {Fragos}, {Tremmel},
  {Jenkins}, {Zezas}, {Lehmer}, {Hornschemeier}, {Kalogera}, {Ptak}, \&
  {Basu-Zych}}]{Tzanavaris13}
{Tzanavaris}, P., {et~al.} 2013, \apj, 774, 136

\bibitem[{{Viero} {et~al.}(2013){Viero}, {Moncelsi}, {Quadri}, {Arumugam},
  {Assef}, {B\'{e}thermin}, {Bock}, {Bridge}, {Casey}, {Conley}, {Cooray},
  {Farrah}, {Glenn}, {Heinis}, {Ibar}, {Ikarashi}, {Ivison}, {Kohno},
  {Marsden}, {Oliver}, {Roseboom}, {Schulz}, {Scott}, {Serra}, {Vaccari},
  {Vieira}, {Wang}, {Wardlow}, {Wilson}, {Yun}, \& {Zemcov}}]{Viero13}
{Viero}, M.~P., {et~al.} 2013, \apj, 779, 32

\bibitem[{{Whitaker} {et~al.}(2014){Whitaker}, {Franx}, {Leja}, {van Dokkum},
  {Henry}, {Skelton}, {Fumagalli}, {Momcheva}, {Brammer}, {Labb\'{e}},
  {Nelson}, \& {Rigby}}]{Whitaker14}
{Whitaker}, K.~E., {et~al.} 2014, \apj, 795, 104

\bibitem[{{Whitaker} {et~al.}(2011){Whitaker}, {Labb\'{e}}, {van Dokkum},
  {Brammer}, {Kriek}, {Marchesini}, {Quadri}, {Franx}, {Muzzin}, {Williams},
  {Bezanson}, {Illingworth}, {Lee}, {Lundgren}, {Nelson}, {Rudnick}, {Tal}, \&
  {Wake}}]{Whitaker11}
---. 2011, \apj, 735, 86

\bibitem[{{Whitaker} {et~al.}(2012){Whitaker}, {van Dokkum}, {Brammer}, \&
  {Franx}}]{Whitaker12}
{Whitaker}, K.~E., {van Dokkum}, P.~G., {Brammer}, G., \& {Franx}, M. 2012,
  \apjl, 754, L29

\bibitem[{{Williams} {et~al.}(2009){Williams}, {Quadri}, {Franx}, {van Dokkum},
  \& {Labb\'{e}}}]{Williams09}
{Williams}, R.~J., {Quadri}, R.~F., {Franx}, M., {van Dokkum}, P., \&
  {Labb\'{e}}, I. 2009, \apj, 691, 1879

\bibitem[{{Wuyts} {et~al.}(2011){Wuyts}, {F\"{o}rster Schreiber}, {Lutz},
  {Nordon}, {Berta}, {Altieri}, {Andreani}, {Aussel}, {Bongiovanni}, {Cepa},
  {Cimatti}, {Daddi}, {Elbaz}, {Genzel}, {Koekemoer}, {Magnelli}, {Maiolino},
  {McGrath}, {P\'{e}rez Garc\'{\i}a}, {Poglitsch}, {Popesso}, {Pozzi},
  {Sanchez-Portal}, {Sturm}, {Tacconi}, \& {Valtchanov}}]{Wuyts11}
{Wuyts}, S., {et~al.} 2011, \apj, 738, 106

\bibitem[{{Wuyts} {et~al.}(2008){Wuyts}, {Labb\'{e}}, {F\"{o}rster Schreiber},
  {Franx}, {Rudnick}, {Brammer}, \& {van Dokkum}}]{Wuyts08}
{Wuyts}, S., {Labb\'{e}}, I., {F\"{o}rster Schreiber}, N.~M., {Franx}, M.,
  {Rudnick}, G., {Brammer}, G.~B., \& {van Dokkum}, P.~G. 2008, \apj, 682, 985

\bibitem[{{Wuyts} {et~al.}(2007){Wuyts}, {Labb\'{e}}, {Franx}, {Rudnick}, {van
  Dokkum}, {Fazio}, {F\"{o}rster Schreiber}, {Huang}, {Moorwood}, {Rix},
  {R\"{o}ttgering}, \& {van der Werf}}]{Wuyts07}
{Wuyts}, S., {et~al.} 2007, \apj, 655, 51

\bibitem[{{Xue} {et~al.}(2011){Xue}, {Luo}, {Brandt}, {Bauer}, {Lehmer},
  {Broos}, {Schneider}, {Alexander}, {Brusa}, {Comastri}, {Fabian}, {Gilli},
  {Hasinger}, {Hornschemeier}, {Koekemoer}, {Liu}, {Mainieri}, {Paolillo},
  {Rafferty}, {Rosati}, {Shemmer}, {Silverman}, {Smail}, {Tozzi}, \&
  {Vignali}}]{Xue11}
{Xue}, Y.~Q., {et~al.} 2011, \apjs, 195, 10

\bibitem[{{Zhang} {et~al.}(2012){Zhang}, {Gilfanov}, \& {Bogd{\'a}n}}]{Zhang12}
{Zhang}, Z., {Gilfanov}, M., \& {Bogd{\'a}n}, {\'A}. 2012, \aap, 546, A36

\end{thebibliography}

}
\end{document}